\documentclass[10pt, oneside]{article}
\usepackage{subscript} 
\usepackage{hyperref}

\usepackage{enumerate}

\usepackage{easylist}  	
\usepackage{graphicx}			
\usepackage{amssymb}
\usepackage[usenames]{color}
\usepackage[dvipsnames]{xcolor}
\usepackage{pst-grad} 
\usepackage{url}
\usepackage{wrapfig}
\usepackage{float} 
\usepackage[abs]{overpic} 

\newenvironment{s-itemize}{
\begin{itemize}
  \setlength{\itemsep}{2pt}
  \setlength{\parskip}{0pt}
  \setlength{\parsep}{0pt}
}{\end{itemize}}

\newenvironment{s-alpha}{
\begin{enumerate}[(a)]
  \setlength{\itemsep}{2pt}
  \setlength{\parskip}{0pt}
  \setlength{\parsep}{0pt}
}{\end{enumerate}}

\newenvironment{s-enumerate}{
\begin{enumerate}[(1)]
  \setlength{\itemsep}{2pt}
  \setlength{\parskip}{0pt}
  \setlength{\parsep}{0pt}
}{\end{enumerate}}

\let\OLDthebibliography\thebibliography
\renewcommand\thebibliography[1]{
  \OLDthebibliography{#1}
  \setlength{\parskip}{.5ex}
  \setlength{\itemsep}{0pt plus 0.3ex}
}

\title{The cellular automaton pulsing model,\\experiments with DDLab%
\thanks{Presented at Summer Solstice 2018 Conference on Discrete Models of Complex Systems.}}%
\author{Andrew Wuensche%
\thanks{andy@ddlab.org,  \url{http://www.ddlab.org}}%
\hspace{2ex}{\it \small  Discrete Dynamics Lab.}\\
Edward Coxon%
\thanks{edward.coxon@act.gov.au}%
\hspace{2ex}{\it \small Dept. of Anaethesia and Pain Medicine,}\\
{\it \small The Canberra Hospital, ACT, Australia.}
}

\date{}	

\begin{document}

\maketitle

\begin{abstract}

  \noindent The cellular automaton (CA) pulsing model\cite{Wuensche-pulsingCA}
  described the surprising phenomenon of spontaneous, sustained and
  robust rhythmic oscillations, pulsing dynamics, when random wiring
  is applied to a 2D ``glider'' rule running in a 3-value totalistic
  CA. Case studies, pulsing measures, possible mechanisms, and
  implications for oscillatory networks in biology were presented.  In
  this paper we summarise the results, extend the entropy-density and
  density-return map plots to include a linked history, look at
  totalistic glider rules with neighborhoods of 3, 4 and 5, as
  well as 6 and 7 studied previously, introduce methods to
  automatically recognise the wavelength, and extend results for
  randomly asynchronous updating.  We show how the model is
  implemented in DDLab to validate results, output data, and allow
  experiments and research by others.

\end{abstract}

\begin{center}
  {\it keywords:  cellular automata, glider dynamics, random wiring, pulsing, bio-oscillations,
    emergence, chaos, complexity, strange attractor, heartbeat}
\end{center}

\vspace{2ex}
\section{Introduction}

\noindent The cellular automaton pulsing model (the CAP
model)\cite{Wuensche-pulsingCA} is a 2D cellular automaton (CA)
subject to a 3-value $k$-totalistic ``glider'' rule, where the local
wiring is randomised. Pulsing dynamics, sustained rhythmic
oscillations of density and entropy measures, emerge spontaneously and
almost inevitably. The characteristic wave-forms are robust and depend
on the specific glider rule applied. If the extent or reach of random
wiring is reduced, pulsing will eventually break down, suggesting a
threshold and phase transition.
\clearpage

The CAP model is a significant phenomena in its own right, posing
questions in CA theory on the mechanisms of glider
dynamics, the mechanisms of pulsing, and how the two are related.
It is also significant in the context of bio-oscillations
ubiquitous at many time/size scales in biology.  Currently there is no
satisfactory theory to explain essential oscillations in various
animal organs, for example heart beat, uterine contractions in
childbirth, and various rhythmic behaviours such as breathing controlled
by the central nervous system.  The CAP model provides a possible
oscillatory bio-mechanism based on long range signalling between
cells/neurons in excitable tissue, following the classical three state
dynamic, Firing, Refractory, and Ready-to-fire, and subject to ``glider
rule'' equivalent logic.

These concepts, ideas and results were defined and documented in
\cite{Wuensche-pulsingCA}.  In this paper we present a summary, extend
the entropy-density and density-return plots to include a linked
history, and look at glider rules with smaller
neighborhoods $k$, of 3, 4 and 5, as well as 6 and 7 studied
previously.  We introduce methods to automatically recognise and
measure wave-length/wave-height and output the data. 
We extend results for randomly asynchronous updating. 
We show how the model is implemented in DDLab to allow
validation of results and further experiments and research by
others, and include pre-assembled collections of glider rules.

\section{Glider dynamics in 2D CA}
\label{Glider dynamics in 2D CA}
\vspace{-3ex}

\begin{figure}[htb]
\textsf{\footnotesize
\begin{center}
\begin{minipage}[c]{.8\linewidth}
\begin{minipage}[c]{.12\linewidth} 
\includegraphics[width=1\linewidth,bb=135 172 206 241, clip=]{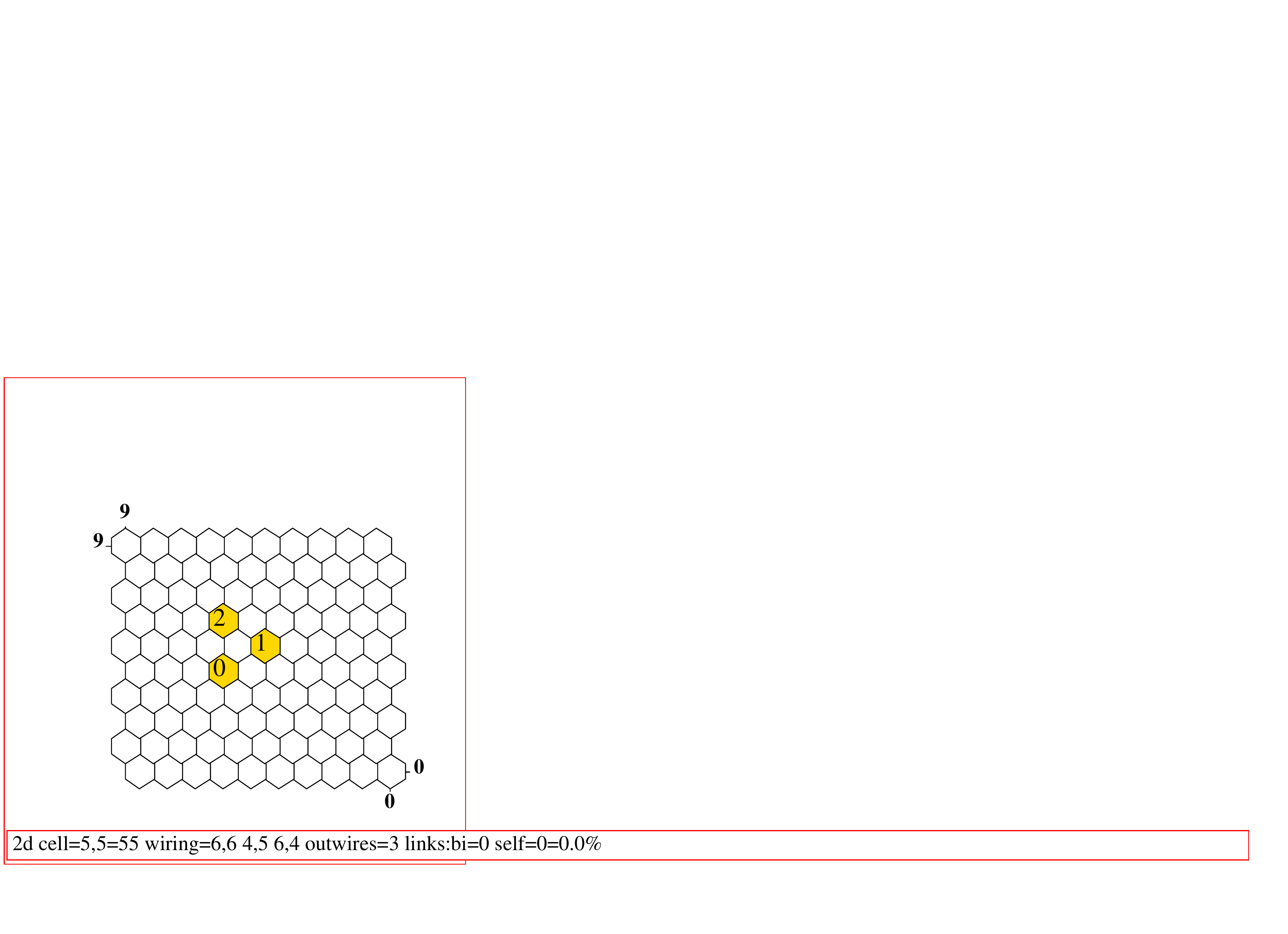}\\[-5ex]
\begin{center}(a)$k$=3\end{center}
\end{minipage}
\hfill
\begin{minipage}[c]{.12\linewidth}
\includegraphics[width=1\linewidth,bb=135 172 206 241, clip=]{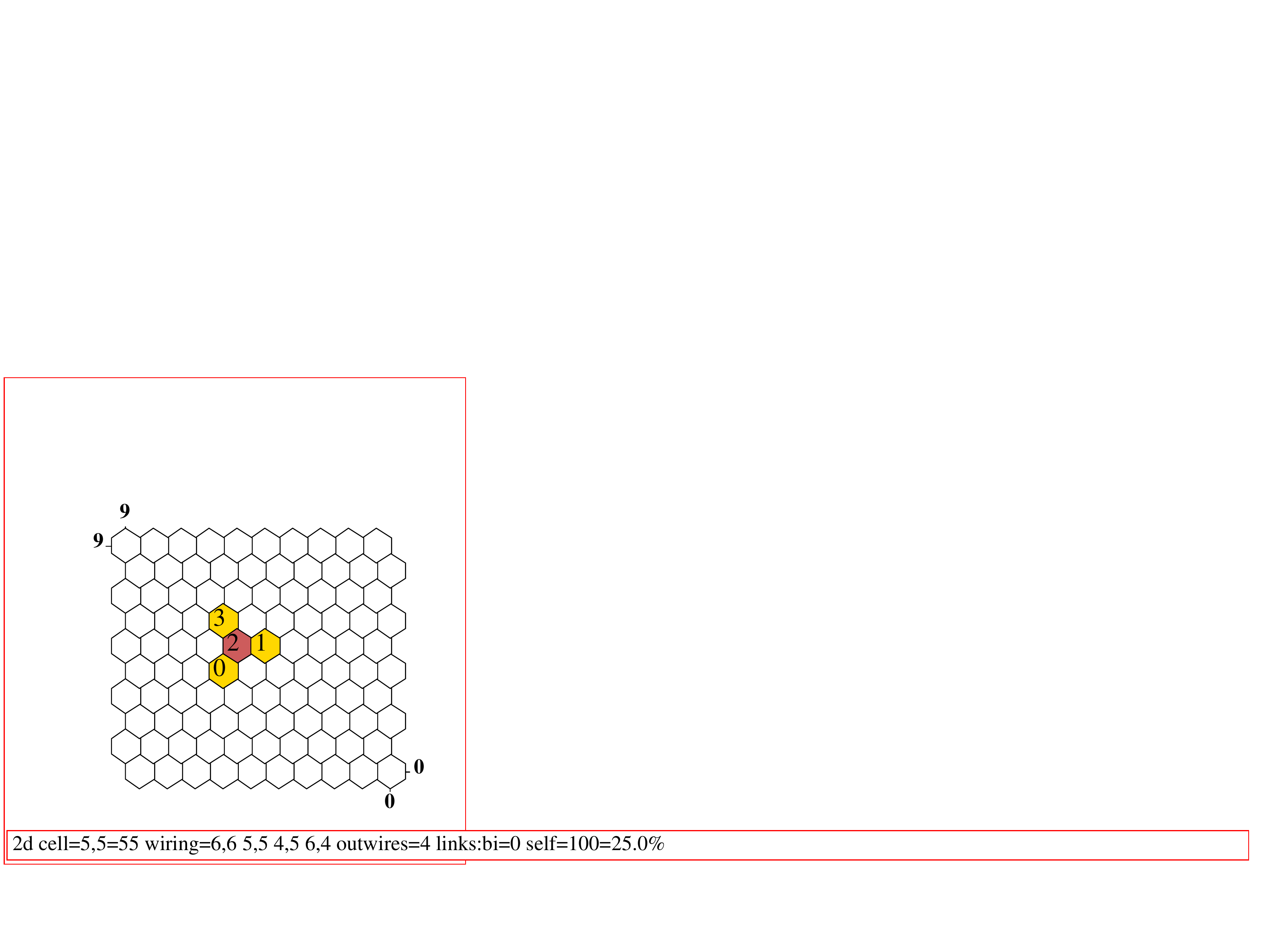}\\[-5ex]
\begin{center}(b)$k$=4t\end{center}
\end{minipage}
\hfill
\begin{minipage}[c]{.12\linewidth} 
\includegraphics[width=1\linewidth,bb=130 178 210 256, clip=]{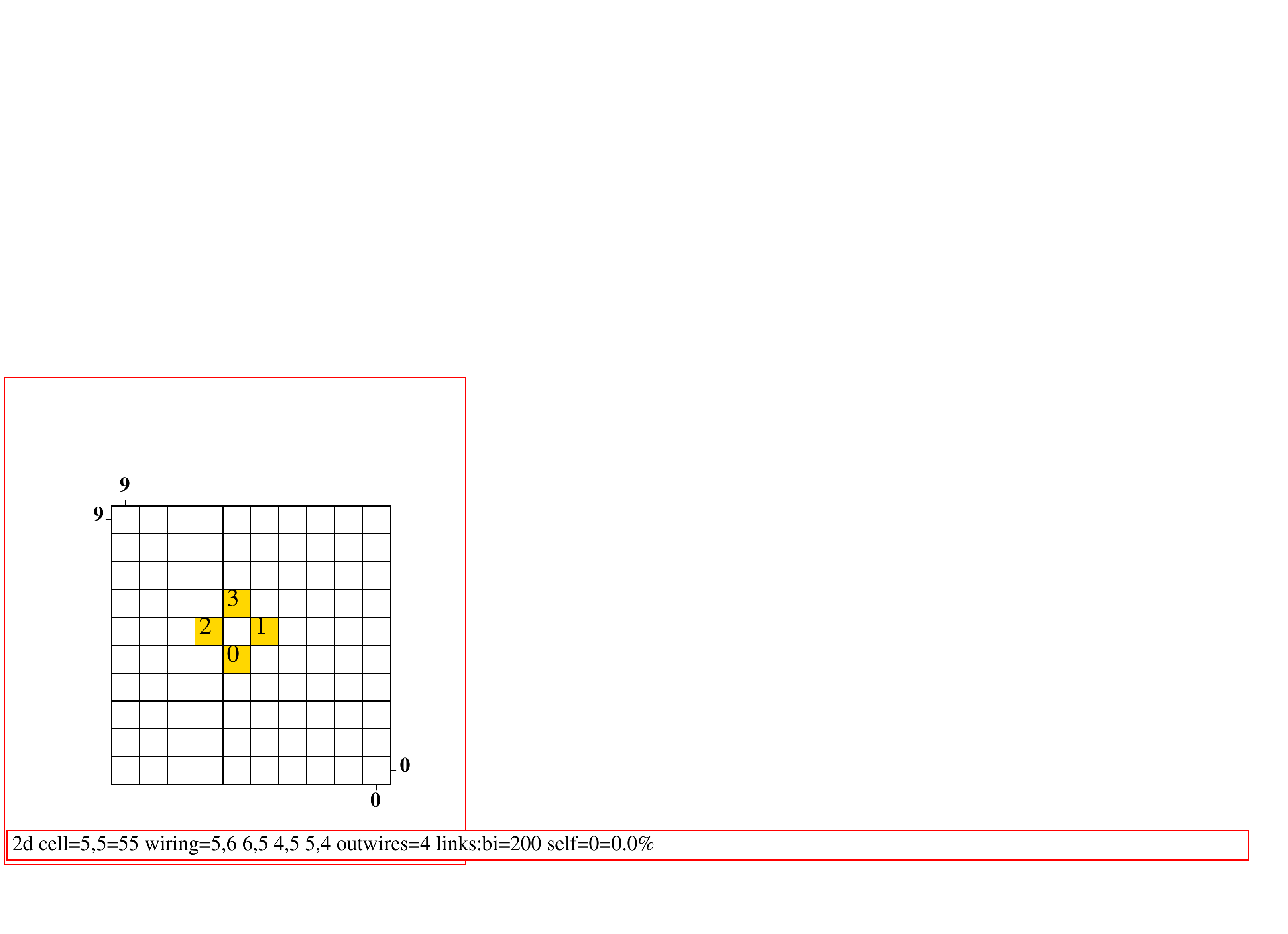}\\[-5ex]
\begin{center}(c)$k$=4s\end{center}
\end{minipage}
\hfill
\begin{minipage}[c]{.12\linewidth} 
\includegraphics[width=1\linewidth,bb=130 178 210 256, clip=]{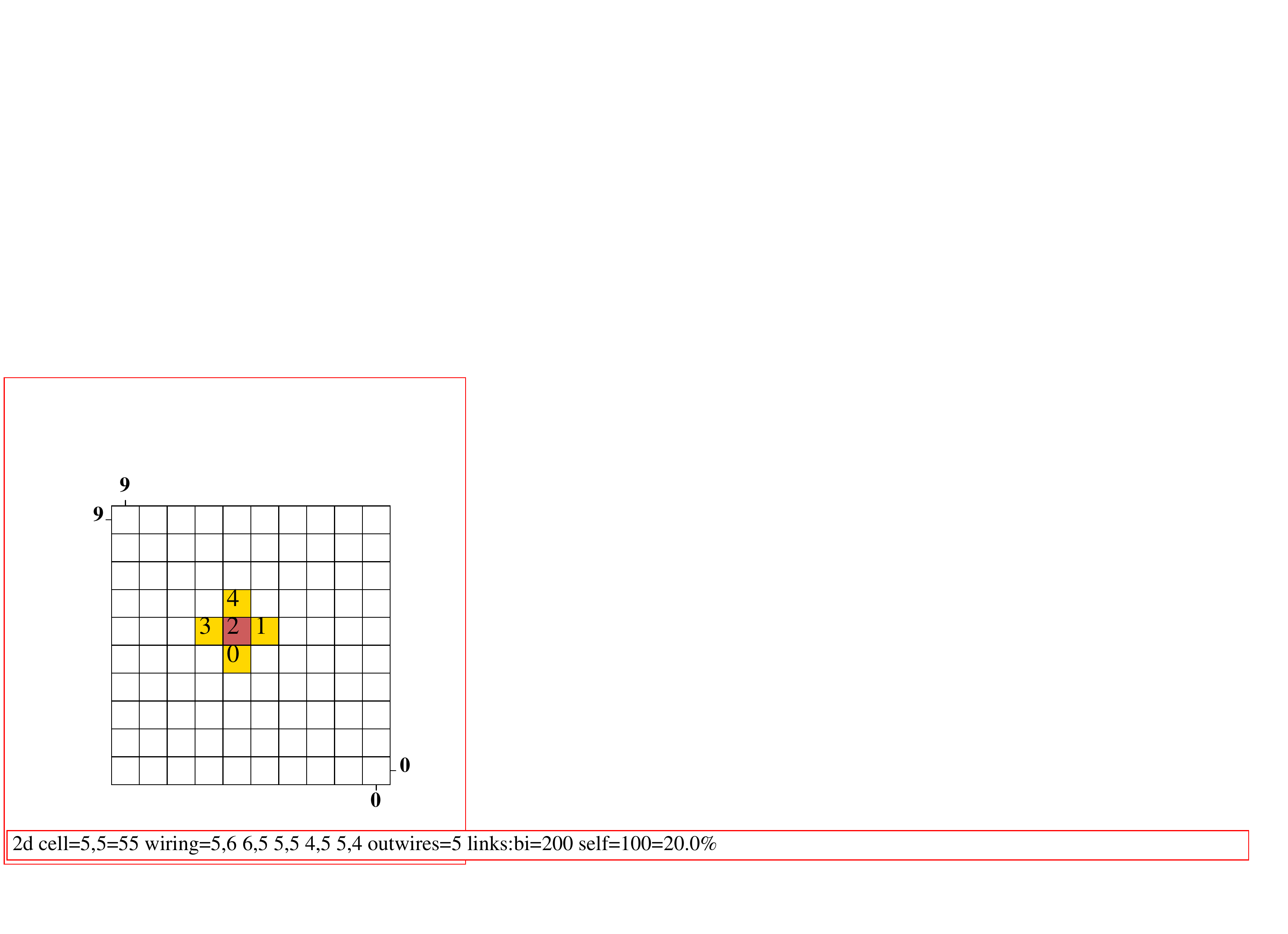}\\[-5ex]
\begin{center}(d)$k$=5\end{center}
\end{minipage}
\hfill
\begin{minipage}[c]{.12\linewidth} 
\includegraphics[width=1\linewidth,bb=135 172 206 241, clip=]{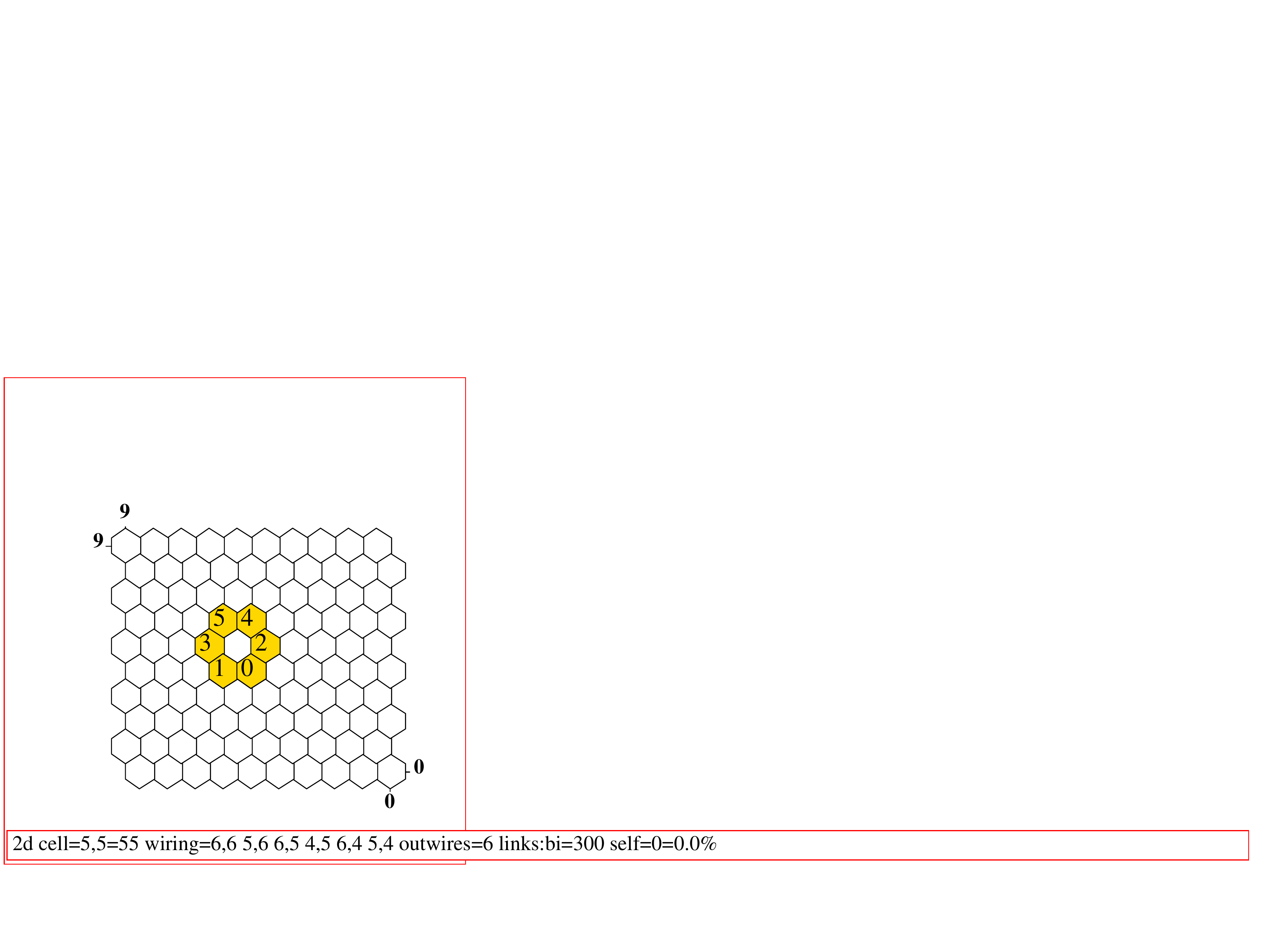}\\[-5ex]
\begin{center}(e)$k$=6\end{center}
\end{minipage}
\hfill
\begin{minipage}[c]{.12\linewidth} 
\includegraphics[width=1\linewidth,bb=135 172 206 241, clip=]{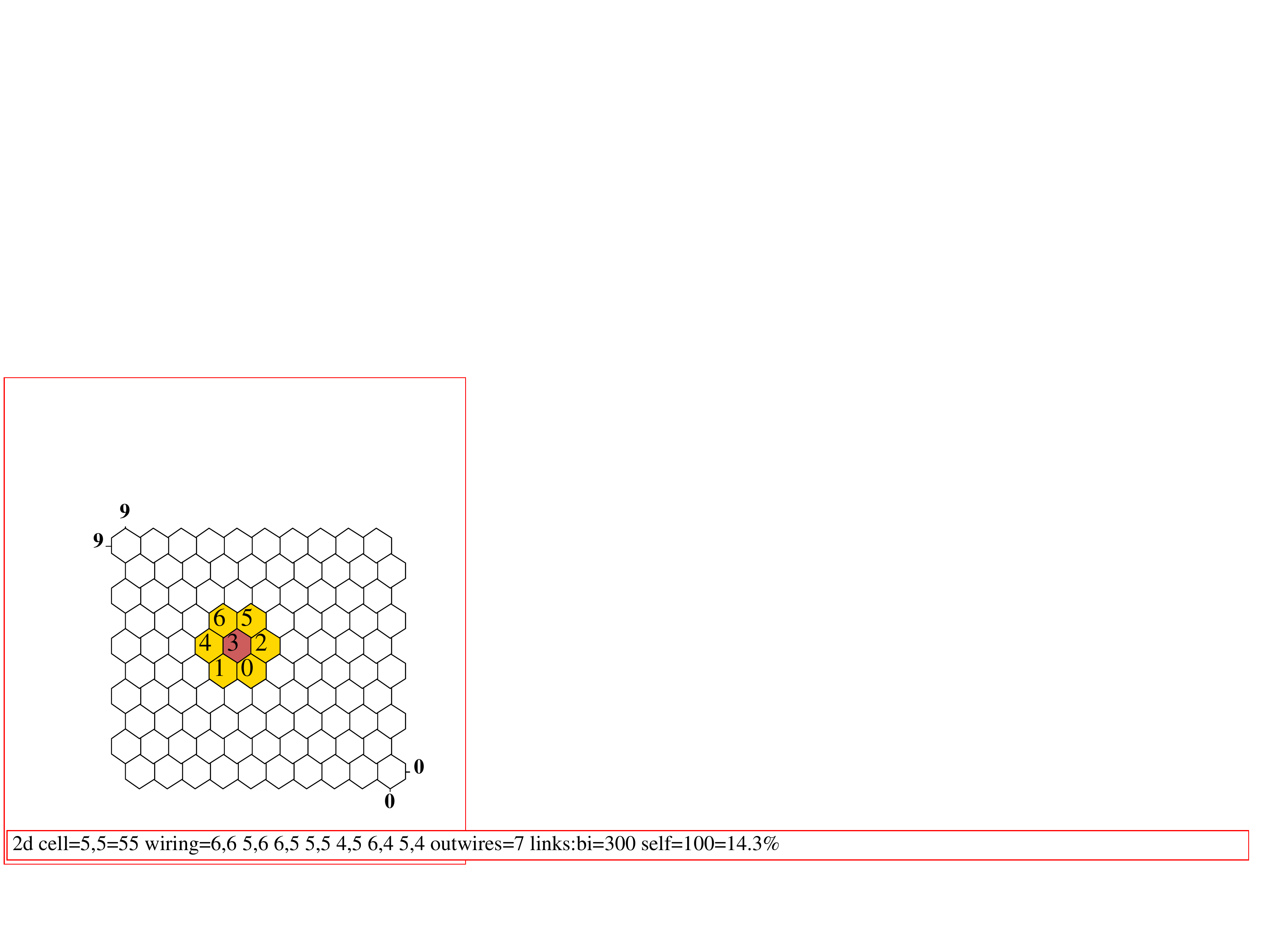}\\[-5ex]
\begin{center}(f)$k$=7\end{center}
\end{minipage}
\end{minipage}
\end{center}
}
\vspace{-2ex}
\caption[2D Neighborhood templates]
{\textsf{2D neighborhood templates ($k$=3 to 7) as defined (and numbered) in DDLab, setting
the lattice geometry, both hexagonal and square. 
The target cell is central even if not part of the template.}}
\label{2D neighborhood templates}
\end{figure}

\begin{figure}[htb]
\textsf{\footnotesize
\begin{center}
\begin{minipage}[c]{1\linewidth}
\begin{minipage}[c]{.3\linewidth}
\includegraphics[width=1\linewidth,bb=276 136 405 284, clip=]{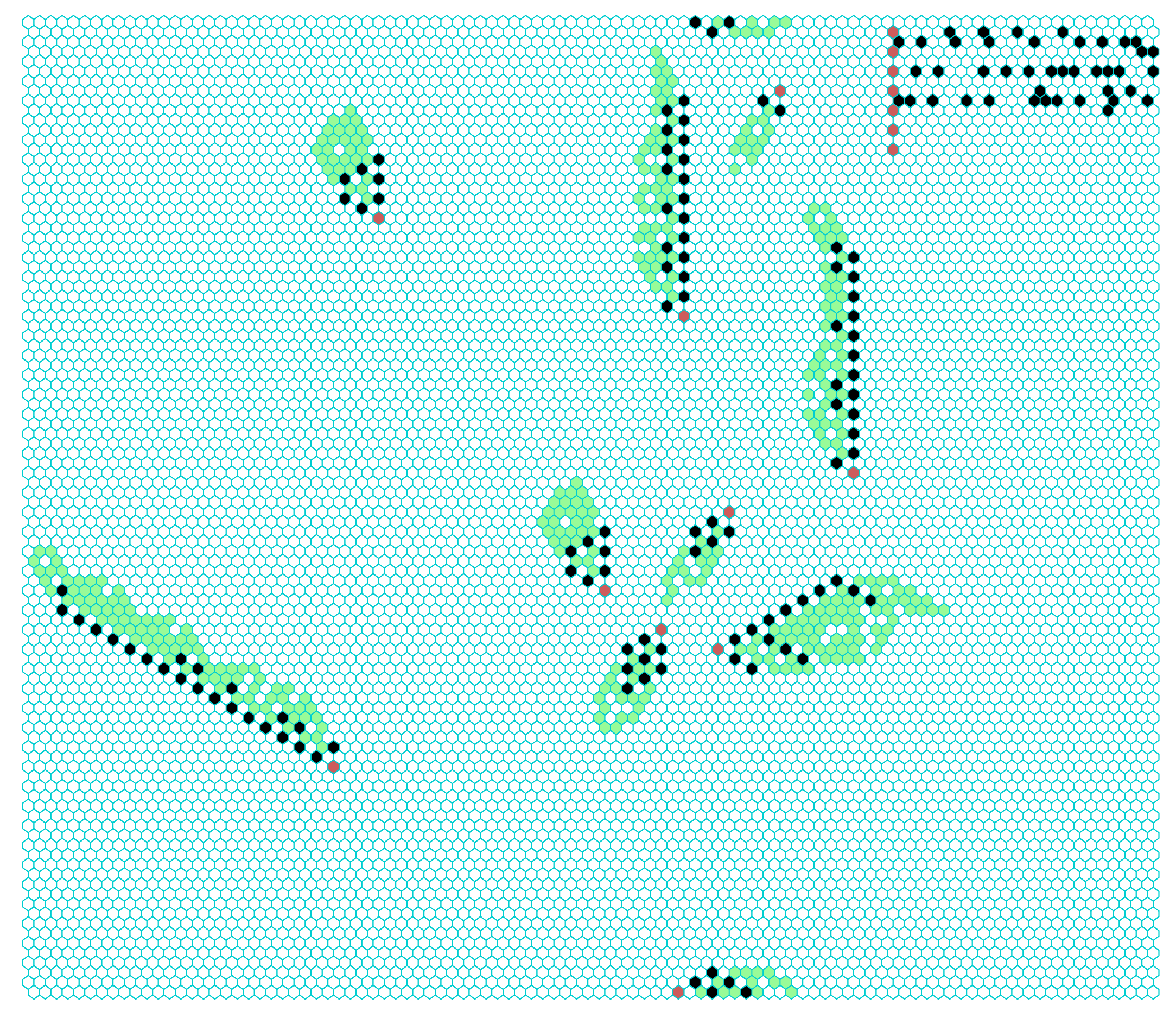}\\[-5ex]
\begin{center}(a) v3k3x1.vco, g1\\(hex)00a864\end{center}
\end{minipage}
\hfill
\begin{minipage}[c]{.3\linewidth}
\includegraphics[width=1\linewidth,bb=240 13 345 135, clip=]{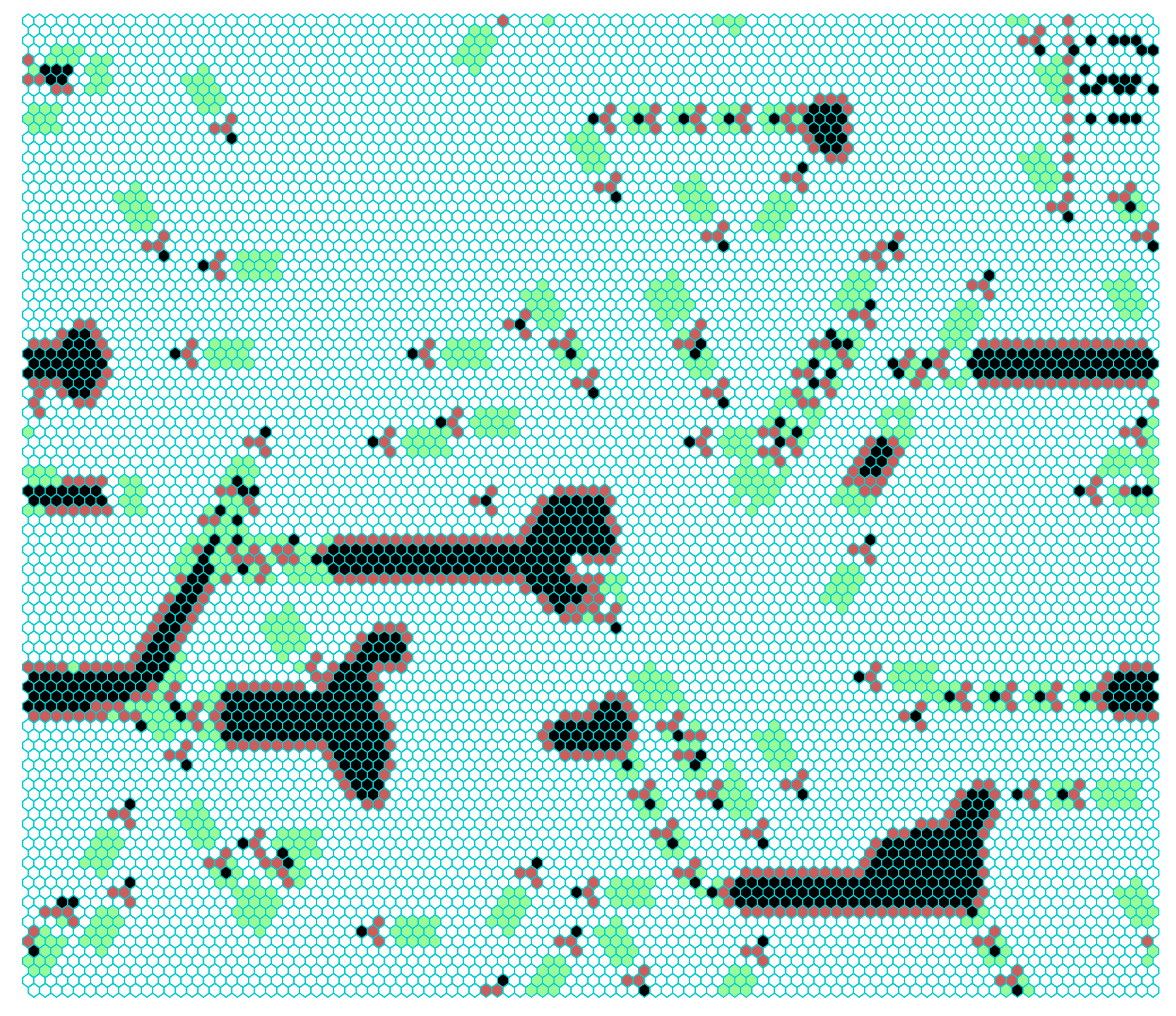}\\[-5ex]
\begin{center}(b) v3k4t1.vco, g1\\(hex)2a945900\end{center}
\end{minipage}
\hfill
\begin{minipage}[c]{.3\linewidth}
\includegraphics[width=1\linewidth,bb=314 99 462 267, clip=]{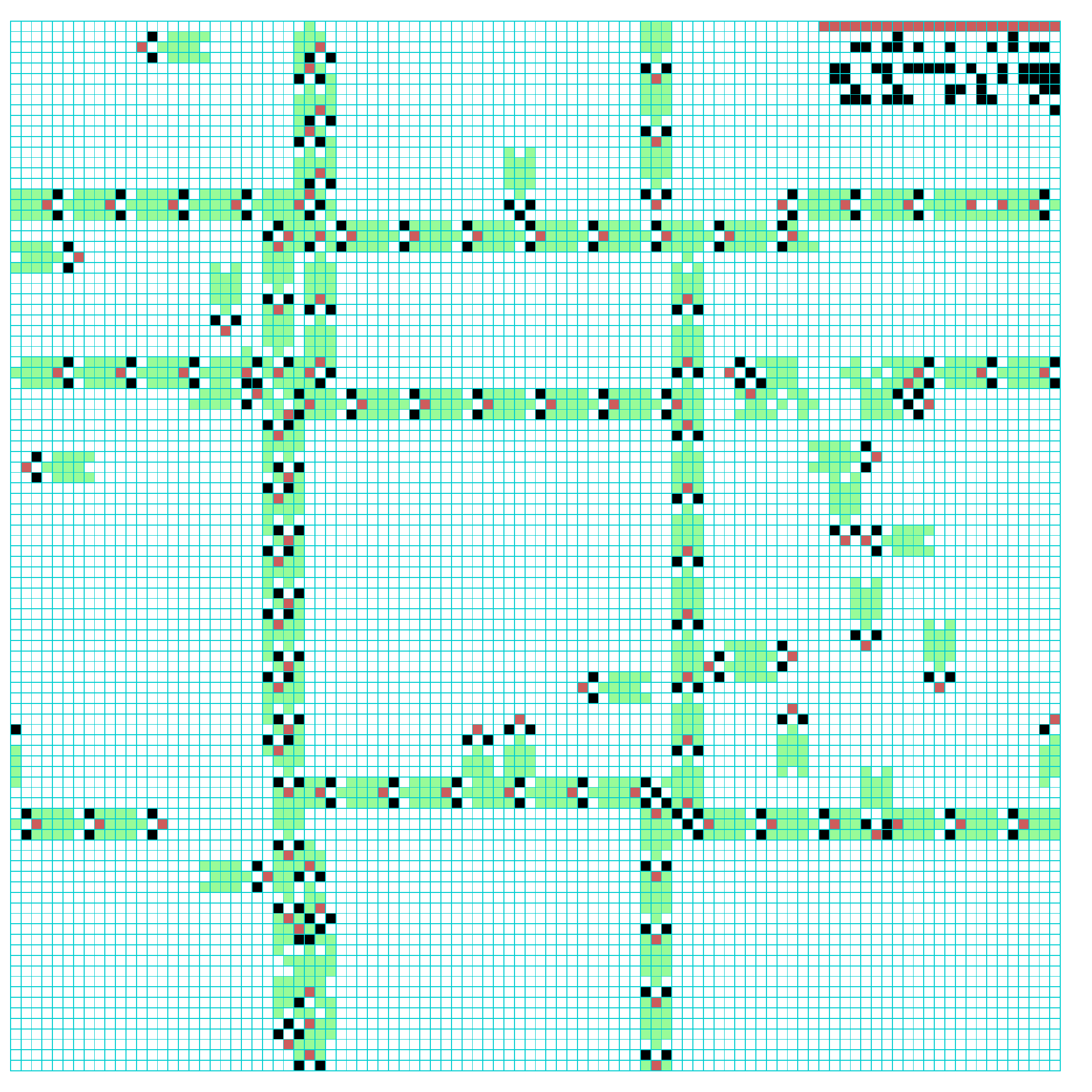}\\[-5ex]
\begin{center}(c) v3k4x1.vco\\(hex)2282a1a4\end{center}
\end{minipage}\\[1ex]
\begin{minipage}[c]{.3\linewidth}
\includegraphics[width=1\linewidth,bb=140 224 268 373, clip=]{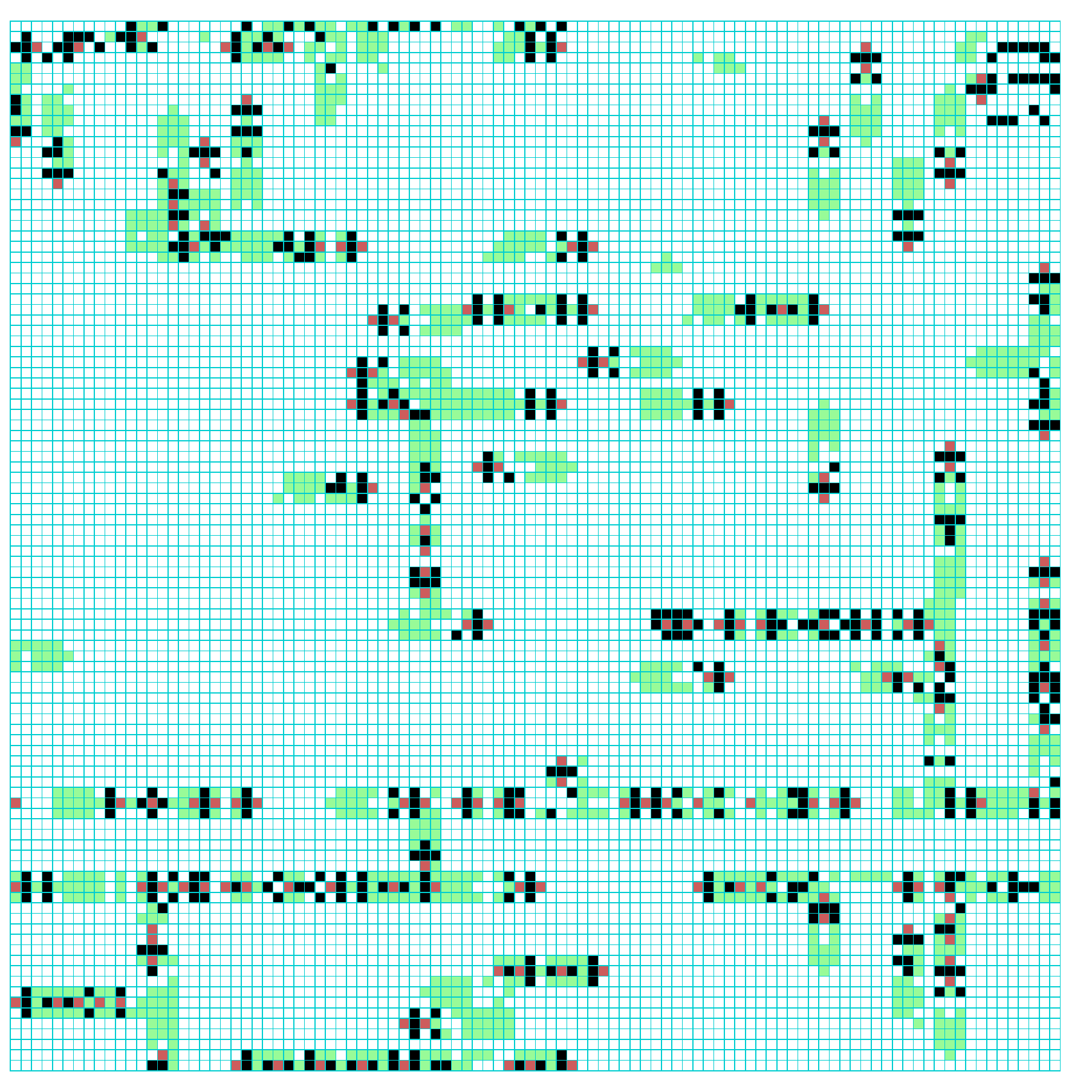}\\[-5ex]
\begin{center}(d) v3k5x1.vco, g1\\(hex)004a8a2a8254\\ \textcolor{white}{xx} \end{center}
\end{minipage}
\hfill
\begin{minipage}[c]{.3\linewidth}
\includegraphics[width=1\linewidth,bb=262 22 392 174, clip=]{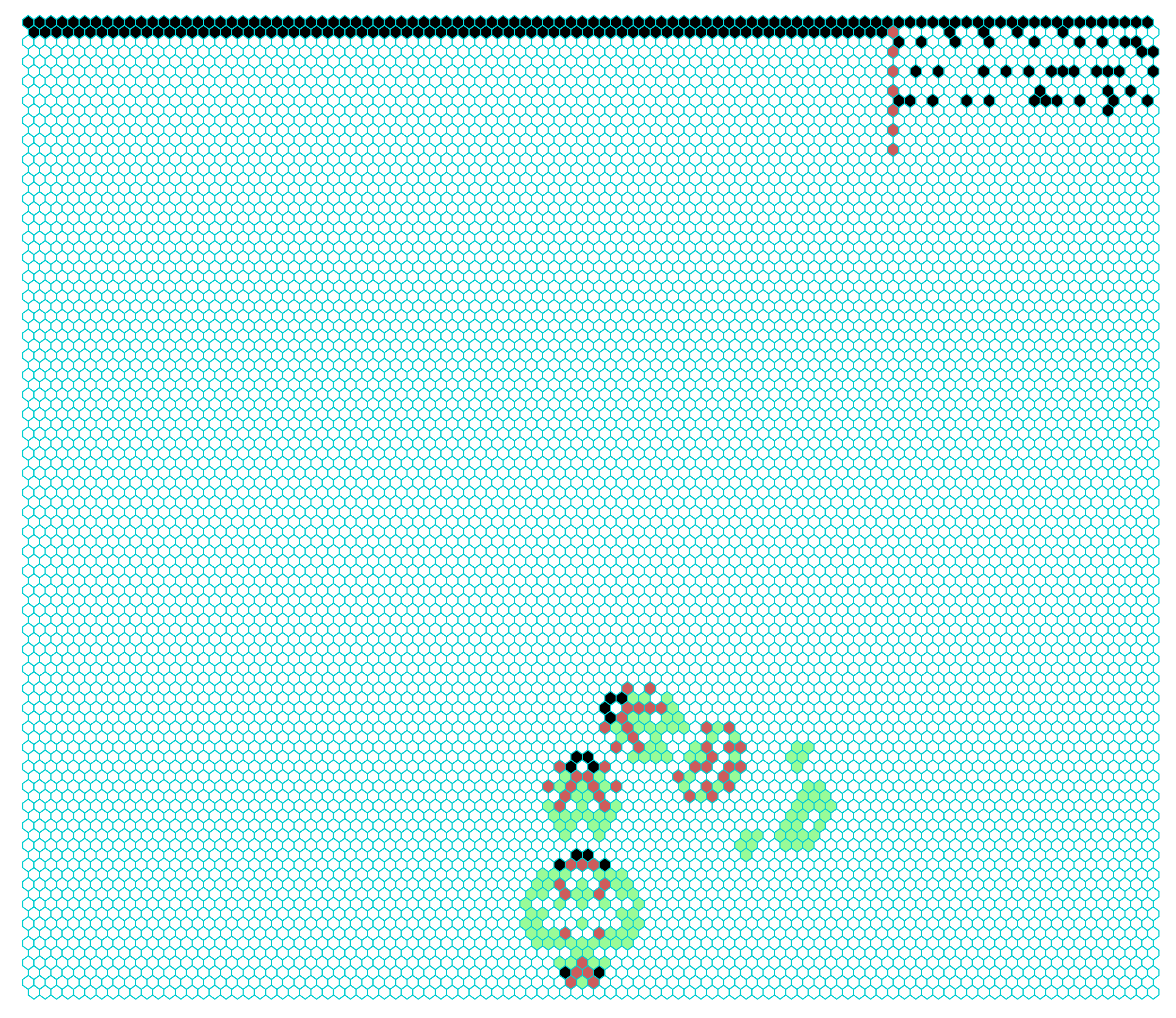}\\[-5ex]
\begin{center}(e) v3k6n6.vco, g16\\(hex)01059059560040\\ \textcolor{white}{xx} \end{center}
\end{minipage}
\hfill
\begin{minipage}[c]{.3\linewidth}
\includegraphics[width=1\linewidth,bb=262 231 380 369, clip=]{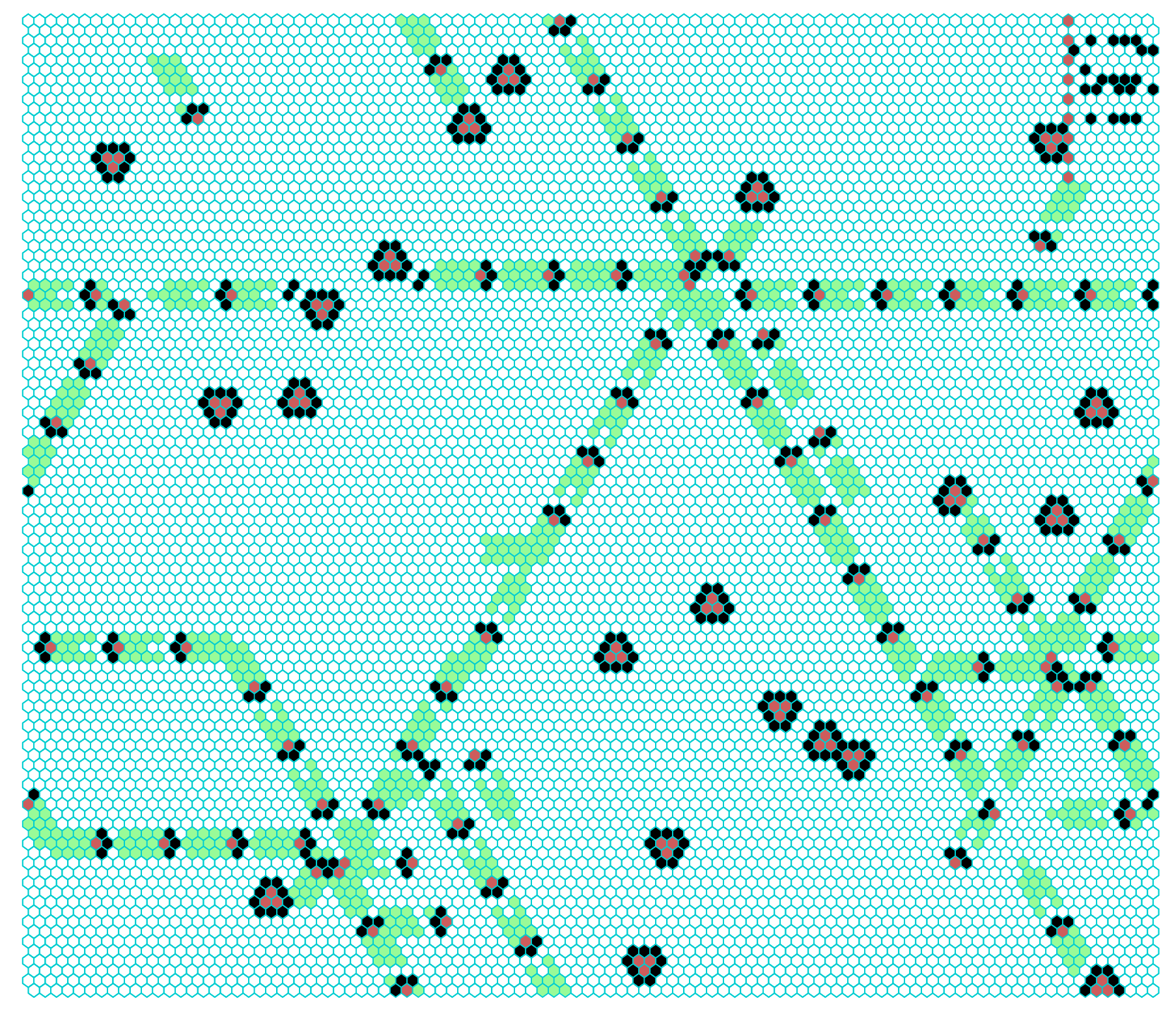}\\[-5ex]
\begin{center}(f) v3k7w1.vco, g1\\(hex)020609a2982a68aa64\\
The Spiral rule\cite{Wuensche&Adamatzky2006}\end{center}
\end{minipage}
\end{minipage}
\end{center}
}
\vspace{-2ex}
\caption[Glider examples]
{\textsf{Examples of glider dynamics for 2D neighborhood 
templates ($k$=3 to 7) in figure~\ref{2D neighborhood templates}.
Cell values: 0=white, 1=red, 2=black.
Green trails of 5 time-steps indicate glider velocity.
Examples b, c, e, and f include glider-guns. The rules can be loaded in DDLab
by their filenames, in hexadecimal, or from the rule collections index g(x).}}
\label{Glider examples}
\vspace{-1ex}
\end{figure}

\noindent The lattice geometry of a 2D CA depends on its neighbourhood template,
and we present templates relevant to this paper in 
figure~\ref{2D neighborhood templates}. Collections of glider rules have been
assembled based on these templates.

Each cell in the lattice updates its value
synchronously\footnote{Although synchronous updating is a necessary
  condition for gliders to emerge, pulsing in the CAP model persists
  for asynchronous and noisy updating\cite{Wuensche-pulsingCA}.}  according to a
homogeneous 3-value $k$-totalistic rule. This determines the
dynamics which can be seen as successive pattern images in the same
way as a series of still images make a movie. Most rules result in
disorder, but we are interested in complex rules characterised by
identifiable mobile features, in particular ``gliders'' or mobile
particles consisting of compact cell-value assemblies moving through
the lattice with a given velocity, comprising a head and tail, and
interacting by collisions with other gliders or stable particles
as in the examples in figure~\ref{Glider examples}.

In DDLab, collections of glider rules are provided,
extracted from automatic samples of complex rules --- not all complex
rules support gliders and pulsing. The older collections for $k$= 6
and 7 relating to \cite{Wuensche-pulsingCA} include complex rules as
well as pulsing rules, whereas for the $k$= 3, 4, and 5 collections,
only pulsing rules have been included. For $k$=3, pulsing results from 
mobile intersecting linear structures, rather than classic
gliders which are less frequent.

While assembling these collections we can confirm that gliders almost inevitably 
imply pulsing, and their absence imply non-pulsing. We should however note
that we have observed a few very rare exceptions. 

\section{Random wiring}
\label{Random wiring section}

\begin{figure}[htb]
\textsf{\small
\begin{center}
\begin{minipage}[c]{.95\linewidth}
\begin{minipage}[c]{.48\linewidth}
\includegraphics[width=1\linewidth,bb=67 103 395 392, clip=]{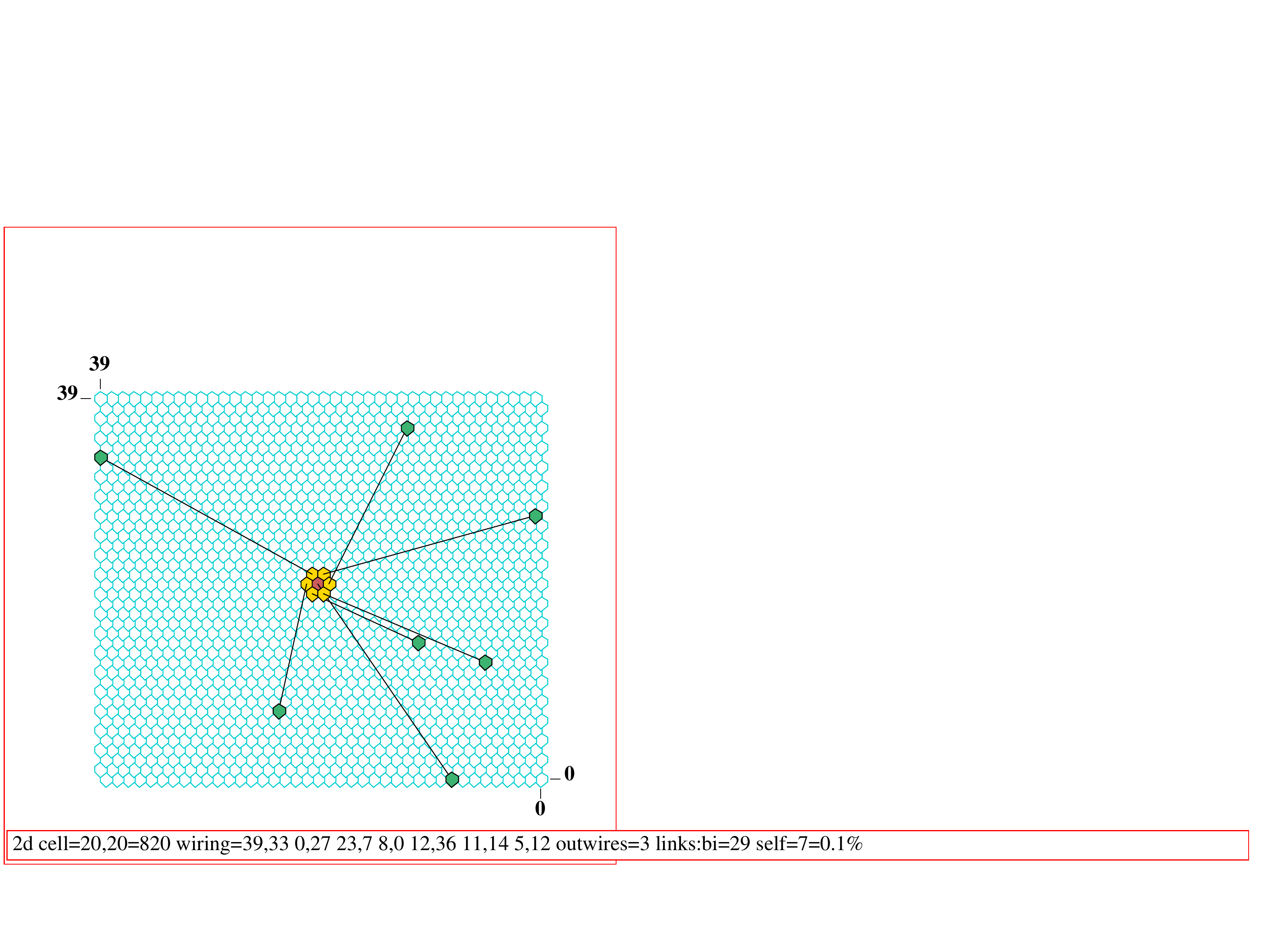}\\[-5ex]
\begin{center}(a) unrestricted\end{center}
\end{minipage}
\hfill
\begin{minipage}[c]{.48\linewidth}
\includegraphics[width=1\linewidth,bb=67 103 395 392, clip=]{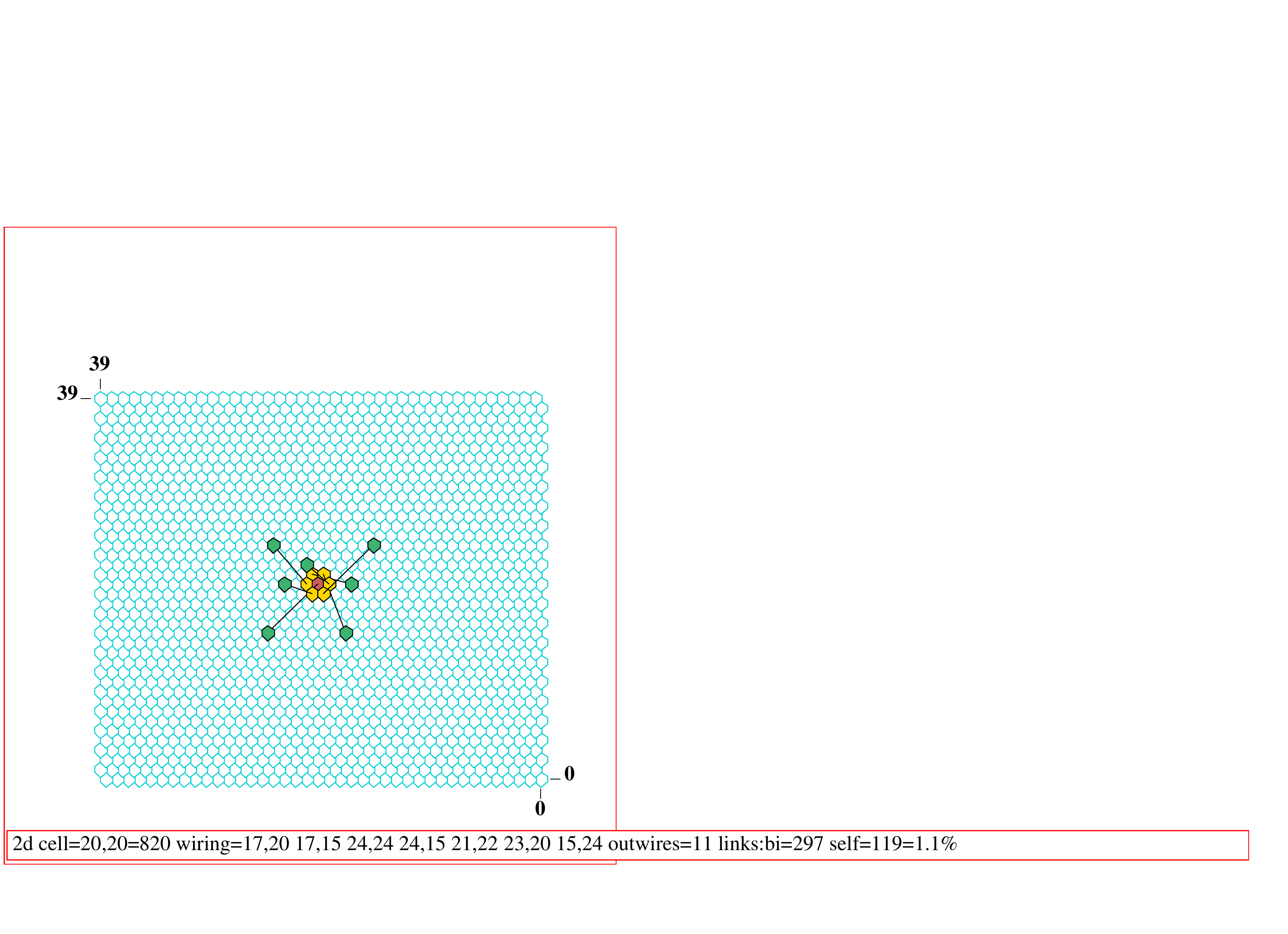}\\[-5ex]
\begin{center}(b) confined in a local zone\end{center}
\end{minipage}\\[1ex]
\begin{minipage}[c]{.48\linewidth}
\includegraphics[width=1\linewidth,bb=67 103 395 392, clip=]{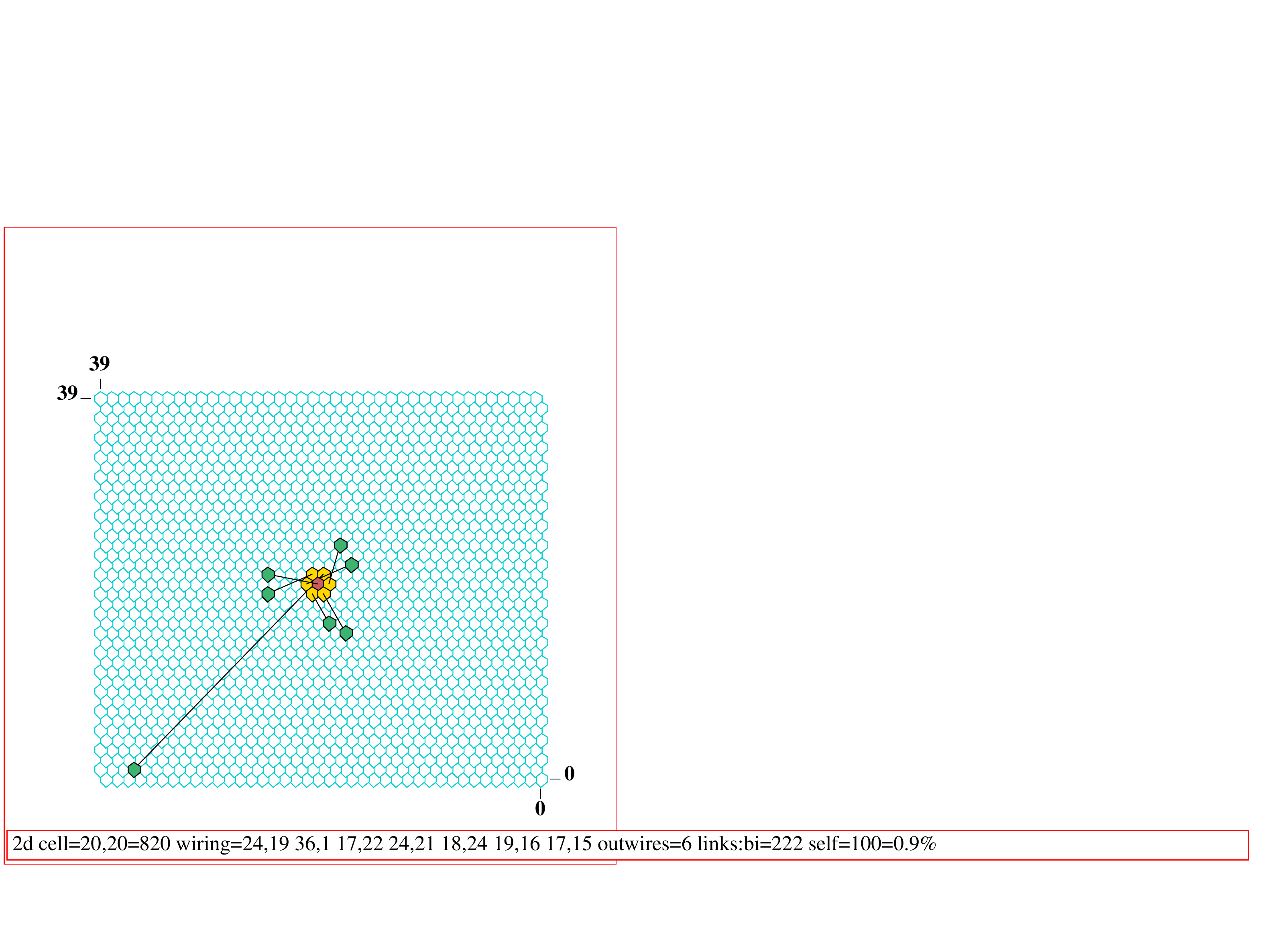}\\[-5ex]
\begin{center}(c) confined but one wire freed\end{center}
\end{minipage}
\hfill
\begin{minipage}[c]{.48\linewidth}
\includegraphics[width=1\linewidth,bb=67 103 395 392, clip=]{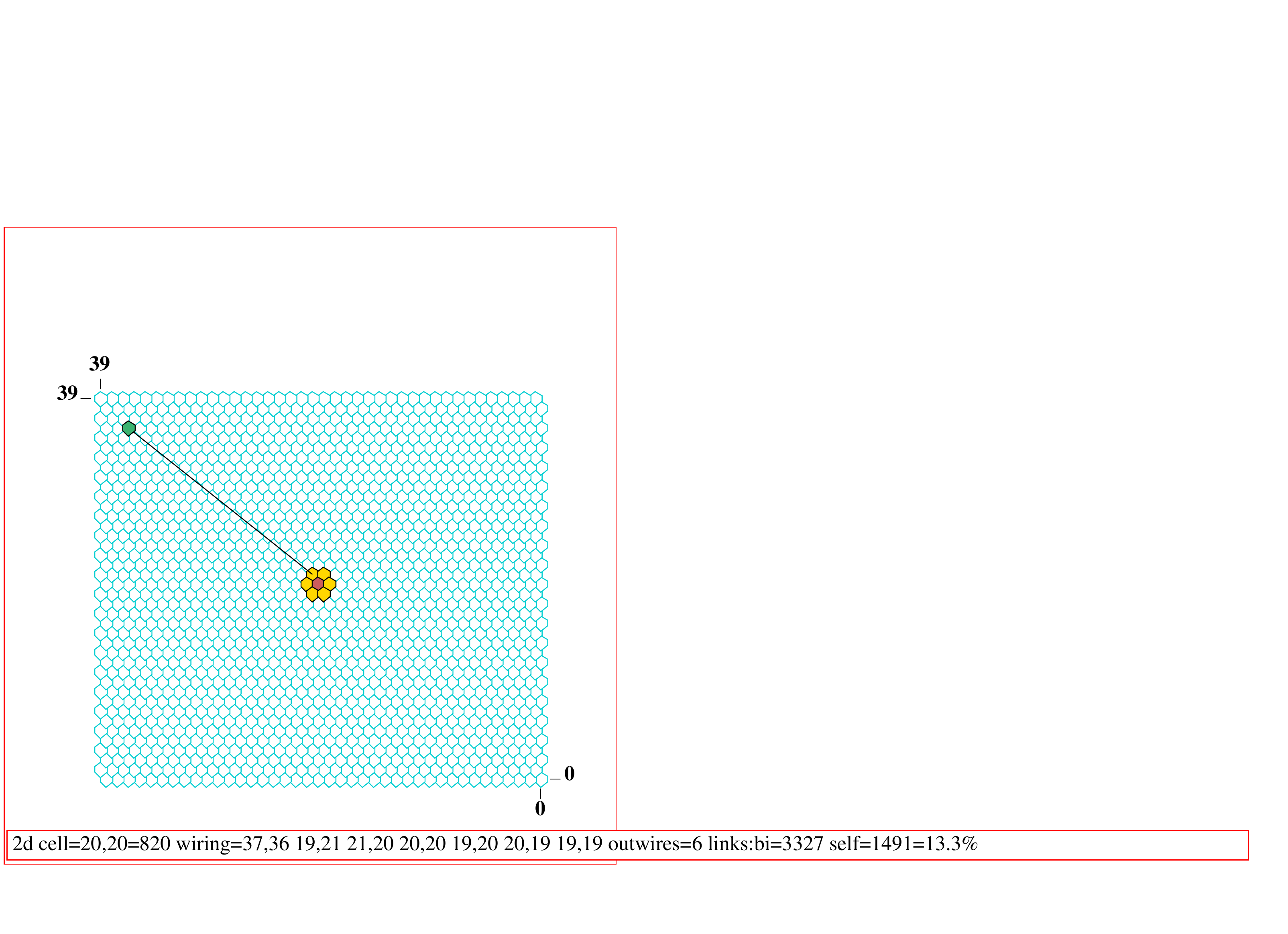}\\[-5ex]
\begin{center}(d) CA but one wire freed\end{center}
\end{minipage}
\end{minipage}
\end{center}
}
\vspace{-2ex}
\caption[Random wiring]
{\textsf{Examples of random wiring in a 40$\times$40 2D lattice. A bigger lattice would
normally be required for robust pulsing, to avoid reaching a uniform value attractor\cite{Wuensche92}.}}
\label{Random wiring}
\end{figure}

\noindent Unrestricted, unbiased, random wiring follows the same connection
approach as Kauffmans's ``Random Boolean Networks''\cite{kauffman69}.
For each target cell, we take $k$ cells at random anywhere in the
lattice and ``wire'' them to distinct cells in the pseudo-neighborhood
template --- ``pseudo'' because the actual template values are
replaced by the values of the random cells, as in 
figure~\ref{Random wiring}(a).  Each target cell is assigned its own random wiring.
We have also studied random wiring with three types of bias
resulting in degrees of degraded pulsing and waveform
signatures\cite{Wuensche-pulsingCA}.
A summary of the consequences relating to 
figure~\ref{Random wiring}(a,b,c,d) are listed below,

\begin{s-alpha}

\item Unbiased random wiring gives the strongest and most robust
  pulsing in the CAP model.

\item If random wiring is confined within a local zone, as the size of
  the zone is reduced, at some stage global pulsing will break down
  into mobile patches of density which may take the form of spiral
  waves.  The threshold properties of this phase transition requires
  further investigation. The waveform signatures are
  recognisable as they degrade.

\item If one wire is ``freed'' from a small local zone, robust pulsing
  with a characteristic waveform is partially restored, and this is
  reinforced as a higher proportion of wires are freed.

\item Starting with a conventional CA, freeing one or more wires
  randomly from each neighborhood template results in the onset of
  pulsing by degrees, also conforming to a characteristic waveform.

\end{s-alpha} 

In these experiments, re-randomising the wiring at each time-step,
as in Derrida's annealed model\cite{Derrida}, makes no significant
difference to the waveform.

A single key press in DDLab enables switching between CA and any
type of preset random wiring, or between stable random wiring and
re-randomising the wiring at each time-step while maintaining the preset bias.

\section{Why 3-value $k$-totalistic rules?}
\label{Why 3-value $k$-totalistic rules?}

\noindent We restrict our investigation to the subset of 3-value $k$-totalistic rules 
for the following reasons:

\begin{s-itemize}
\item The discovery of pulsing in the CAP model, and that 
no pulsing is evident in an equivalent 2-value system.

\item Compared to a general CA, 
the $k$-totalistic rule-table is relatively short\cite[\#13.6.1]{Wuensche2016}
and thus tractable for displaying aspects of the pulsing waveform. 

\item The dynamics are isotropic so closer to nature --- the same output for
neighborhood template rotation or reflection, though $k$-totalistic rules are restricted
beyond isotropy giving smaller rule-spaces than just isotropic rules.

\item The rules can be reinterpreted as reaction-diffusion systems with
inhibitor-activator reagents in a chemical 
medium\cite{Adamatzky&Wuensche&Cosello2006,Adamatzky&Wuensche2006,Wuensche&Adamatzky2006}, 
where the three CA values are seen as: Activator, Inhibitor, and Substrate.

\item The availability of short-lists of glider rules, extracted from large samples of
complex rules that are found (and sorted automatically) by the variability of
input-entropy\cite{Wuensche99,Wuensche05,Wuensche2016}.

\item The CAP model can be appled to bio-oscillations in excitable tissue
according to classical 3-state neuronal dynamics: Firing, Refractory, and Ready to Fire. 

\end{s-itemize}

\section{Definition of 3-value $k$-totalistic rules}
\label{Definition of 3-value $k$-totalistic rules}

\noindent The properties and definition of 3-value $k$-totalistic rules are summarised as follows:

\begin{s-itemize}
\item  The target cell at time step $t$ depends on
the combination of $k$ totals, or frequencies, 
of the values in the neighborhood template at $t$-1.

\item  Each combination of totals make up the rule-table (named ``kcode''), 
for example the 3-value ($v$=3) $k$=5 rule v3k5x1.vco
in figure~\ref{Glider examples}(d),

{\baselineskip2ex \scriptsize
\begin{verbatim}
               20...................0 <--kcode index
                |                   |
           > 2: 544333222211111000000  < frequency strings       5    0
v=3 values > 1: 010210321043210543210  < of 2s, 1s, 0s,     from 0 to 0 
           > 0: 001012012301234012345  < shown vertically        0    5
                |||||||||||||||||||||
                010222022022220021110  <--rule table (kcode), outputs [0,1,2]
\end{verbatim}}

In DDLab, kcode can be expressed in hexadecimal for compactness, in this case 
004a8a2a8254, also shown in figure~\ref{Glider examples}(d).

\item kcode size $S = (v + k - 1)! / (k! \times (v-1)!)$,
which increases arithmetically; kcode-space=$v^S$. A general rule-table
(rcode) has $v^k$ entries increasing exponentially; rcode-space=$v^{v^k}$.
For $v$=3 and $k$=3 to 7, the size of the kcode and rcode strings are as follows,

\footnotesize{
\begin{center}
\begin{tabular}[t]{c|r|r|r|r|r}
$k$& 3&4&5&6&7\\
\hline
vcode &10 &15 &21 &28 &36\\
\hline
rcode &27 &64 &125 &216 &343
\end{tabular}
\end{center}}
\end{s-itemize}

\section{Signs of pulsing}
\label{Signs of pulsing}

\begin{figure}[htb]
\textsf{\small
\begin{center}
\begin{minipage}[c]{1\linewidth} 
\begin{minipage}[c]{.13\linewidth}
\includegraphics[width=1\linewidth]{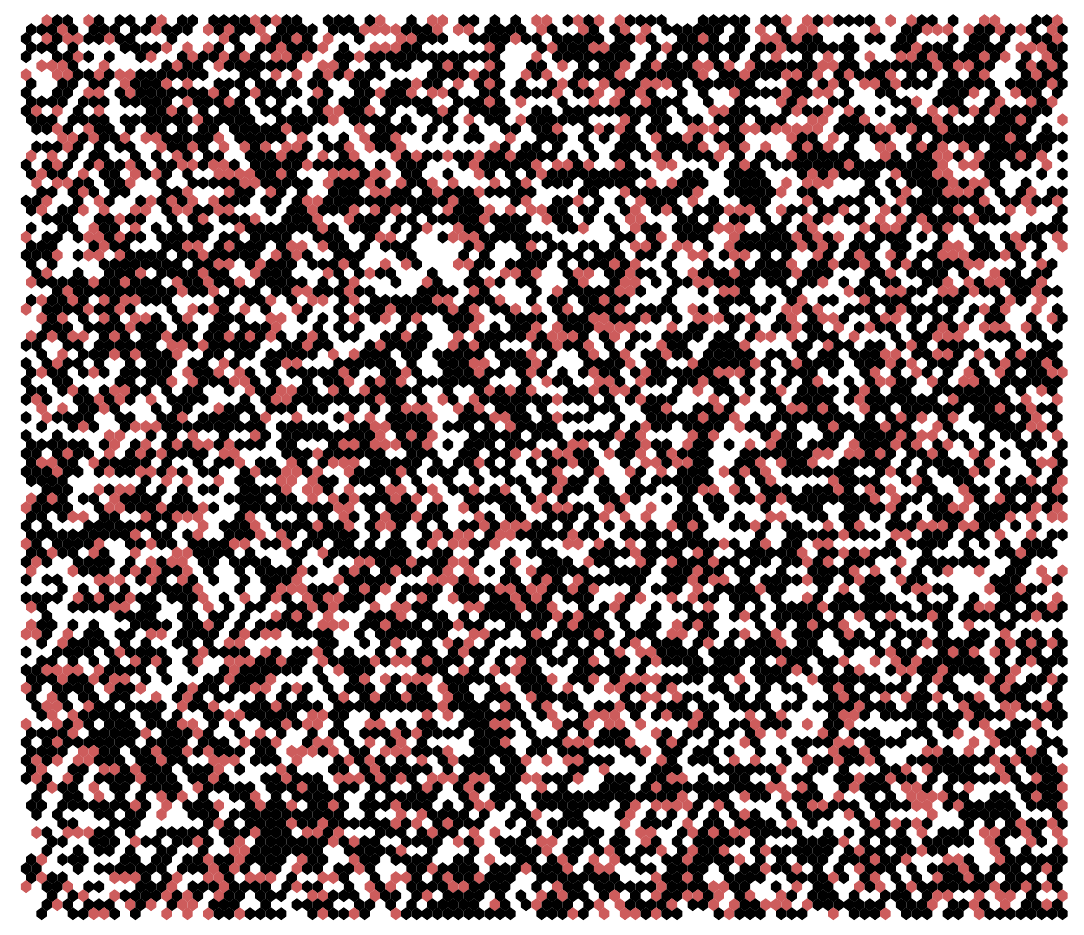}\\[2ex]
\includegraphics[width=1\linewidth]{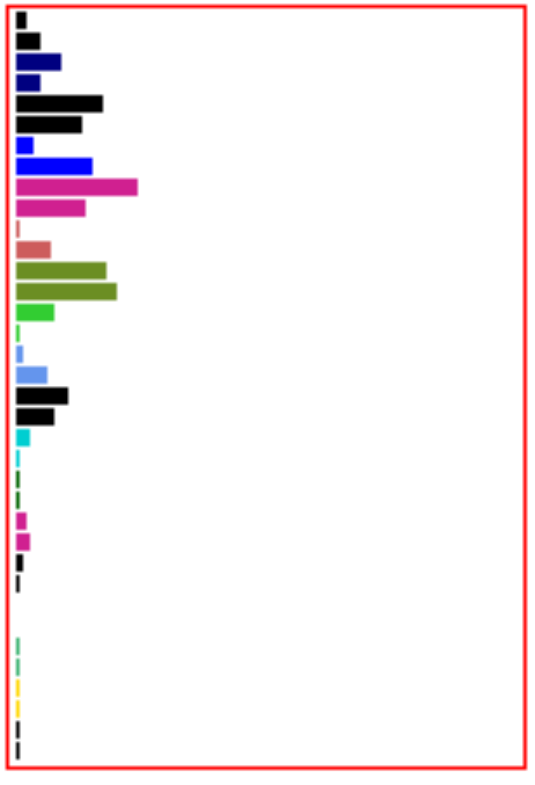}\\
1
\end{minipage}
\hfill
\begin{minipage}[c]{.13\linewidth}
\includegraphics[width=1\linewidth]{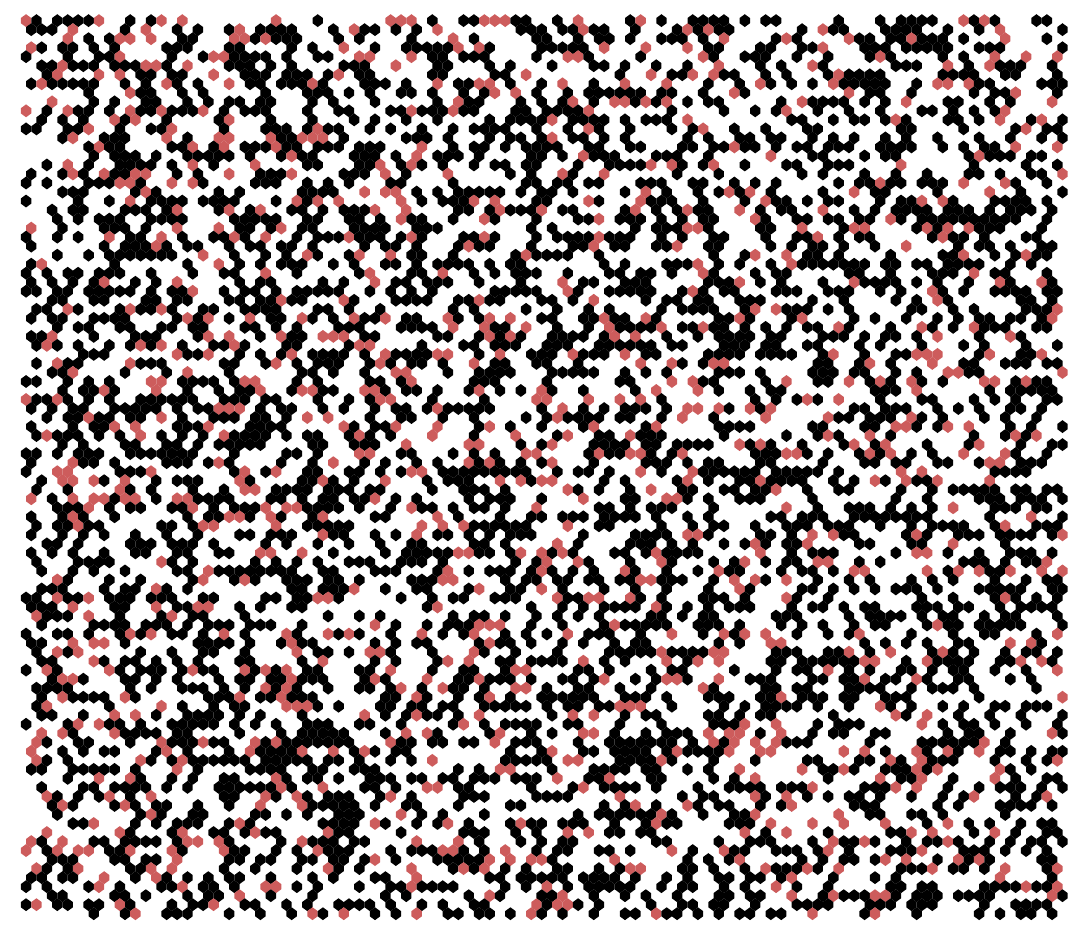}\\[2ex]
\includegraphics[width=1\linewidth]{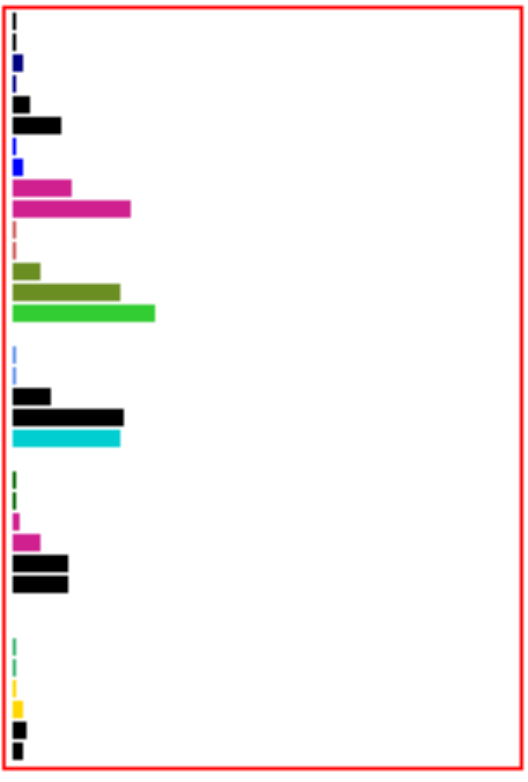}\\
2
\end{minipage}
\hfill
\begin{minipage}[c]{.13\linewidth}
\includegraphics[width=1\linewidth]{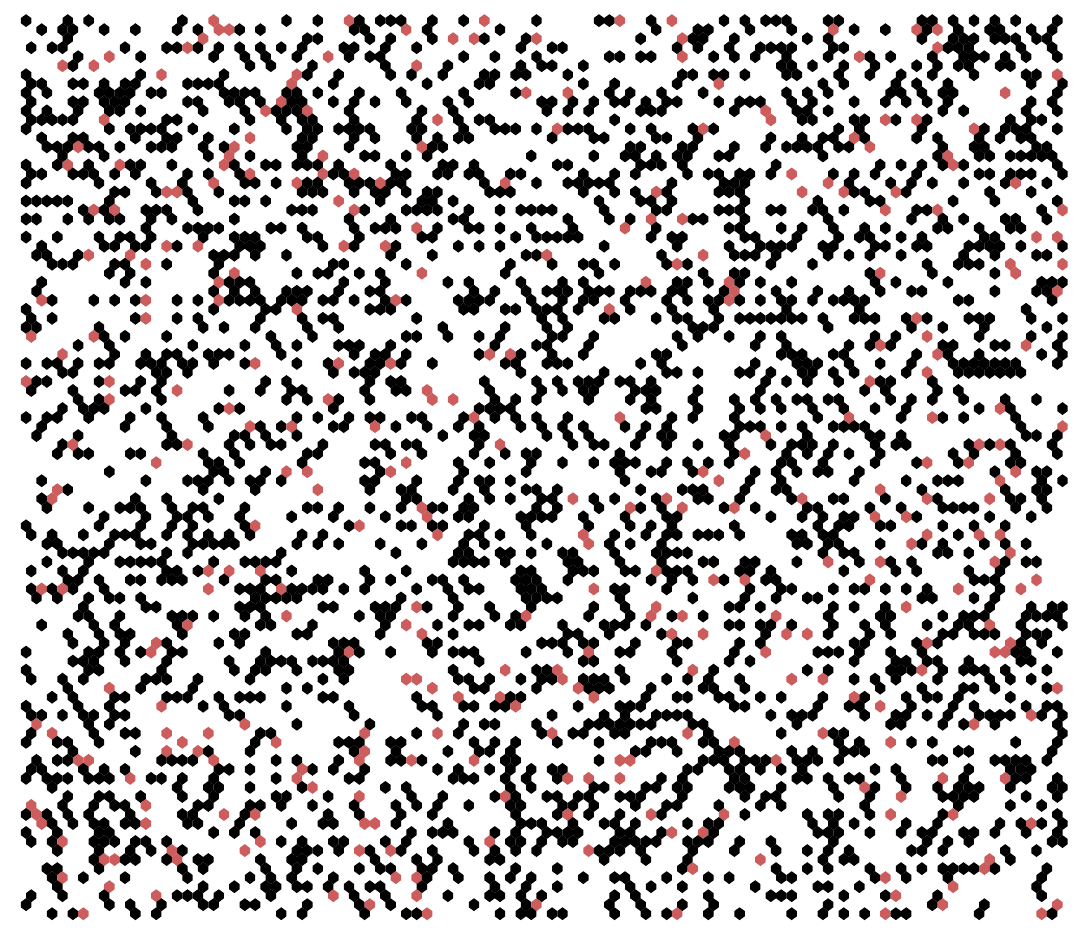}\\[2ex]
\includegraphics[width=1\linewidth]{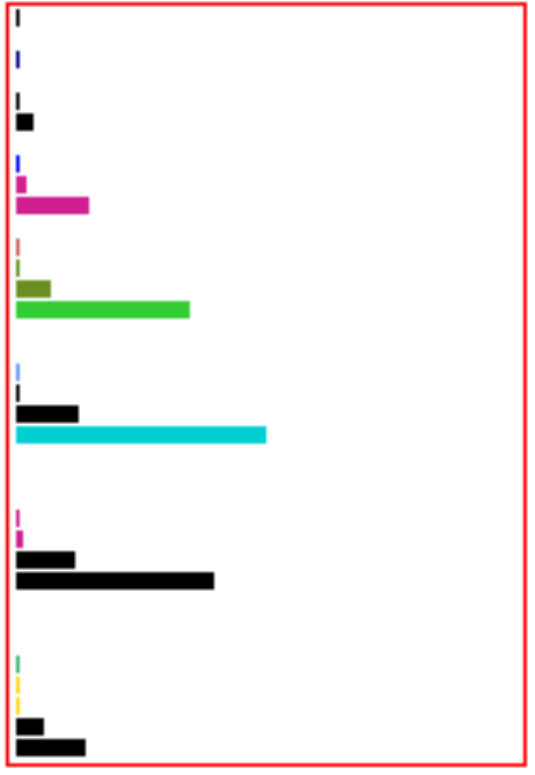}\\
3
\end{minipage}
\hfill
\begin{minipage}[c]{.13\linewidth}
\includegraphics[width=1\linewidth]{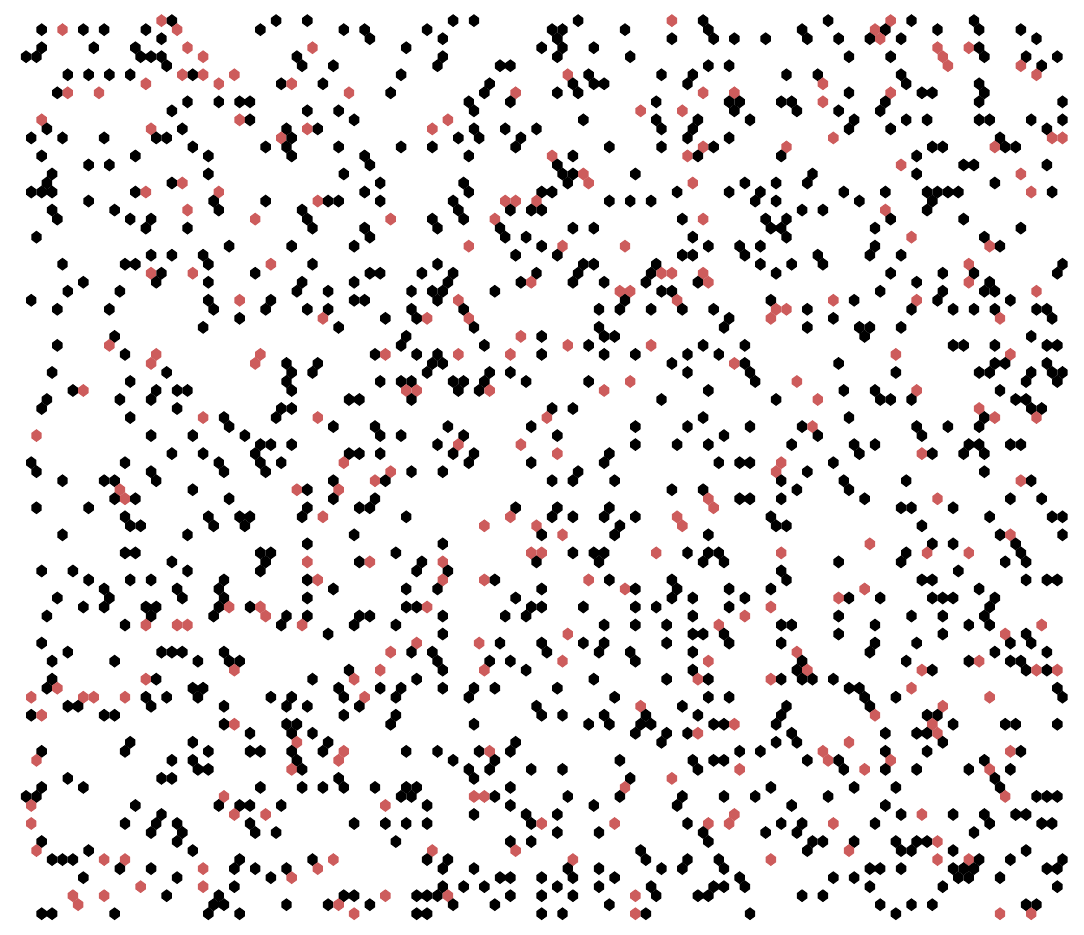}\\[2ex]
\includegraphics[width=1\linewidth]{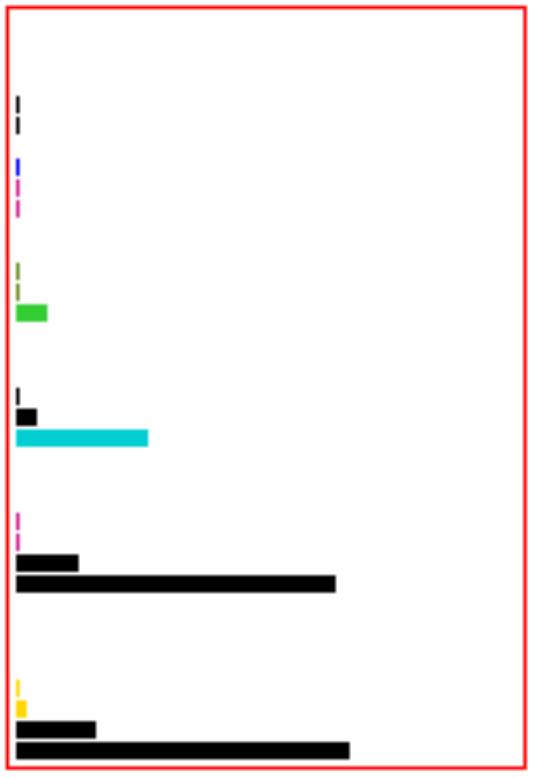}\\
4
\end{minipage}
\hfill
\begin{minipage}[c]{.13\linewidth}
\includegraphics[width=1\linewidth]{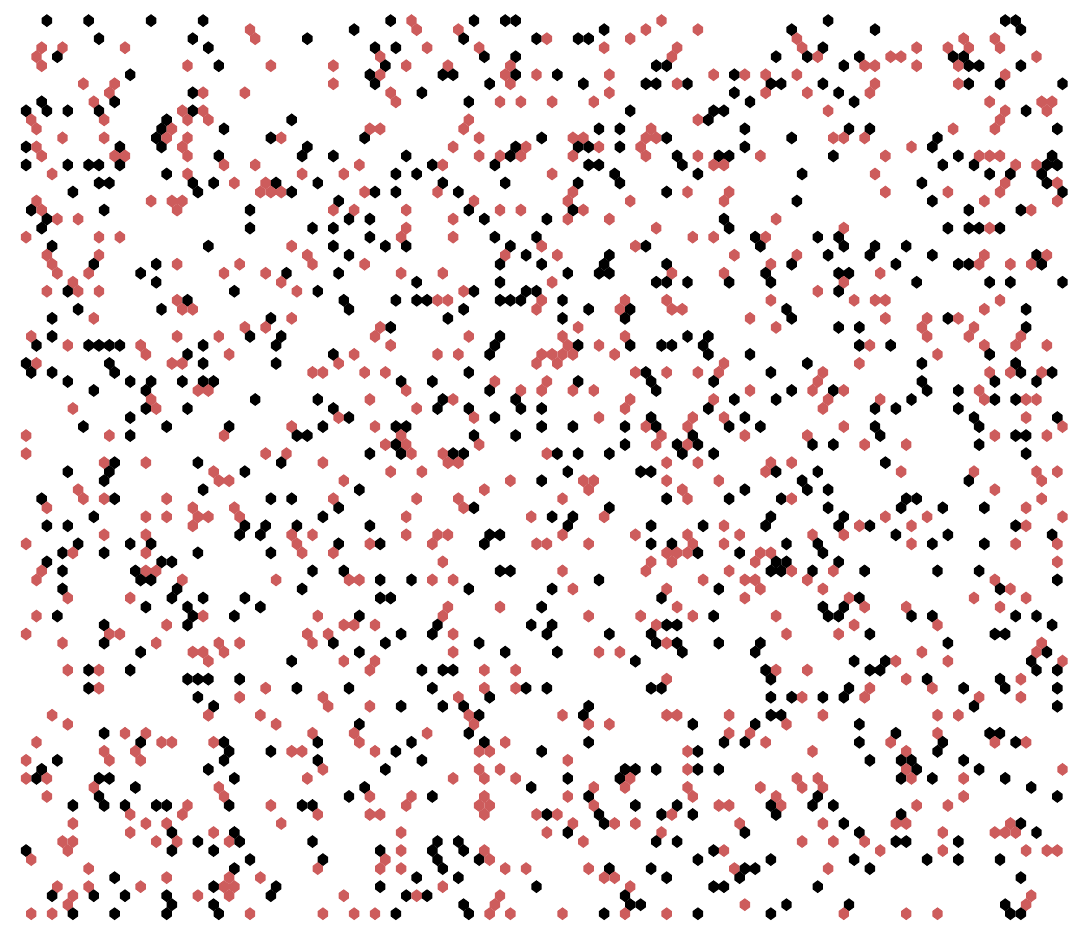}\\[2ex]
\includegraphics[width=1\linewidth]{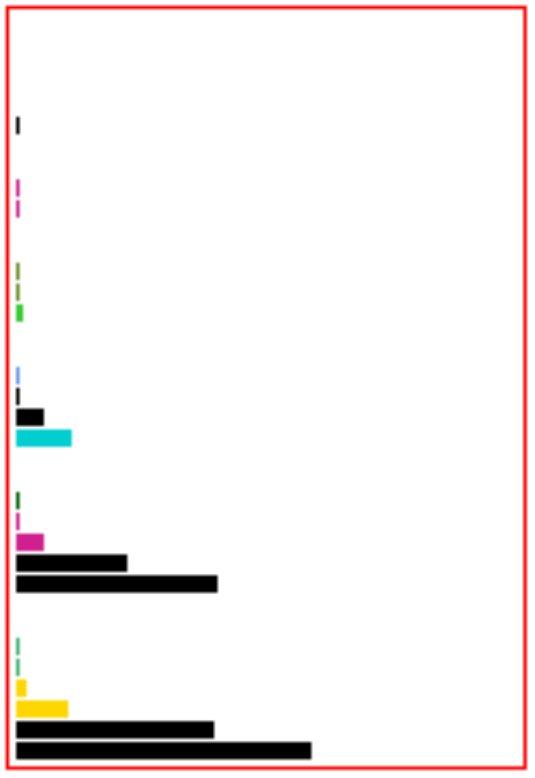}\\
5
\end{minipage}
\hfill
\begin{minipage}[c]{.13\linewidth}
\includegraphics[width=1\linewidth]{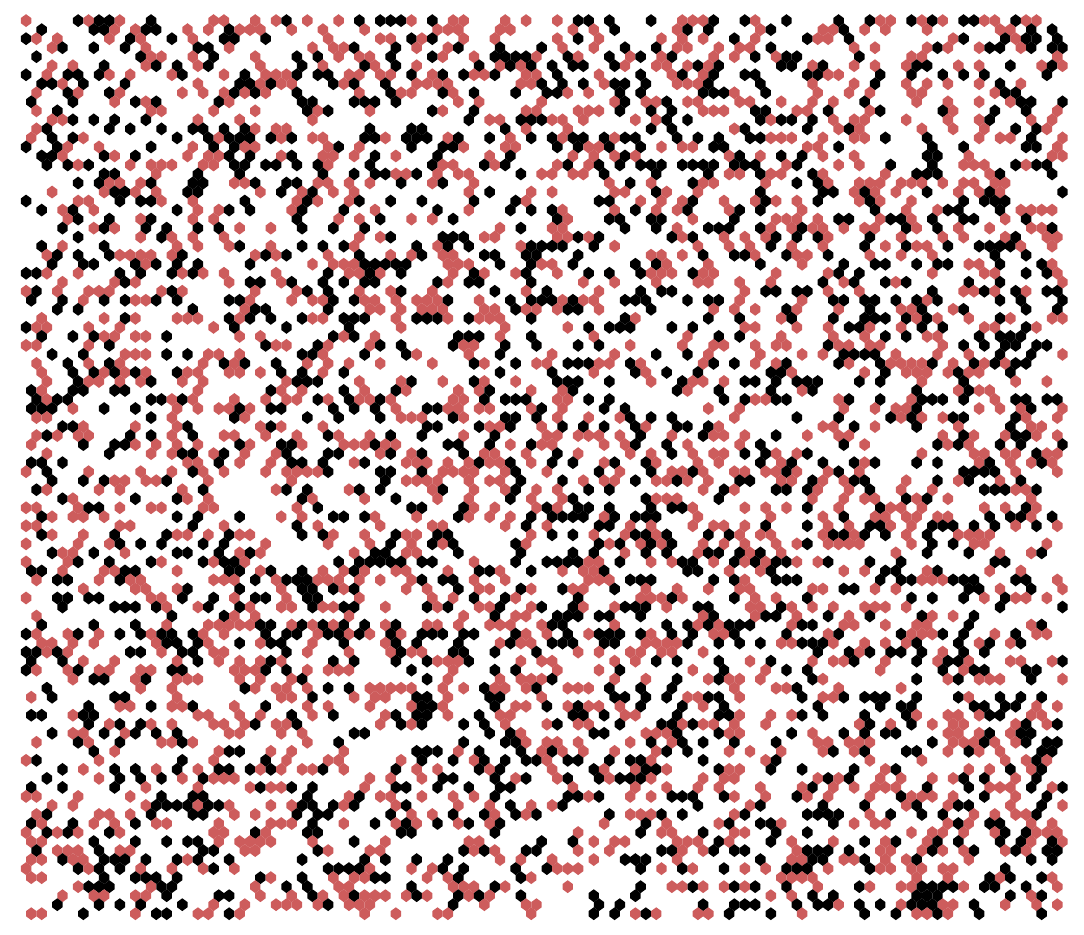}\\[2ex]
\includegraphics[width=1\linewidth]{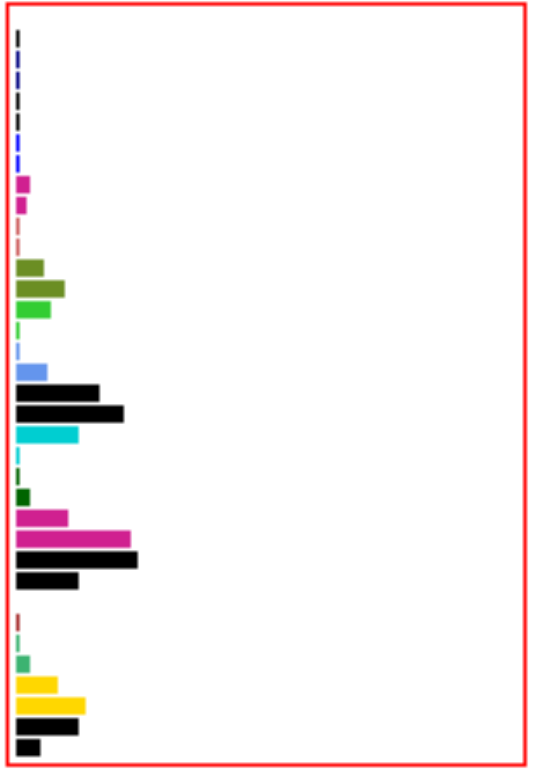}\\
6
\end{minipage}
\hfill
\begin{minipage}[c]{.13\linewidth}
\includegraphics[width=1\linewidth]{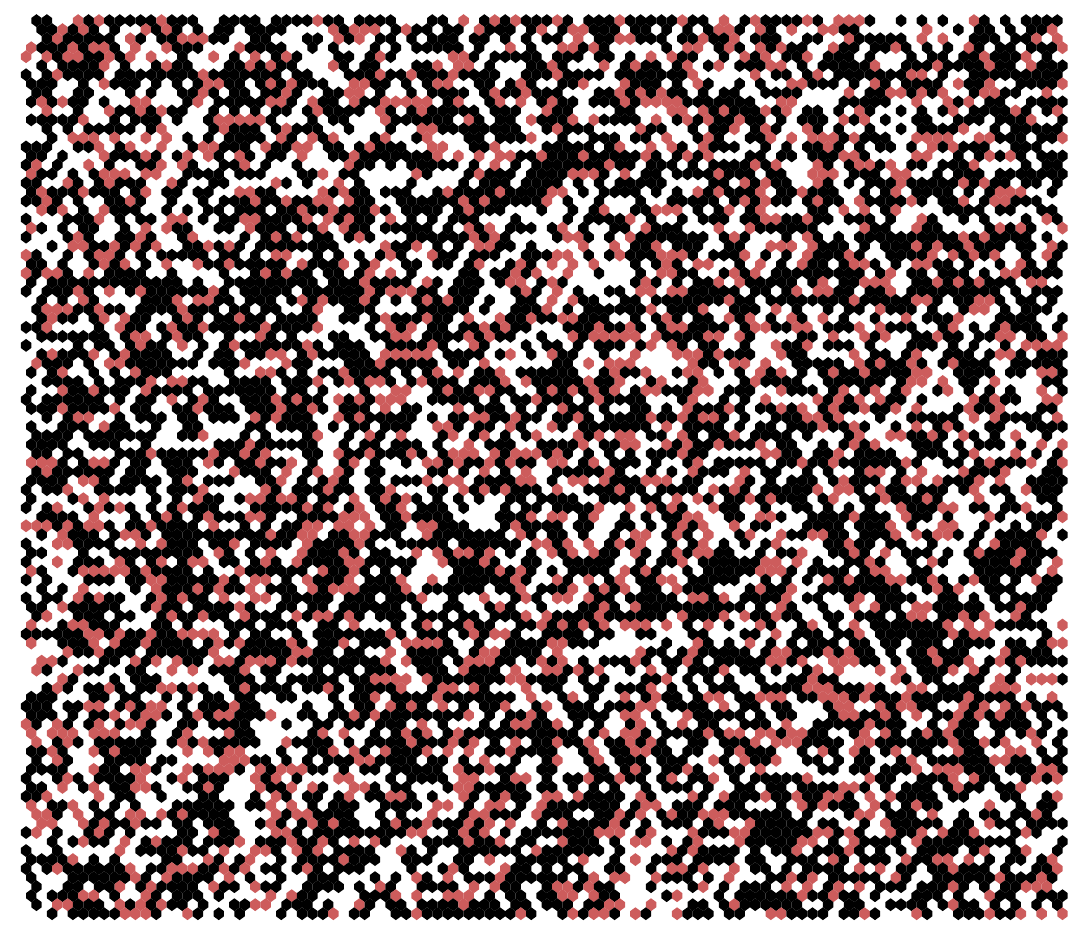}\\[2ex]
\includegraphics[width=1\linewidth]{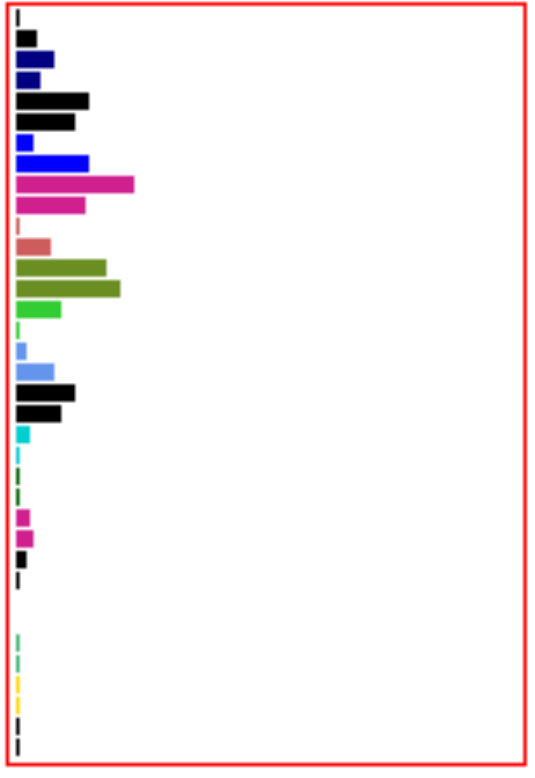}\\
7 repeat
\end{minipage}
\end{minipage}
\end{center}
}
\vspace{-1ex}
\caption[input-histograms]
{\textsf{
\noindent 
Dynamic graphics in DDLab show up pulsing
in the $v3k7$ Spiral rule\cite{Wuensche&Adamatzky2006} from figure~\ref{Glider examples}(f),
with a period of 7 time-steps.
Each  100$\times$100 pattern in a typical cycle is shown above
its input-histogram,  
where horizontal bars represent the lookup-frequency of 36 neighborhoods,
(all-2s at the top) for each corresponding time-step.
}}
\label{input-histograms}
\end{figure}

\noindent When pulsing occurs, it is clearly evident by eye 
when the system is run in DDLab (fig~\ref{input-histograms}).
Space-time pattern density will exhibit a steady rhythmic periodic beat,
and this will be reflected by the input-histogram.
The input-histogram tracks how frequently the
different entries in a rule-table are actually looked~up at each time-step,
or the changing 2D block-frequency, where the ``blocks''
are the alternative patterns within the neighborhood template.
\vspace{2ex}
\begin{figure}[htb]
\begin{center}
\textsf{\small
\begin{minipage}[c]{.95\linewidth} 
\begin{minipage}[c]{.20\linewidth}
\includegraphics[width=.85\linewidth]{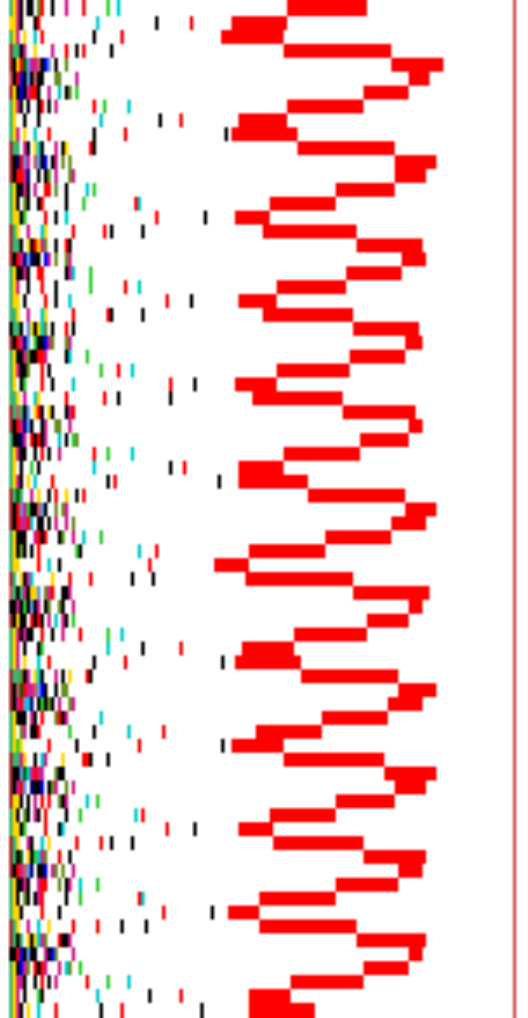}\\[1ex]
(a)
\end{minipage}
\hfill
\begin{minipage}[c]{.35\linewidth}
\includegraphics[width=1\linewidth]{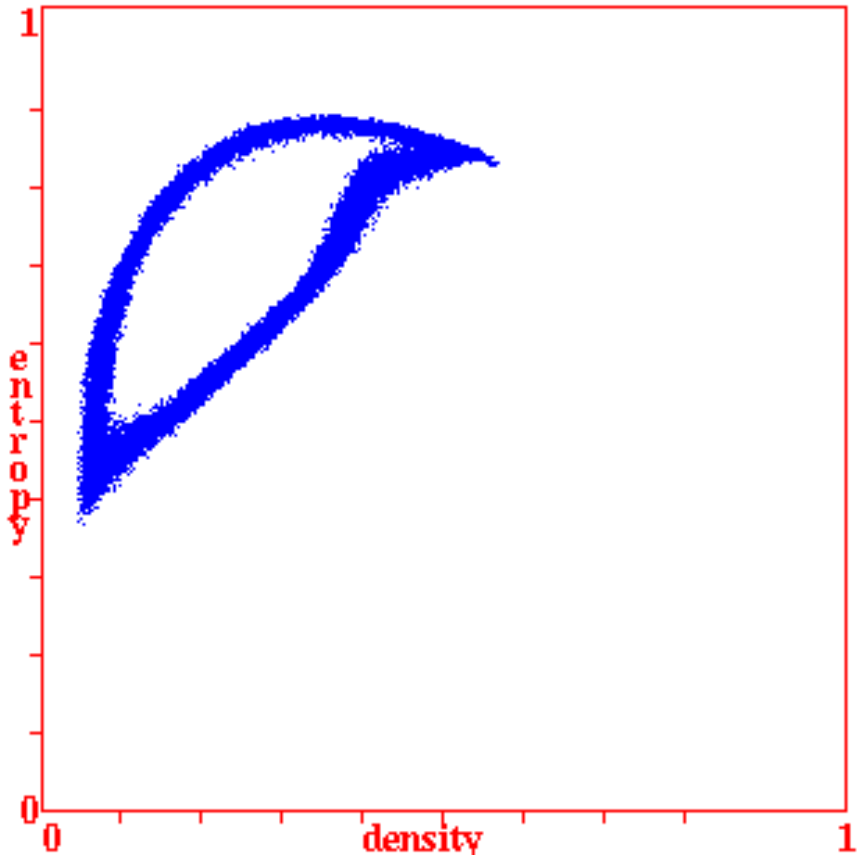}\\
(b)
\end{minipage}
\hfill
\begin{minipage}[c]{.35\linewidth}
\includegraphics[width=1\linewidth]{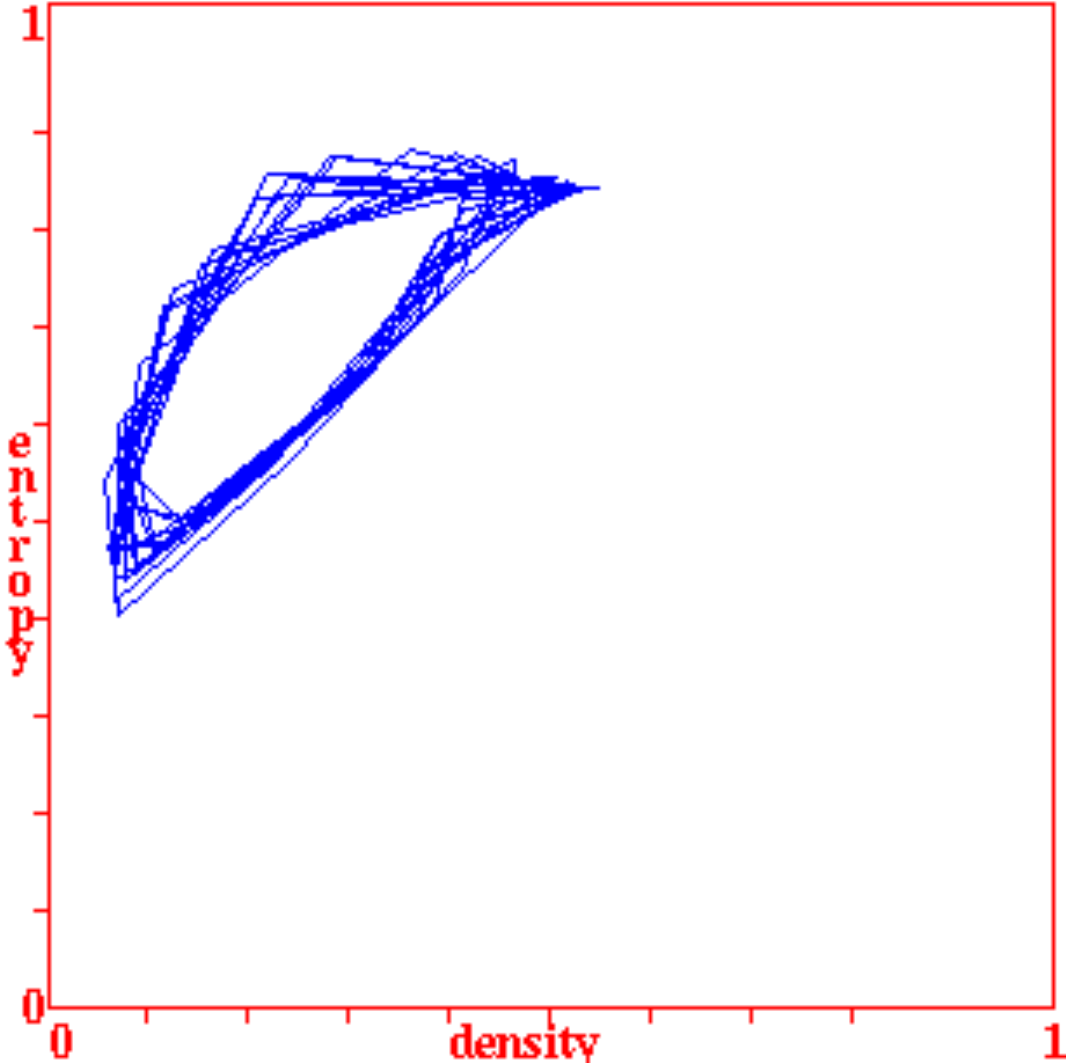}\\
(c)
\end{minipage}
\end{minipage}
}
\end{center}
\vspace{-2ex}
\caption[input-entropy]
{\textsf{For the $v3k7$ Spiral rule from figure~\ref{Glider examples}(f),
(a) Input-entropy oscillations with time (y-axis, stretched), $wl\approx7$ time-steps, 
$wh\approx0.4$.
Left edge: superimposed histogram values plots.
(b) The entropy-density scatter plot --- input-entropy (y-axis)
against the non-zero density (x-axis), for about 33000 time-steps.
(c) The same plot for just a few pulsing cycles, but linking successive dots giving a time-history.
}}
\label{input-entropy}
\end{figure}

\begin{figure}[htb]
\begin{center}
\textsf{\small
\begin{minipage}[c]{.95\linewidth} 
\begin{minipage}[c]{.20\linewidth}
\includegraphics[width=.85\linewidth]{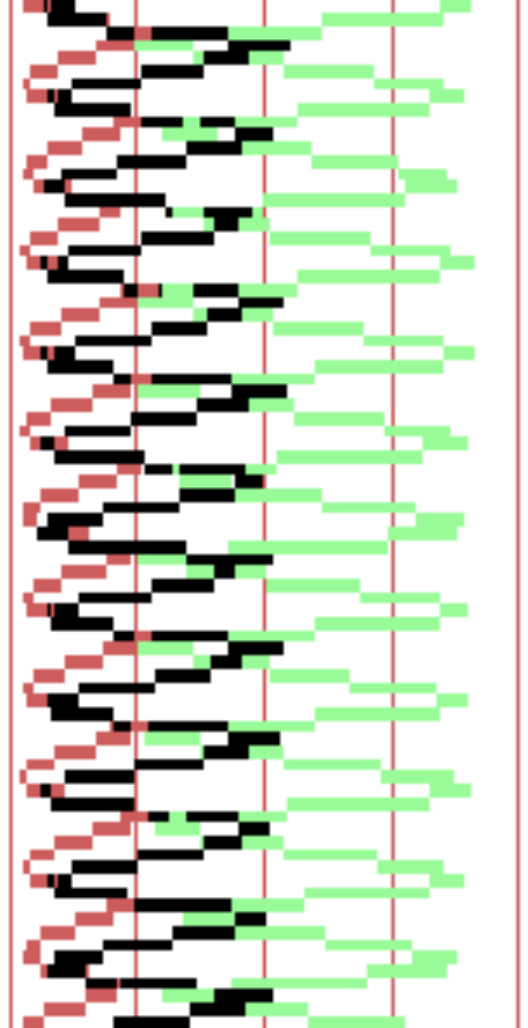}\\[1ex]
(a)
\end{minipage}
\hfill
\begin{minipage}[c]{.35\linewidth}
\includegraphics[width=1\linewidth]{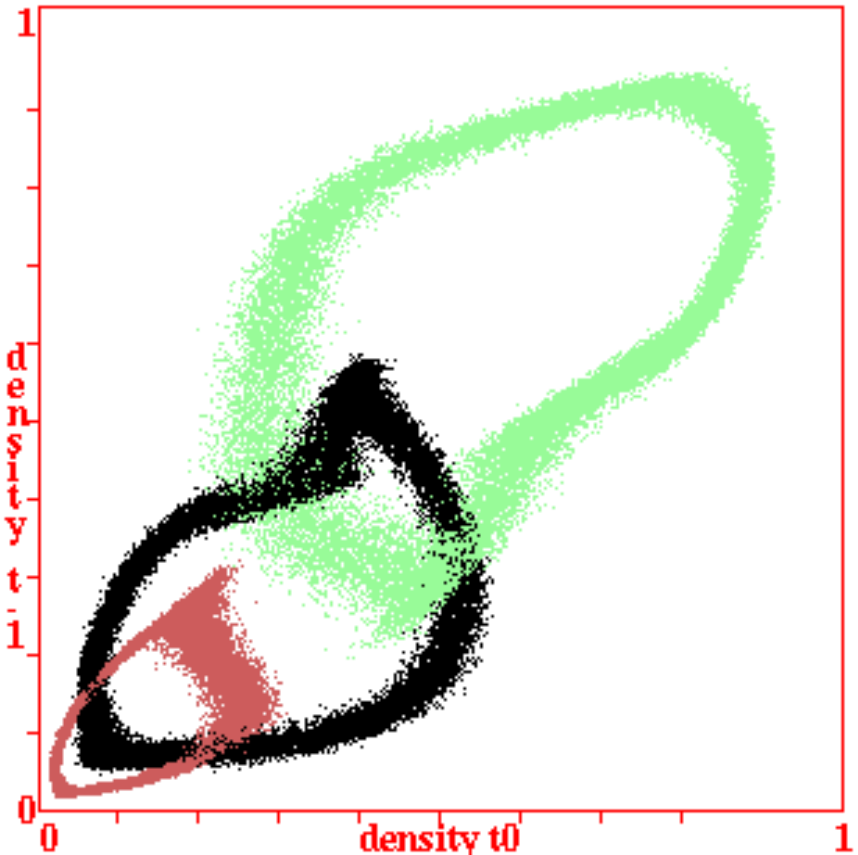}\\
(b)
\end{minipage}
\hfill
\begin{minipage}[c]{.35\linewidth}
\includegraphics[width=1\linewidth]{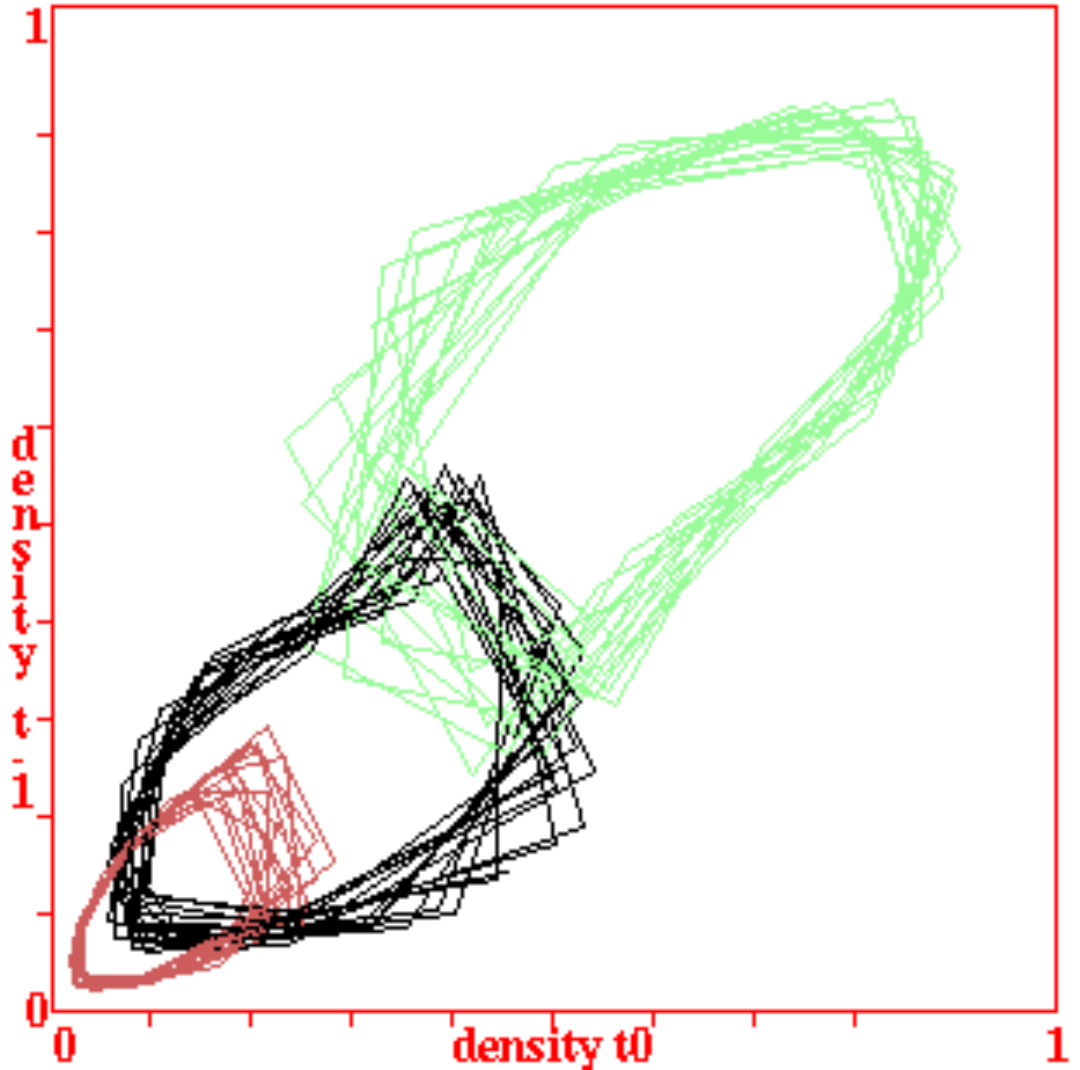}\\
(c)
\end{minipage}
\end{minipage}
}
\end{center}
\vspace{-2ex}
\caption[value-density]
{\textsf{For the $v3k7$ Spiral rule  from figure~\ref{Glider examples}(f),
(a) Value-density oscillations with time (y-axis, stretched). 0=green, 1=brown, 2=black. 
(b) The density return-map scatter plot --- the density of each value  at $t_0$ (x-axis) 
against its density at $t_1$, plotted as colored dots as above.
for about 33000 time-steps.
(c) The same plot for just a few pulsing cycles, but linking successive dots giving a time-history.
}}
\label{value-density}
\end{figure}
\clearpage

Further time-plot measures can be derived and presented graphically as the system iterates.
The entropy of the input-histogram can be calculated\footnote{The input-entropy is the 
Shannon entropy $H$ of the input-histogram.
For one time-step,
$H^t = -\sum_{i=0}^{S-1}\left( {Q_{i}^{t}}/{n} \times
  log_{2}\left( {Q_{i}^{t}}/{n}\right)\right)$,
where $Q_{i}^{t}$ is the lookup-frequency of neighborhood $i$ at time~$t$. $S$ is the
  rule-table size, and $n$ is the network size.
The normalised Shannon entropy $H_N$ is
  a value between 0 and 1,  $H_N=H^t/log_{2}n$,
  which measures the heterogeneity of the histogram ---
  ``entropy'' in this paper refers to $H_N$.}
 and plotted with its characteristic
wavelength ($wl$), wave-height ($wh$, twice amplitude), 
and waveform (its shape or phase), which in turn can generate
an entropy-density scatter plot\cite{Wuensche99} (fig~\ref{input-entropy}).
From space-time patterns, the density or proportion of each value, 
(0, 1, 2) if $v$=3, can be plotted, and this can
generate a density return-map scatter plot\cite{Wuensche2016} (fig~\ref{value-density}).
The scatter plots have the characteristics of chaotic strange attractors,
and successive dots can be connected to create a linked history --- this option
is much faster to produce the characteristic plot because just a few time-step are needed. 
The $wl$ and $wh$ data can be recognised and output automatically.

We will use the term ``waveform'' to sum up these pulsing measures.
Each glider rule in the CAP model maintains its
distinctive waveform signature, reflecting the distinctive glider
dynamics.  It was shown in \cite{Wuensche-pulsingCA} that the
underlying waveform signature is independent of the network size $n$,
becoming more focused as $n$ increases towards infinity,  
but reducing $n$ makes reaching a uniform attractor\cite{Wuensche92} more likely,
where the system would freeze.

Waveform measures and plots are usually averaged over a moving window of $w$ time-steps,
$w$=10 to classify rules by the variability of input-entropy\cite{Wuensche99},
but to observe pulsing dynamics most effectively we take the measures over each time-step
where  $w$=1.
However, when measuring the wave-length $wl$ automatically, $w$$\geq$2
can sometimes be preferable. Figures~\ref{input-entropy} and \ref{value-density}
show examples of the measures for the $k$=7 
Spiral rule\cite{Wuensche&Adamatzky2006} on a 100$\times$100 lattice, with $w$=1.

\section{Wave-length and wave-height data}
\label{Wave-length and wave-height data}

\begin{figure}[hp]
{\baselineskip2ex \scriptsize
\begin{center}
\begin{minipage}[c]{.8\linewidth}
\begin{minipage}[c]{.176\linewidth}
\includegraphics[width=1\linewidth]{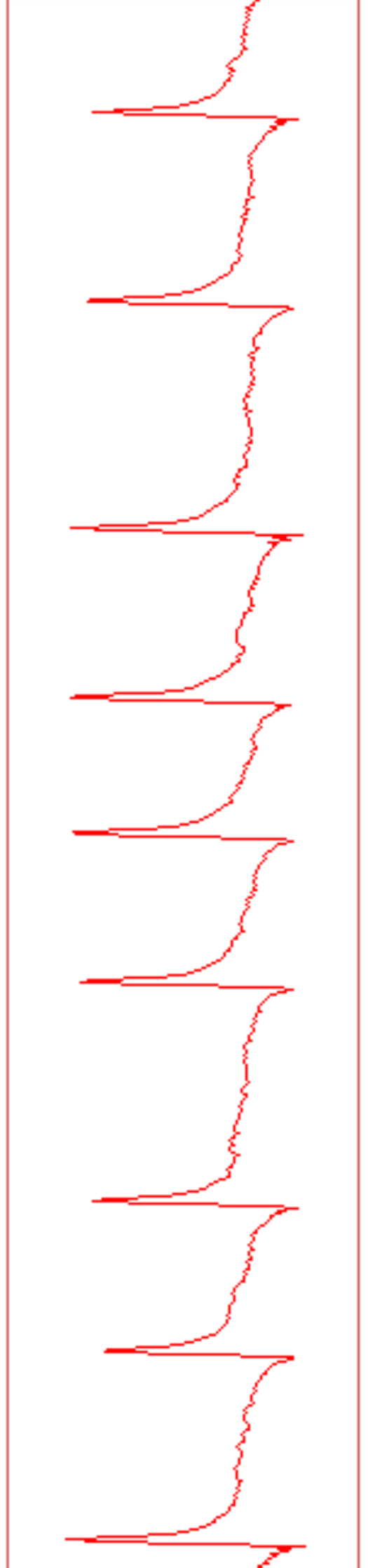}\\
{\normalsize {\textsf (a) $w$=1}}
\end{minipage}
\hfill
\begin{minipage}[c]{.176\linewidth}
\includegraphics[width=1\linewidth]{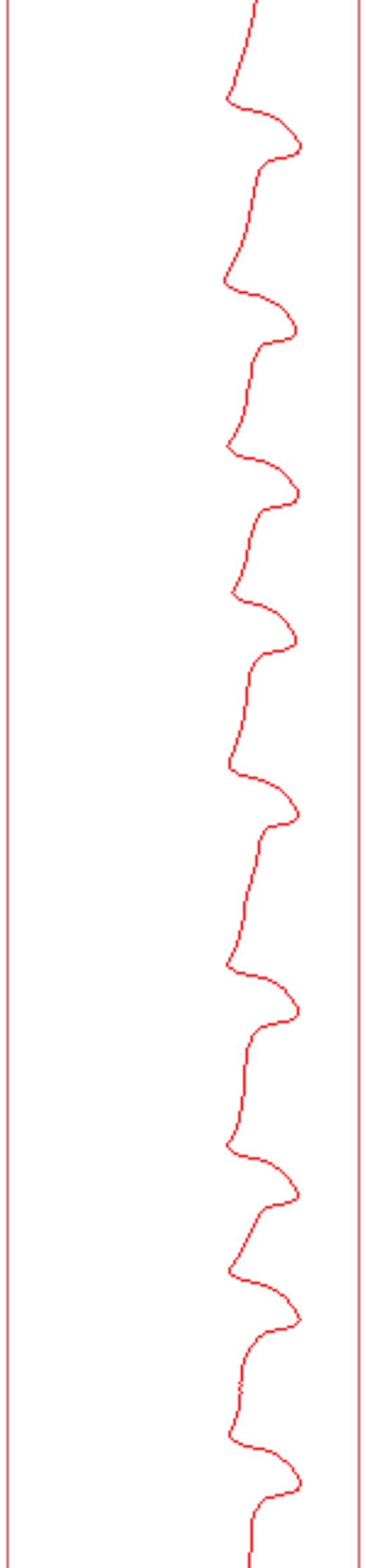}\\
{\normalsize {\textsf (b) $w$=20}}
\end{minipage}
\hfill
\begin{minipage}[c]{.45\linewidth}
\begin{verbatim}
v3k7 kcodeSize=35
(hex)806a22a29a12182a84 w=20
min=691 wl=101 wh==0.206
max=711 wl=101
min=781 wl=90 wh==0.202
max=800 wl=89
min=893 wl=112 wh==0.210
max=913 wl=113
min=969 wl=76 wh==0.199
max=989 wl=76
min=1059 wl=90 wh==0.211
max=1078 wl=89
min=1133 wl=74 wh==0.211
max=1152 wl=74
min=1195 wl=62 wh==0.193
max=1214 wl=62
min=1258 wl=63 wh==0.203
max=1278 wl=64
min=1349 wl=91 wh==0.195
max=1369 wl=91
min=1427 wl=78 wh==0.199
max=1446 wl=77
min=1482 wl=55 wh==0.198
av-wl=79.14, av-wh=0.201, sample=50
\end{verbatim}
\vspace{-2ex}
{\small {\textsf (c) $w$=20}}
\end{minipage}
\end{minipage}
\end{center}
}
\vspace{-2ex}
\caption[Wave-length variable+jagged]
{\textsf{Wave-length ($wl$) for rule $v3k7$ g35 in \cite{Wuensche-pulsingCA},
with steady but variable $wl$ oscillations,
and jagged stretches on the downslope of the entropy plot profile (a) 
can give false min/max results, but this is fixed by
increasing the time-step window ($w$=20) to smooth the plot (b). 
(c) shows typical data, with $wl$ between 62 and 113 time-steps,
though the actual range is slightly greater.
The last line shows average values.
}}
\label{Wave-length variable+jagged}
\end{figure}

\noindent A new method is available in DDLab to automatically recognise and measure the
wave-length ($wl$) and wave-height ($wh$) of entropy oscillations in
the CAP model.
The output appears in the terminal.
The data can be activated while the density-entropy plot is active.
The algorithm is effective for well developed steady (but possibly variable) 
entropy oscillations --- the examples in figure~\ref{Wave-length examples} relate to the rules in
figs~\ref{k3 CAP plots}, \ref{k4t CAP plots}, and \ref{k5 CAP plots}.
Entropy oscillations with jagged stretches or transient min/max values in the plot profile,
as in figures~\ref{Wave-length variable+jagged} and \ref{k5 CAP plots}(a), can disturb the
wave-length ($wl$) measures, but this can be smoothed out by making the
time-step window $w$$\geq$2 without effecting ($wl$), though ($wh$)
would be reduced. Figure~\ref{Wave-length variable+jagged} gives an example
of variable $wl$, and  with jagged stretches resolved by making $w$=20.  
Section~\ref{Space-time patterns on-the-fly options} includes
step-by-step instructions for the method.

\begin{figure}[hp]
{\baselineskip2ex \scriptsize
\begin{center}
\begin{minipage}[c]{1\linewidth}
\begin{minipage}[c]{.3\linewidth}
\begin{verbatim}
v3k3 kcodeSize=10 
(hex)00a864 w=1
min=112 wl=10 wh==0.268
max=115 wl=9
min=121 wl=9 wh==0.260
max=125 wl=10
min=131 wl=10 wh==0.269
max=135 wl=10
min=140 wl=9 wh==0.211
max=144 wl=9
\end{verbatim}
\vspace{-2ex}
{\small {\textsf (a) $w$=1}, fig~\ref{k3 CAP plots}}
\end{minipage}
\hfill
\begin{minipage}[c]{.3\linewidth}
\begin{verbatim}
v3k4 kcodeSize=15 
(hex)2a945900 w=1
min=107 wl=6 wh==0.294
max=110 wl=7
min=114 wl=7 wh==0.332
max=117 wl=7
min=121 wl=7 wh==0.270
max=123 wl=6
min=127 wl=6 wh==0.355
max=129 wl=6
\end{verbatim}
\vspace{-2ex}
{\small {\textsf (b) $w$=1,  fig~\ref{k4t CAP plots}}}
\end{minipage}
\hfill
\begin{minipage}[c]{.3\linewidth}
\begin{verbatim}
v3k5 kcodeSize=20 
(hex)004a8a2a8254 w=2
min=116 wl=11 wh==0.428
max=118 wl=11
min=126 wl=10 wh==0.420
max=129 wl=11
min=138 wl=12 wh==0.457
max=141 wl=12
min=148 wl=10 wh==0.442
max=151 wl=10
\end{verbatim}
\vspace{-2ex}
{\normalsize {\textsf (c)  $w$=2, fig~\ref{k5 CAP plots}}}
\end{minipage}
\end{minipage}
\end{center}
}
\vspace{-2ex}
\caption[Wave-length examples]
{\textsf{Wave-length ($wl$) and wave-height ($wh$) examples showing
data for a typical sequence of 4 pulsing cycles. Data is output continuously
in the terminal, and average values so far if interrupted.
The rules, shown at the top, relate to waveform figures indicated.
The algorithm identifies the time-step at the minimum and
maximum values of each oscillation to calculate ($wl$) and ($wh$).
The size of the time-step window $w$
is shown; usually $w$=1, but for example (c) $w$=2 to smooth out a jagged stretch
at the maximum part of the plot profile in figure~\ref{k5 CAP plots}(a).
}}
\label{Wave-length examples}
\end{figure}

\section{CAP model plots for $k$=3, 4, 5, and 6}
\label{CAP model plots for $k$=3, 4, 5, and 6}

\begin{figure}[!h]
\begin{center}
\textsf{\small
\begin{minipage}[c]{.9\linewidth}
\begin{minipage}[c]{.16\linewidth}
\includegraphics[width=.85\linewidth]{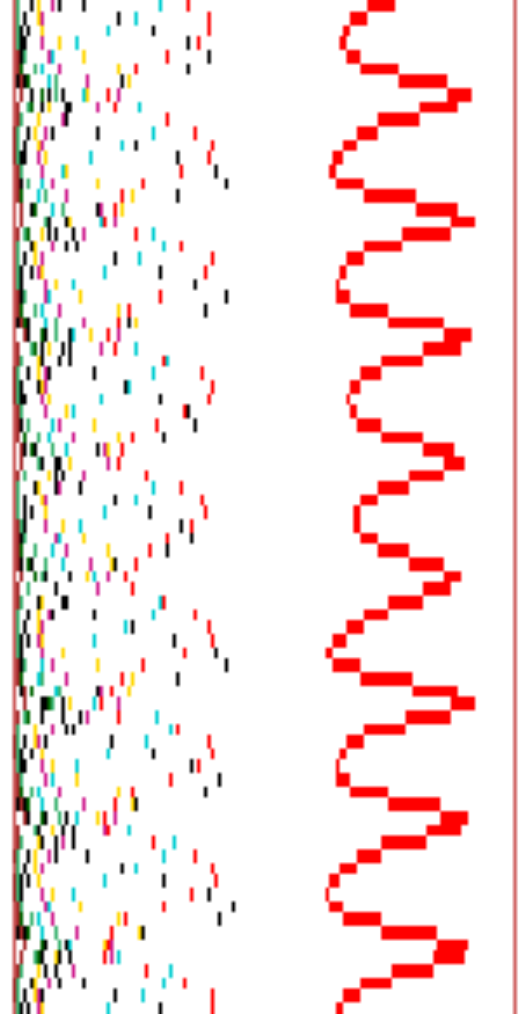}\\[1ex]
(a)
\end{minipage}
\hfill
\begin{minipage}[c]{.28\linewidth}
\includegraphics[width=1\linewidth]{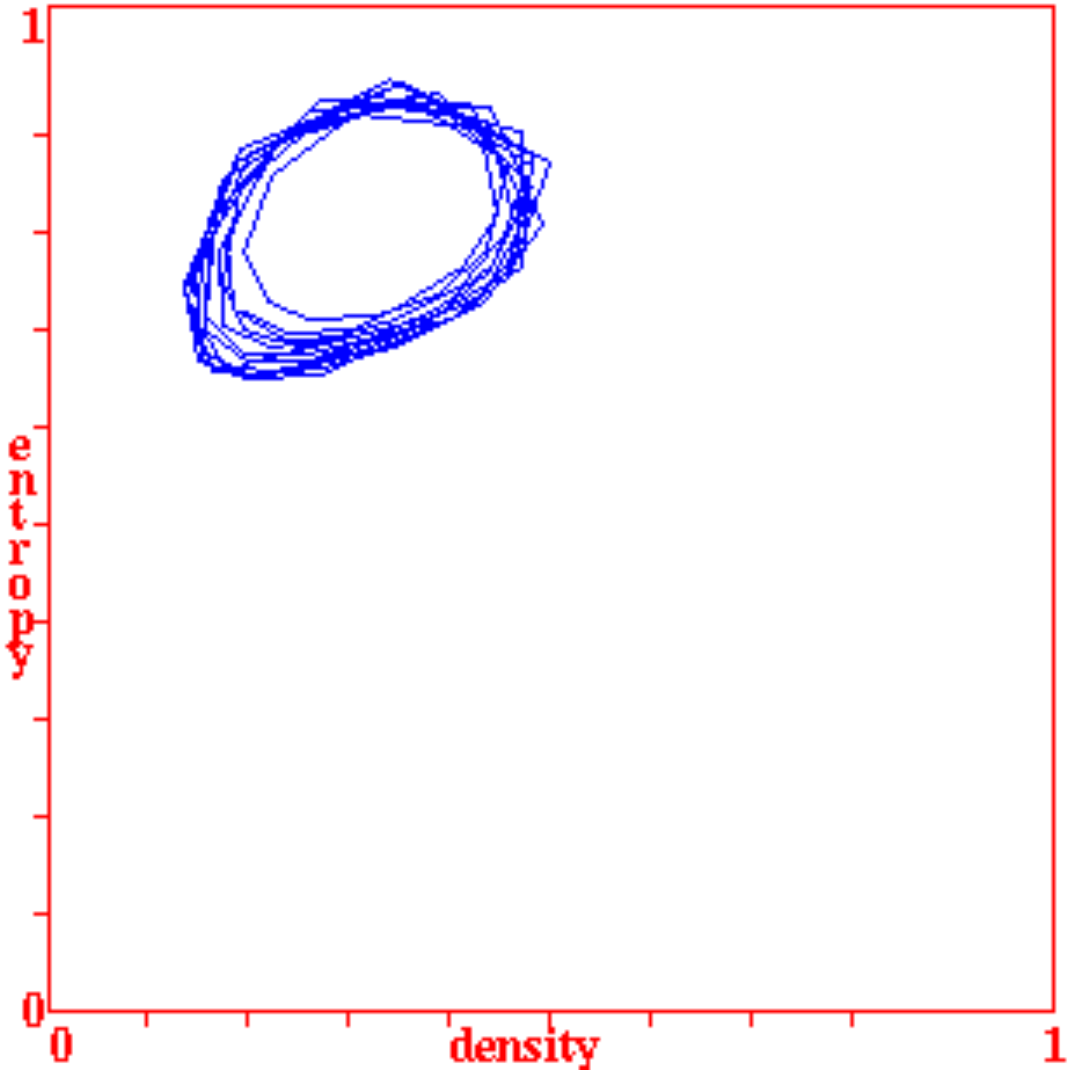}\\
(b)
\end{minipage}
\hfill
\begin{minipage}[c]{.16\linewidth}
\includegraphics[width=.85\linewidth]{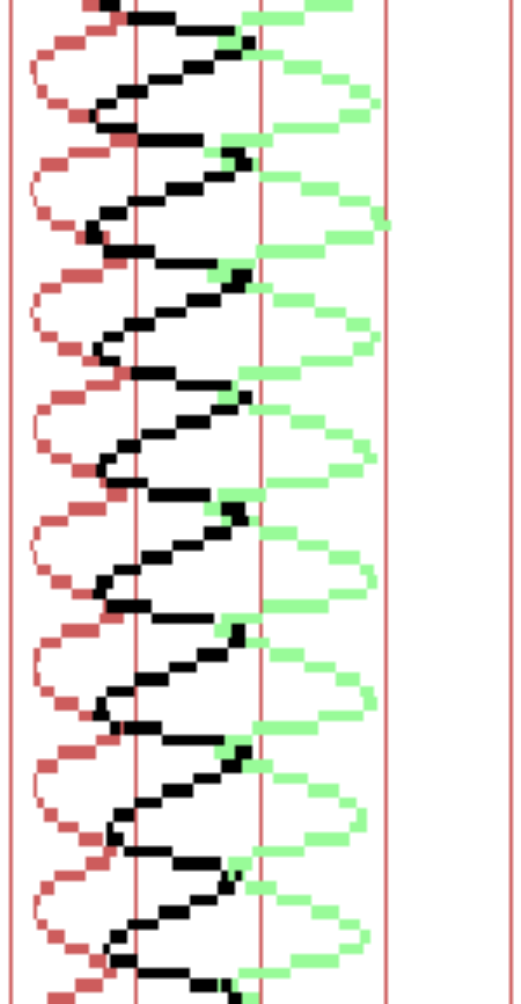}\\[1ex]
(c)
\end{minipage}
\hfill
\begin{minipage}[c]{.28\linewidth}
\includegraphics[width=1\linewidth]{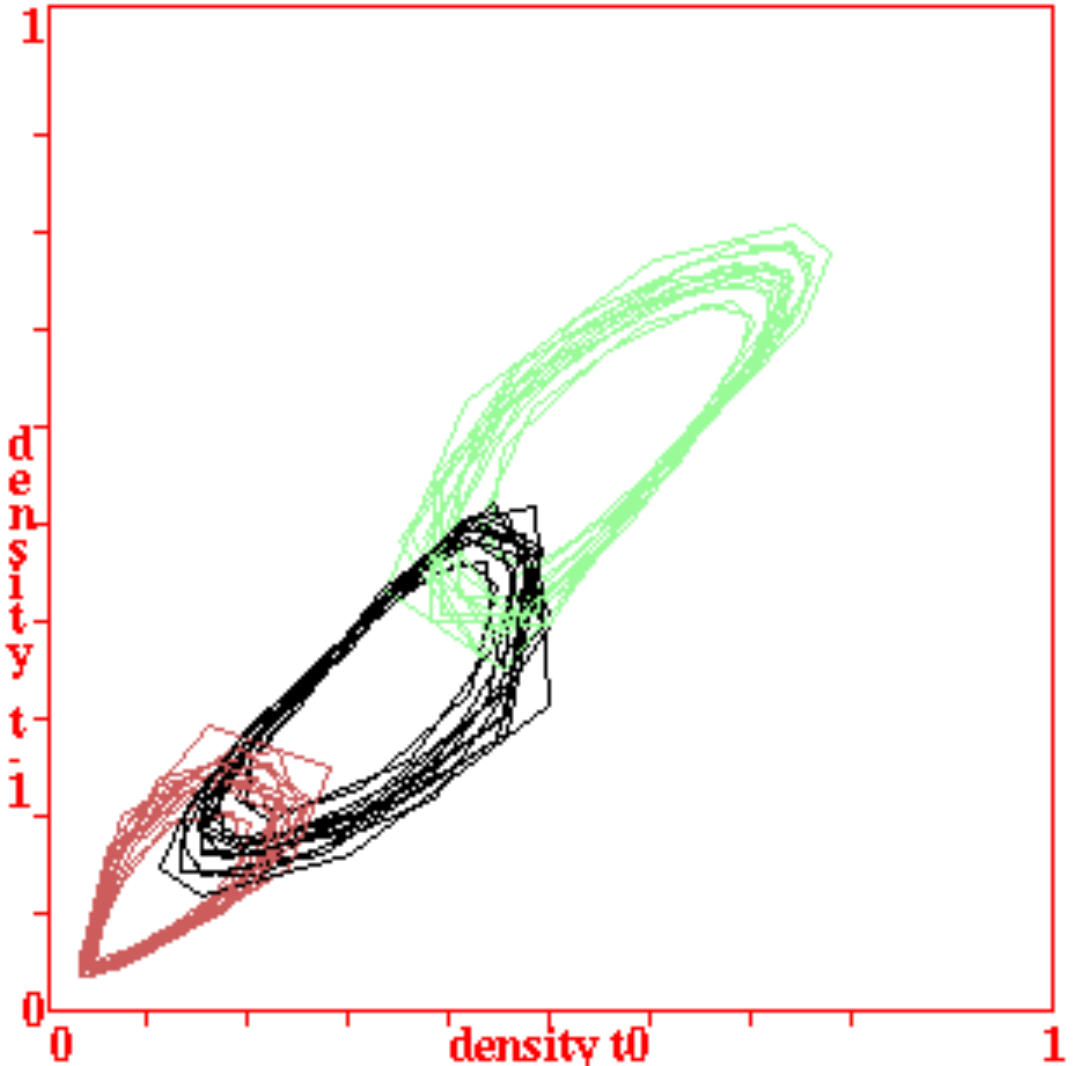}\\
(d)
\end{minipage}\\
\end{minipage}
}
\end{center}
\vspace{-3ex}
\caption[k3 CAP plots]
{\textsf{
k3 CAP plots, v3k3x1.vco, (hex)00a864, g(1).
}}
\label{k3 CAP plots}
\end{figure}

\begin{figure}[!h]
\begin{center}
\textsf{\small
\begin{minipage}[c]{.9\linewidth}
\begin{minipage}[c]{.16\linewidth}
\includegraphics[width=.85\linewidth]{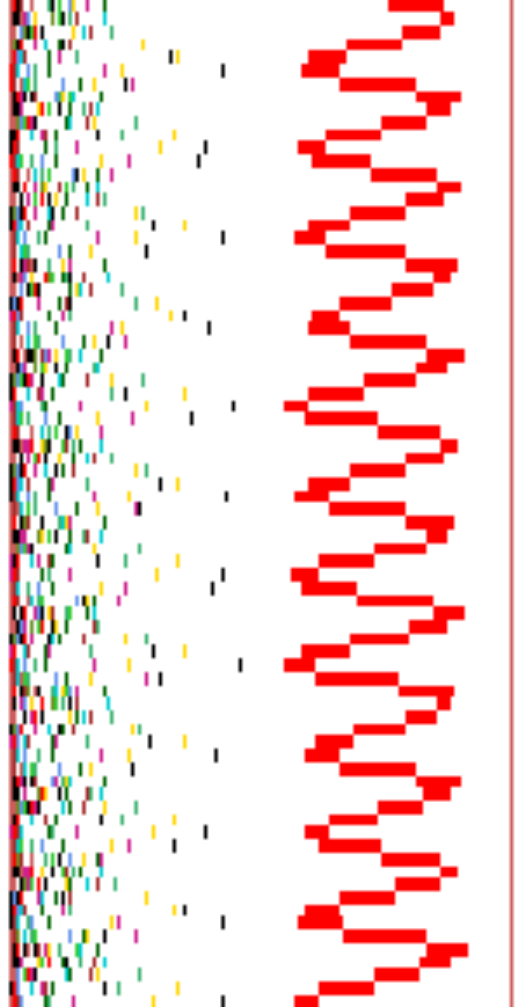}\\[1ex]
(a)
\end{minipage}
\hfill
\begin{minipage}[c]{.28\linewidth}
\includegraphics[width=1\linewidth]{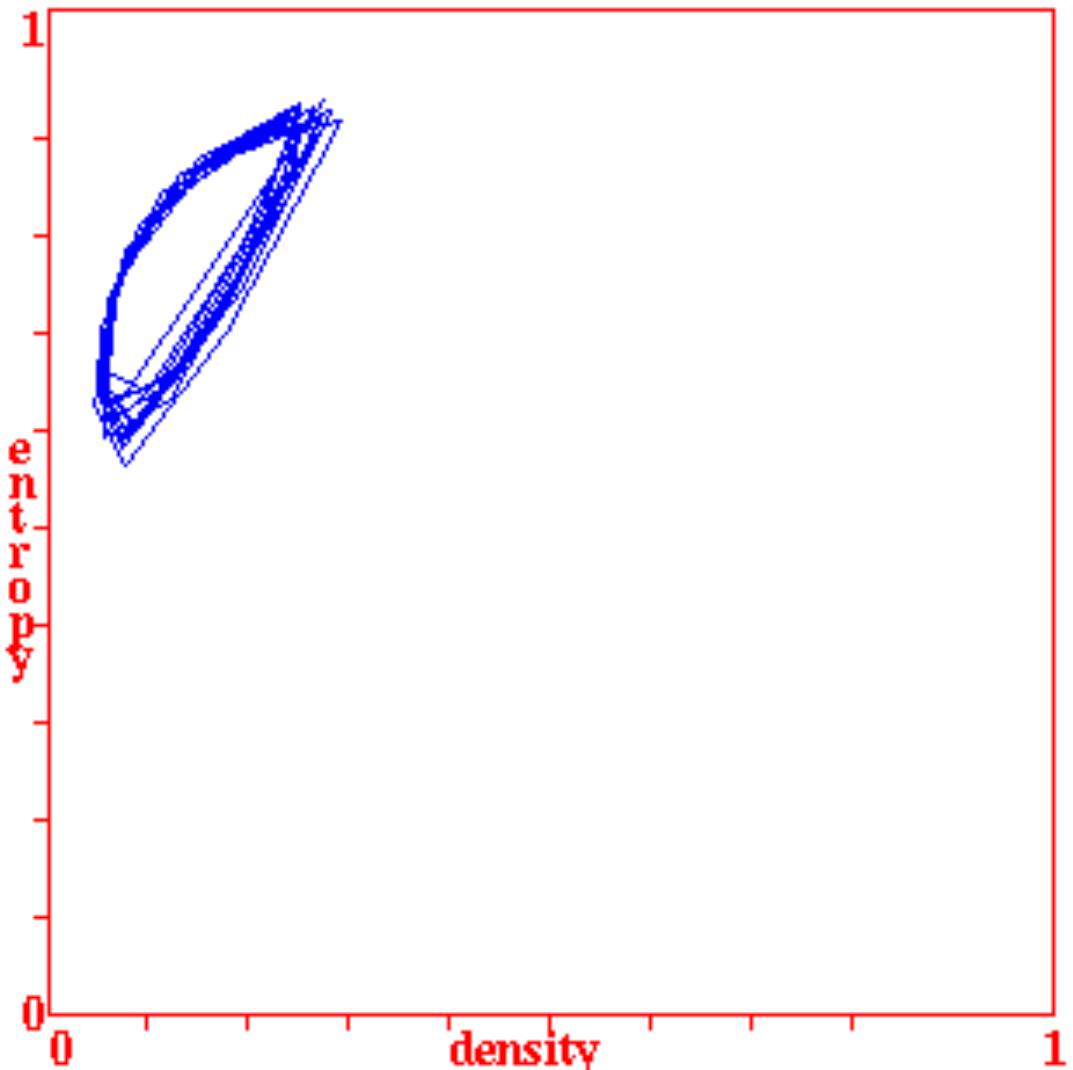}\\
(b)
\end{minipage}
\hfill
\begin{minipage}[c]{.16\linewidth}
\includegraphics[width=.85\linewidth]{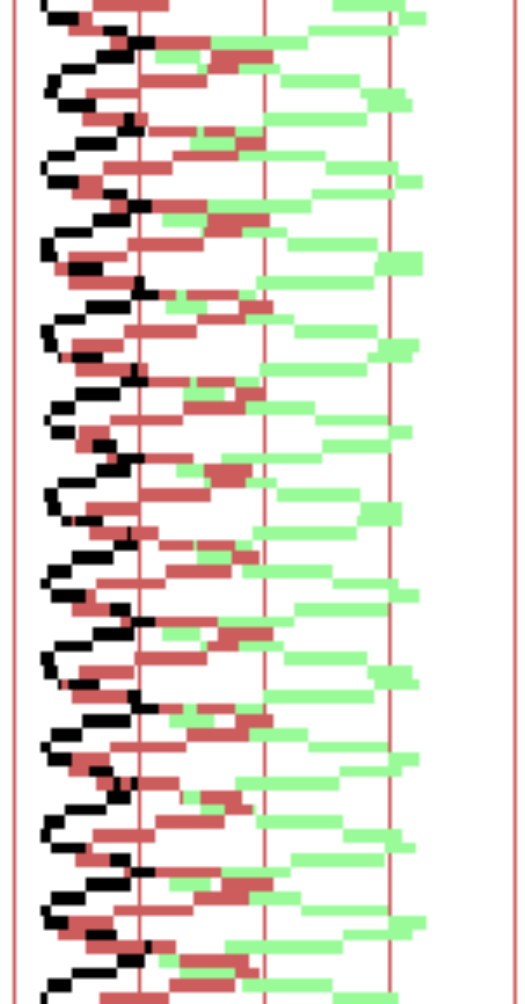}\\[1ex]
(c)
\end{minipage}
\hfill
\begin{minipage}[c]{.28\linewidth}
\includegraphics[width=1\linewidth]{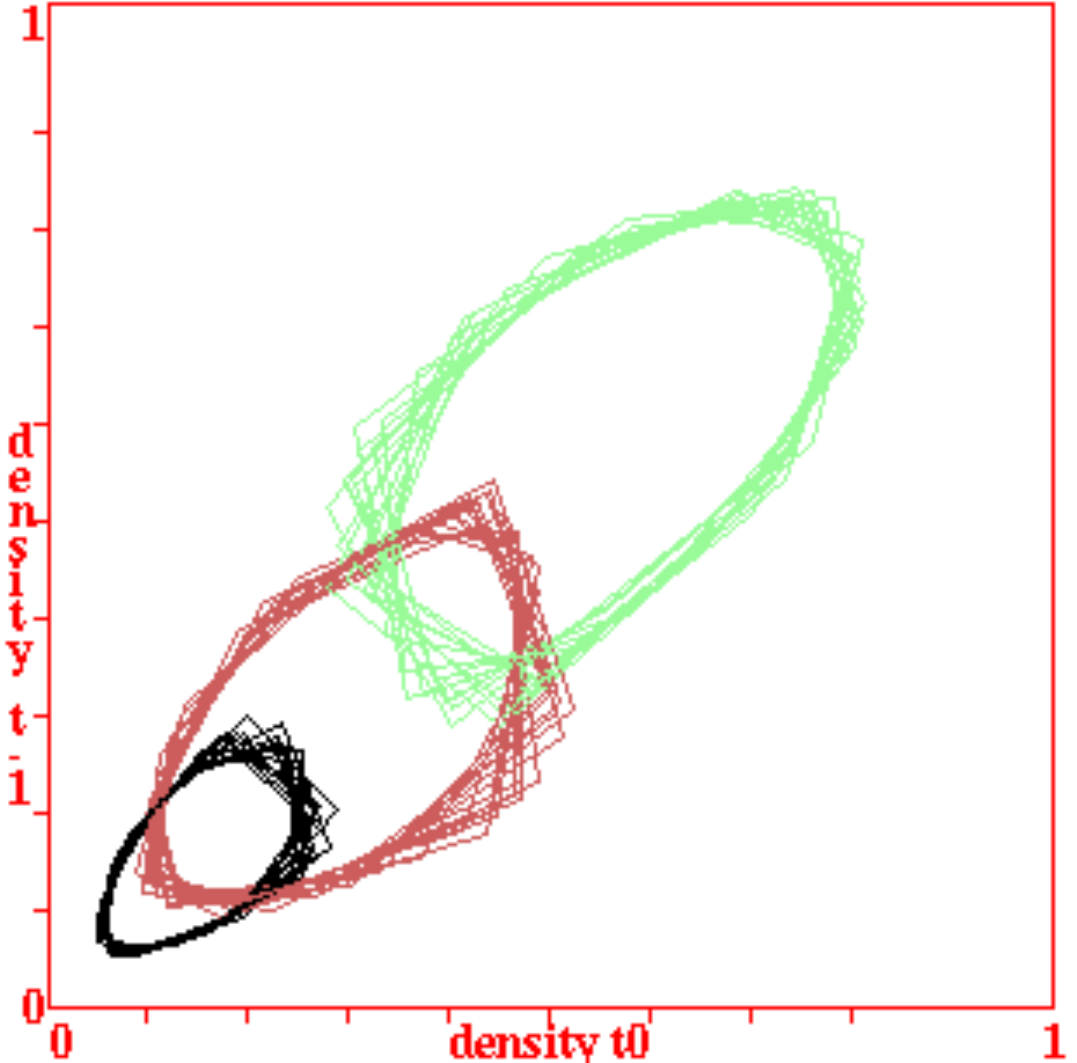}\\
(d)
\end{minipage}\\
\end{minipage}
}
\end{center}
\vspace{-3ex}
\caption[k4 (triangular) CAP plots]
{\textsf{
k4 (triangular) CAP plots, v3k4t1.vco, (hex)2a945900, g(1).
}}
\label{k4t CAP plots}
\end{figure}

\begin{figure}[!h]
\begin{center}
\textsf{\small
\begin{minipage}[c]{.9\linewidth}
\begin{minipage}[c]{.16\linewidth}
\includegraphics[width=.85\linewidth]{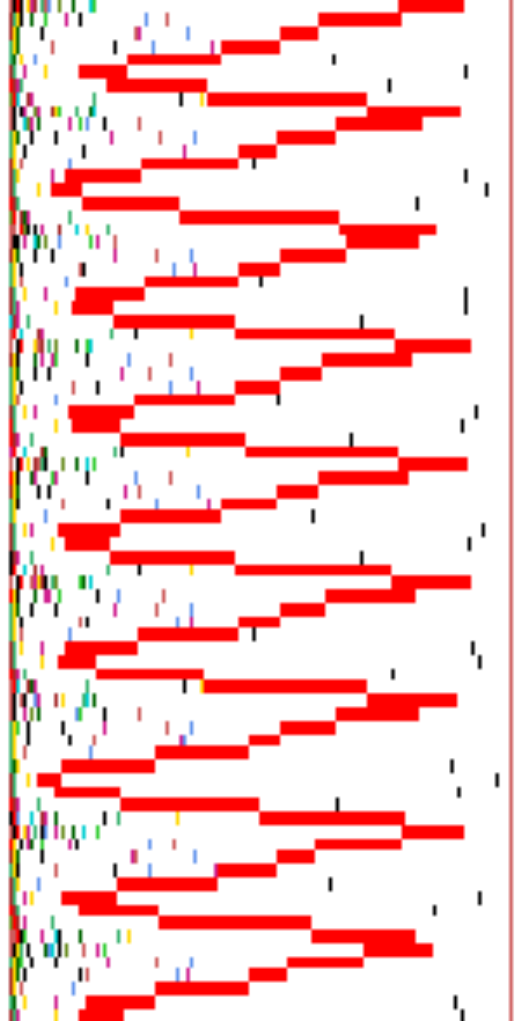}\\[1ex]
(a)
\end{minipage}
\hfill
\begin{minipage}[c]{.28\linewidth}
\includegraphics[width=1\linewidth]{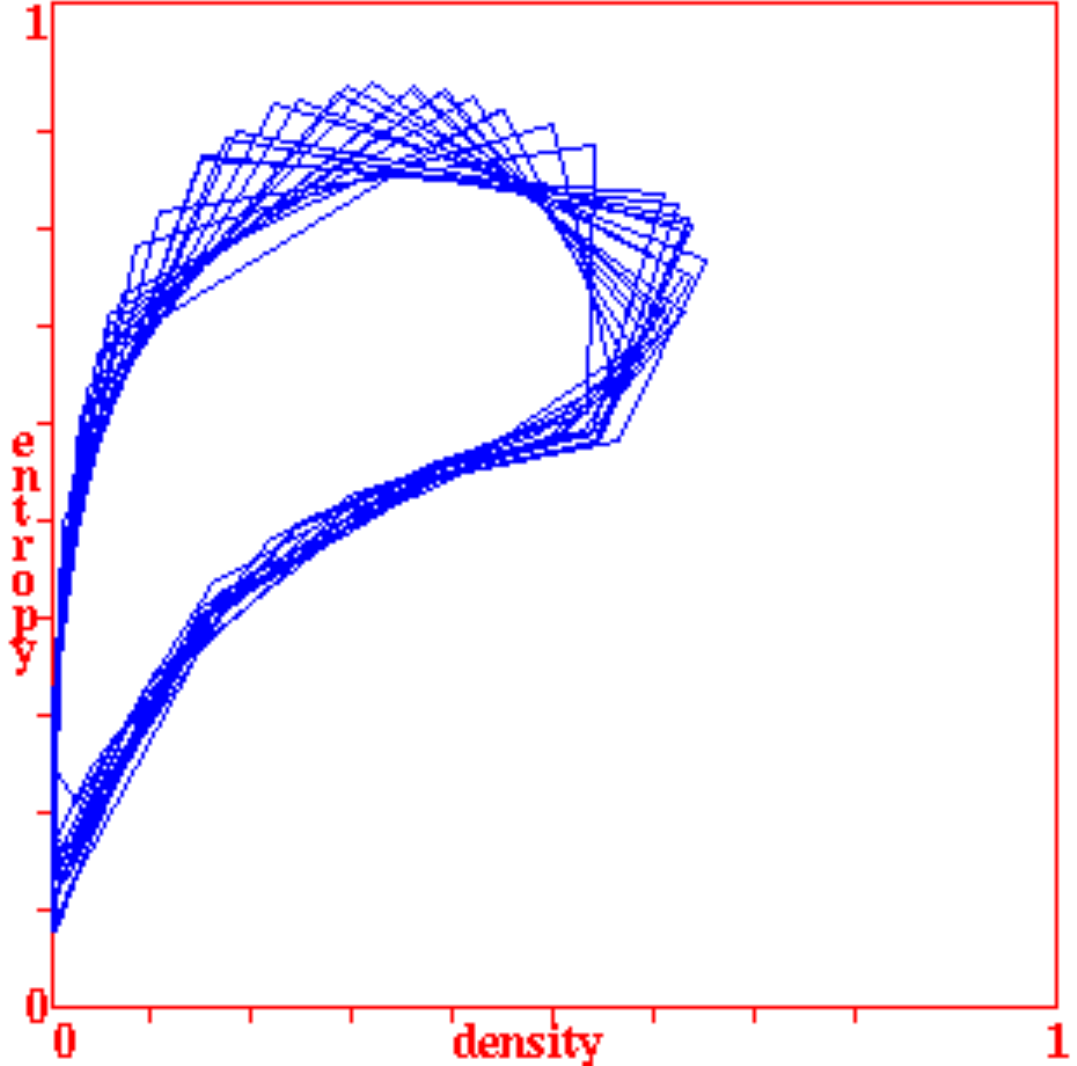}\\
(b)
\end{minipage}
\hfill
\begin{minipage}[c]{.16\linewidth}
\includegraphics[width=.85\linewidth]{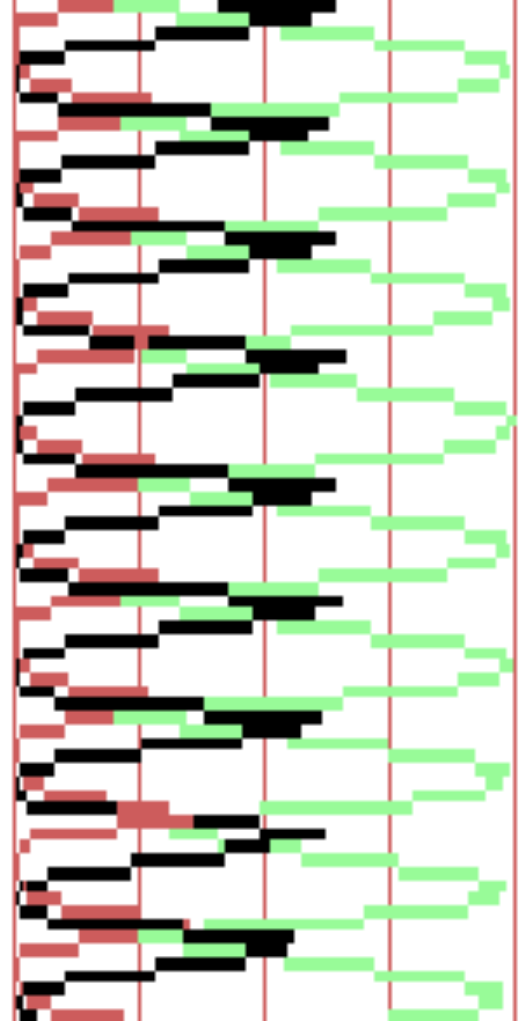}\\[1ex]
(c)
\end{minipage}
\hfill
\begin{minipage}[c]{.28\linewidth}
\includegraphics[width=1\linewidth]{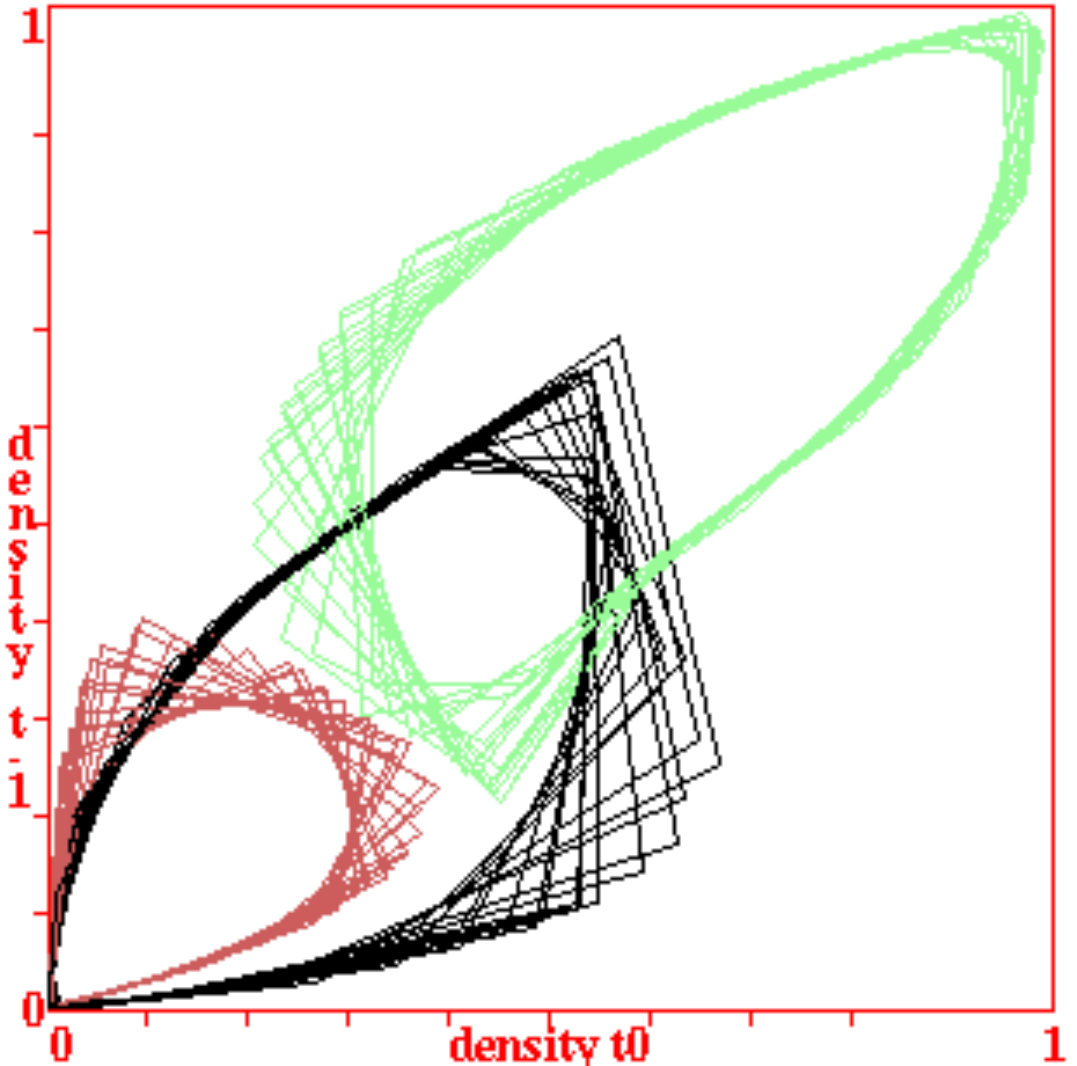}\\
(d)
\end{minipage}
\end{minipage}
}
\end{center}
\vspace{-3ex}
\caption[k4 (orthogonal) CAP plots]
{\textsf{
k4 (orthogonal) CAP plots, v3k4x1.vco, (hex)2282a1a4
}}
\label{k4s CAP plots}
\end{figure}

\begin{figure}[!h]
\begin{center}
\textsf{\small
\begin{minipage}[c]{.9\linewidth}
\begin{minipage}[c]{.16\linewidth}
\includegraphics[width=.85\linewidth]{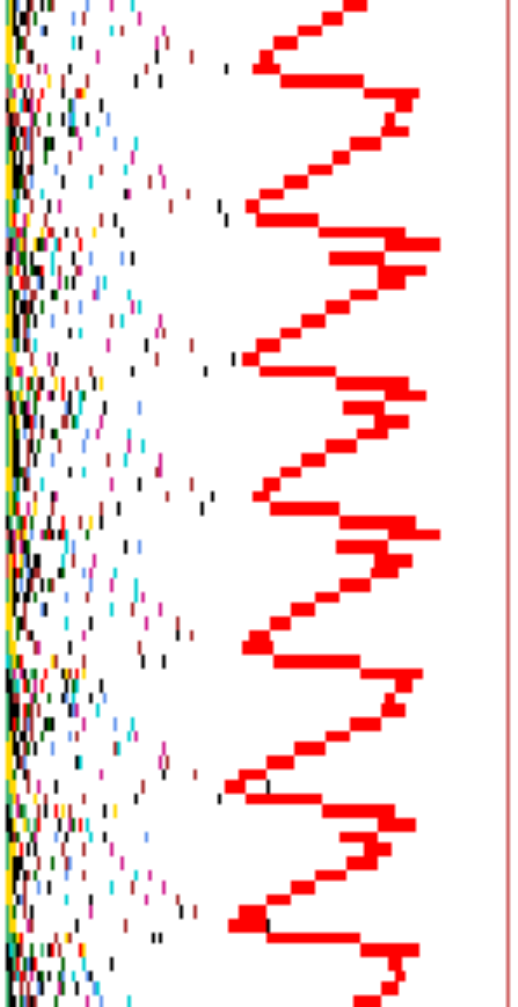}\\[1ex]
(a)
\end{minipage}
\hfill
\begin{minipage}[c]{.28\linewidth}
\includegraphics[width=1\linewidth]{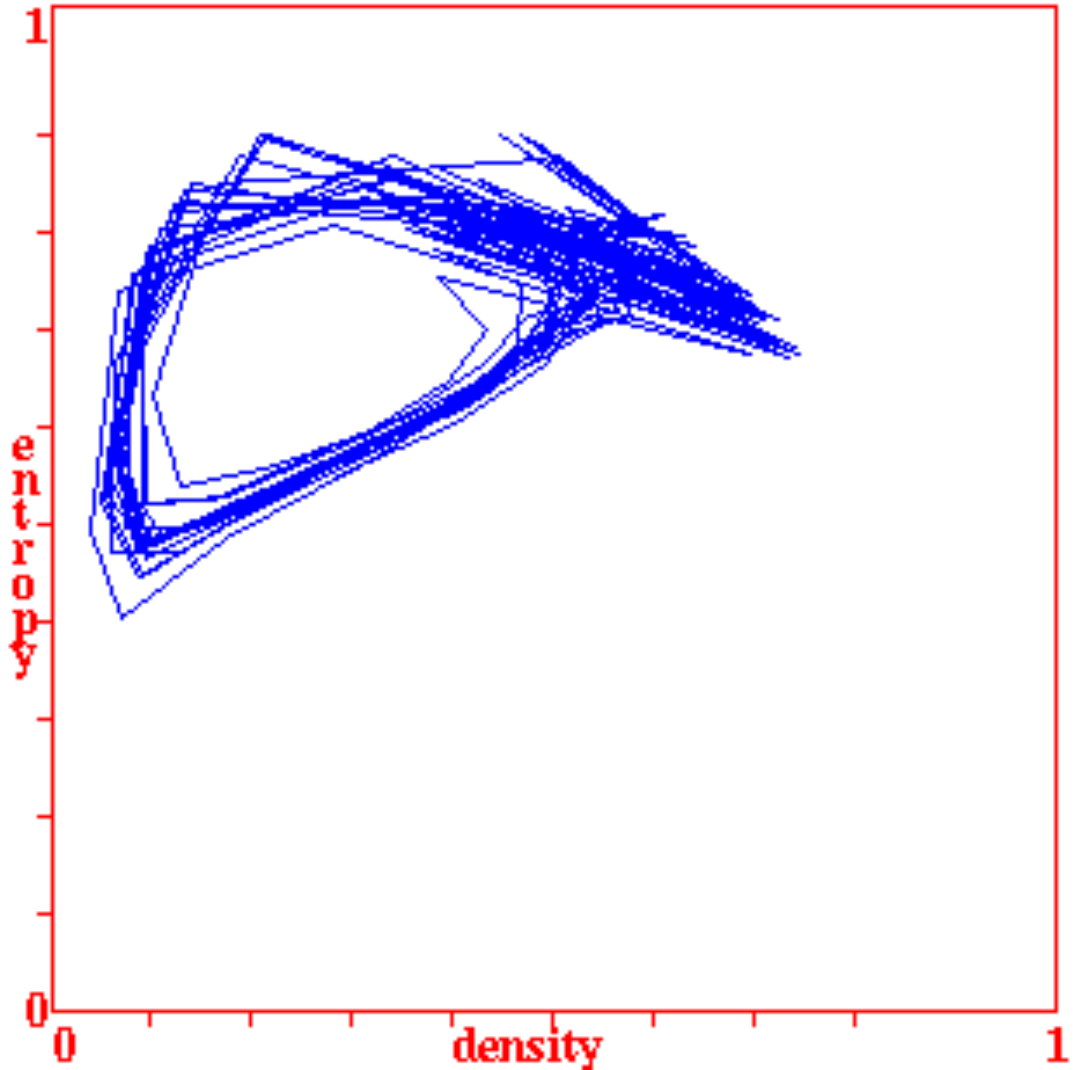}\\
(b)
\end{minipage}
\hfill
\begin{minipage}[c]{.16\linewidth}
\includegraphics[width=.85\linewidth]{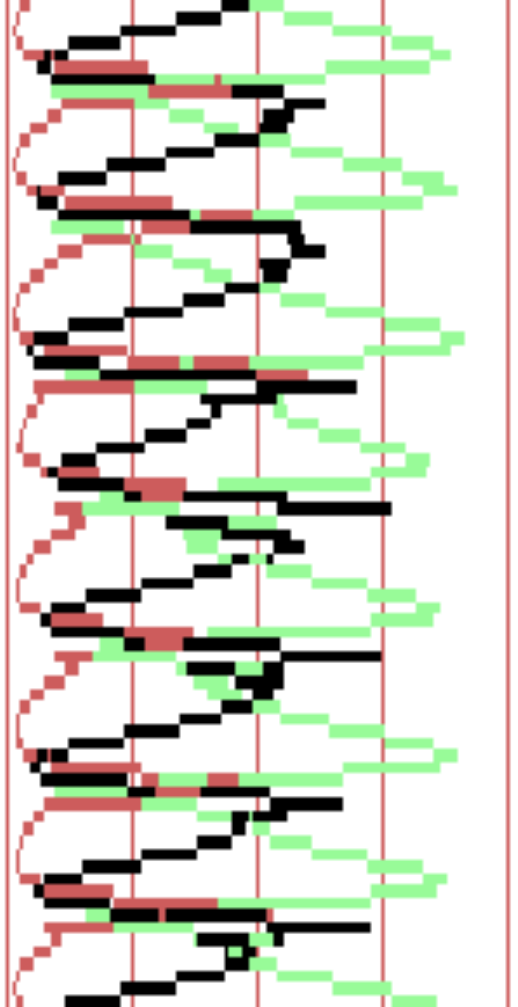}\\[1ex]
(c)
\end{minipage}
\hfill
\begin{minipage}[c]{.28\linewidth}
\includegraphics[width=1\linewidth]{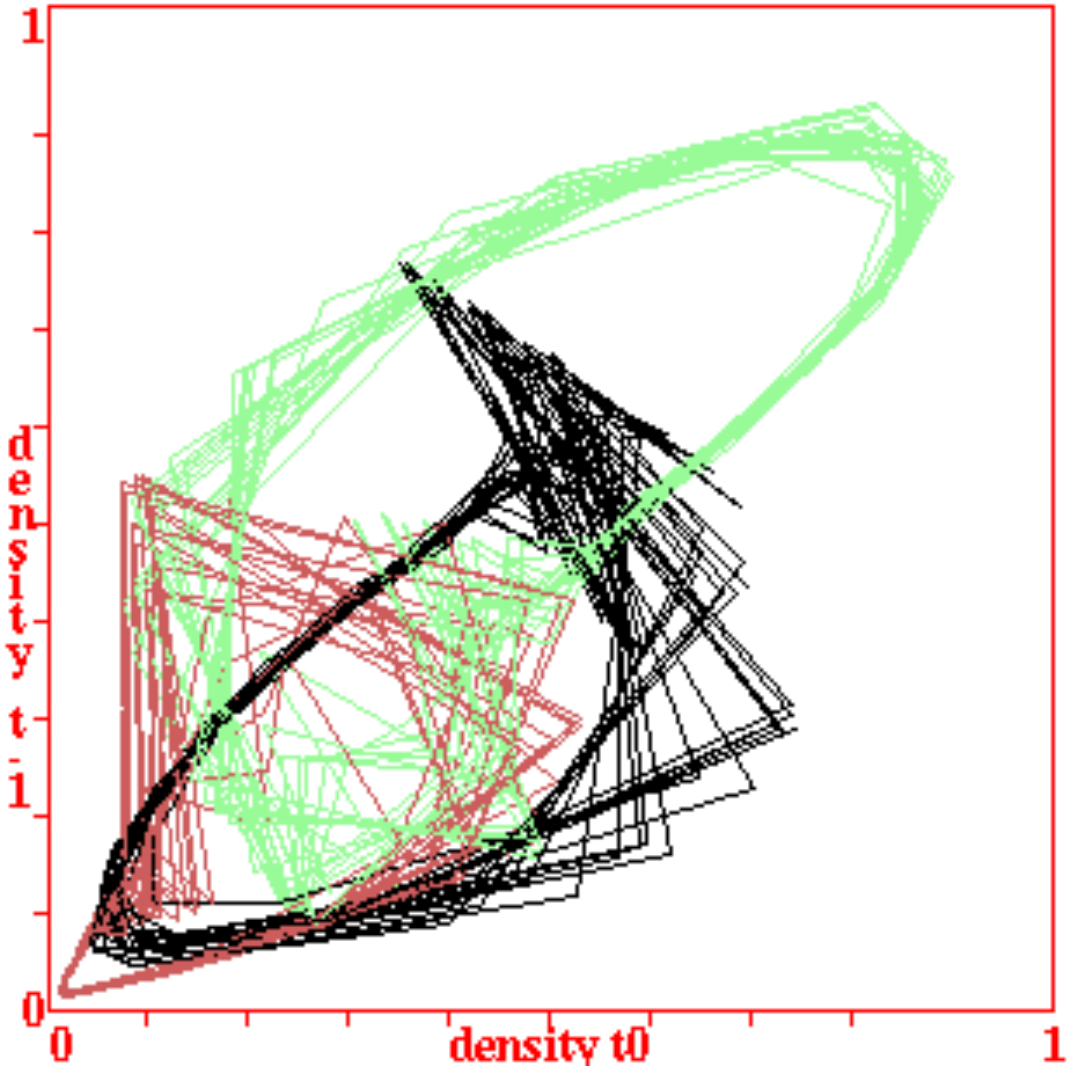}\\
(d)
\end{minipage}\\
\end{minipage}
}
\end{center}
\vspace{-3ex}
\caption[k5 CAP plots]
{\textsf{
k5 CAP plots, v3k5x1.vco, (hex)004a8a2a8254, g(1)
}}
\label{k5 CAP plots}
\end{figure}

\begin{figure}[!h]
\begin{center}
\textsf{\small
\begin{minipage}[c]{.9\linewidth}
\begin{minipage}[c]{.16\linewidth}
\includegraphics[width=.85\linewidth]{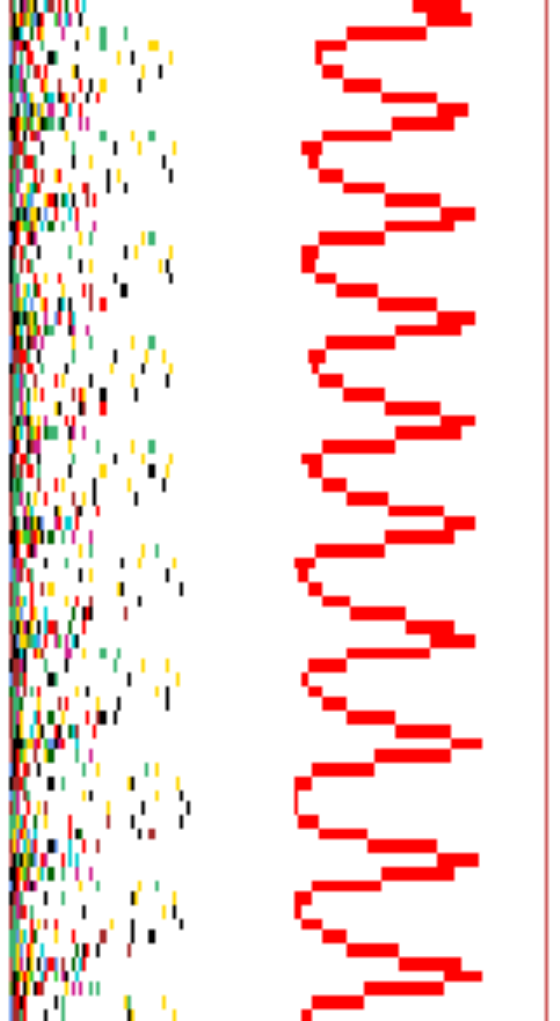}\\[1ex]
(a)
\end{minipage}
\hfill
\begin{minipage}[c]{.28\linewidth}
\includegraphics[width=1\linewidth]{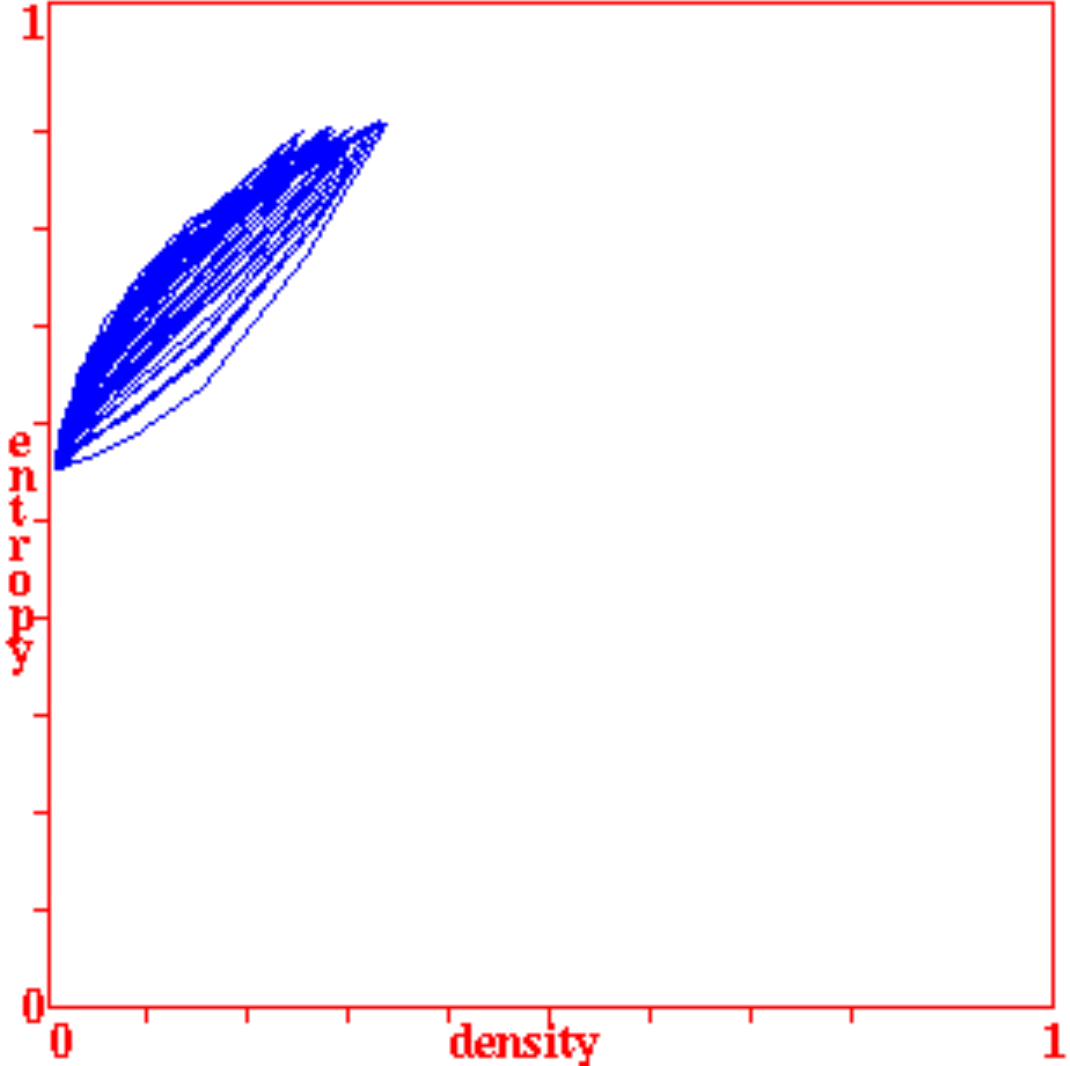}\\
(b)
\end{minipage}
\hfill
\begin{minipage}[c]{.16\linewidth}
\includegraphics[width=.85\linewidth]{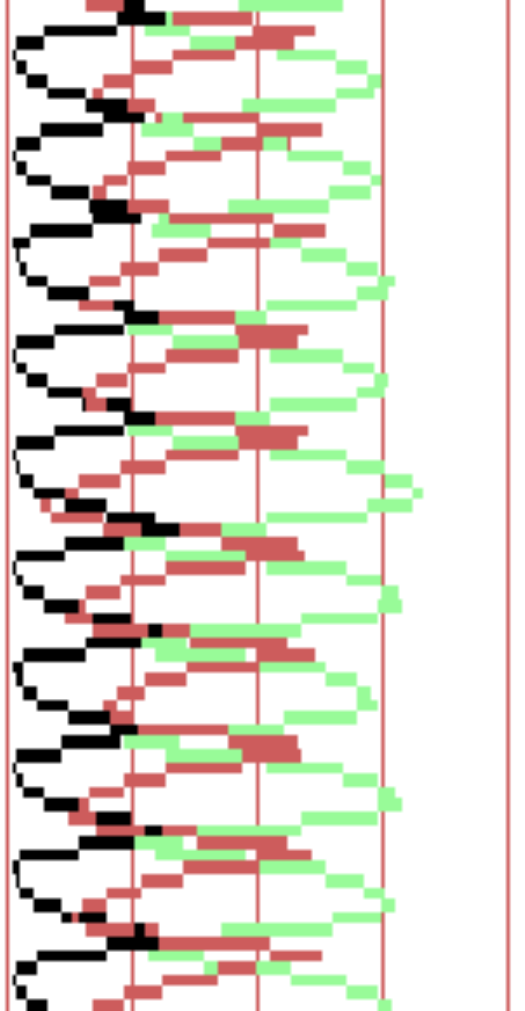}\\[1ex]
(c)
\end{minipage}
\hfill
\begin{minipage}[c]{.28\linewidth}
\includegraphics[width=1\linewidth]{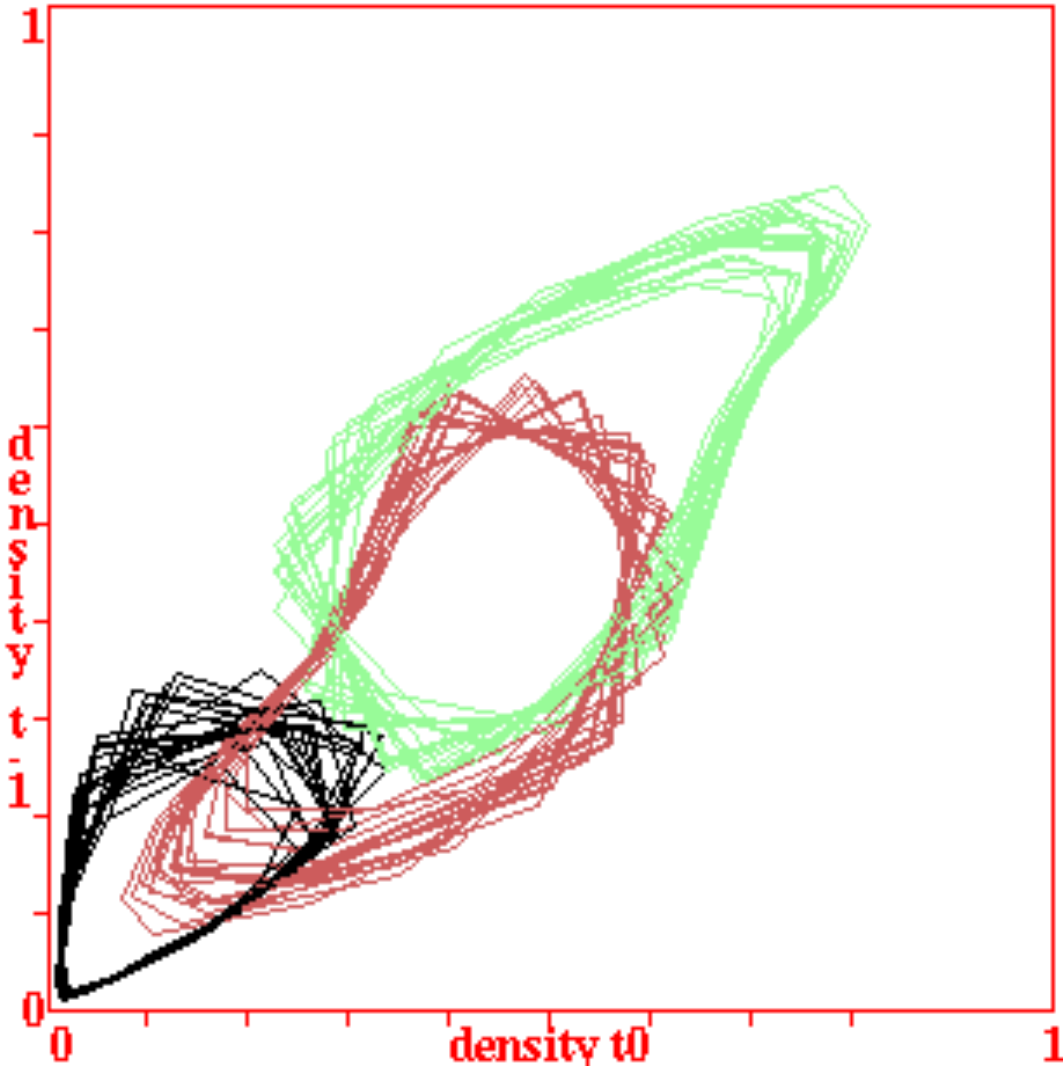}\\
(d)
\end{minipage}\\
\end{minipage}
}
\end{center}
\vspace{-3ex}
\caption[k6 CAP plots]
{\textsf{
k6 CAP plot, v3k6n6.vco, (hex)01059059560040, g(16)
}}
\label{k6 CAP plots}
\end{figure}

The CAP model input-entropy and value-density plots,
as in figs~\ref{input-entropy} and \ref{value-density},
for the $v3k7$ Spiral rule from figure~\ref{Glider examples}(f),
on a 100$\times$100 lattice,
are shown here for the rules in section~\ref{Glider dynamics in 2D CA},
for $k$=3, $k$=4 (triangular and orthogonal), $k$=5, 
and $k$=6. For each rule, four plots (a, b, c, d) described below,
are shown in figs~\ref{k3 CAP plots} to \ref{k6 CAP plots}.
The rules can be loaded in DDLab in various ways including 
by their filenames or in hexadecimal, but to explore the range of pulsing behaviours
most effectively, from the rule collections index g(x).

\begin{s-alpha}

\item Input-entropy oscillations with time (y-axis, stretched).
Left edge: superimposed histogram values plots.

\item The entropy-density scatter plot --- input-entropy (y-axis)
against the non-zero density (x-axis), for just a few pulsing cycles, 
and linking successive dots giving a time-history.

\item Value-density oscillations with time (y-axis, stretched). 0=green, 1=brown, 2=black. 
 
\item The density return-map scatter plot --- the density of each value  at $t_0$ (x-axis) 
against its density at $t_1$, plotted as colored dots for just a few pulsing cycles, 
and linking successive dots giving a time-history.

\end{s-alpha}

\section{Randomly asynchronous updating}
\label{Randomly asynchronous updating}

\begin{figure}[htb]
\begin{center}
\textsf{\small
\begin{minipage}[c]{.95\linewidth} 
\begin{minipage}[c]{.45\linewidth}
\includegraphics[width=1\linewidth]{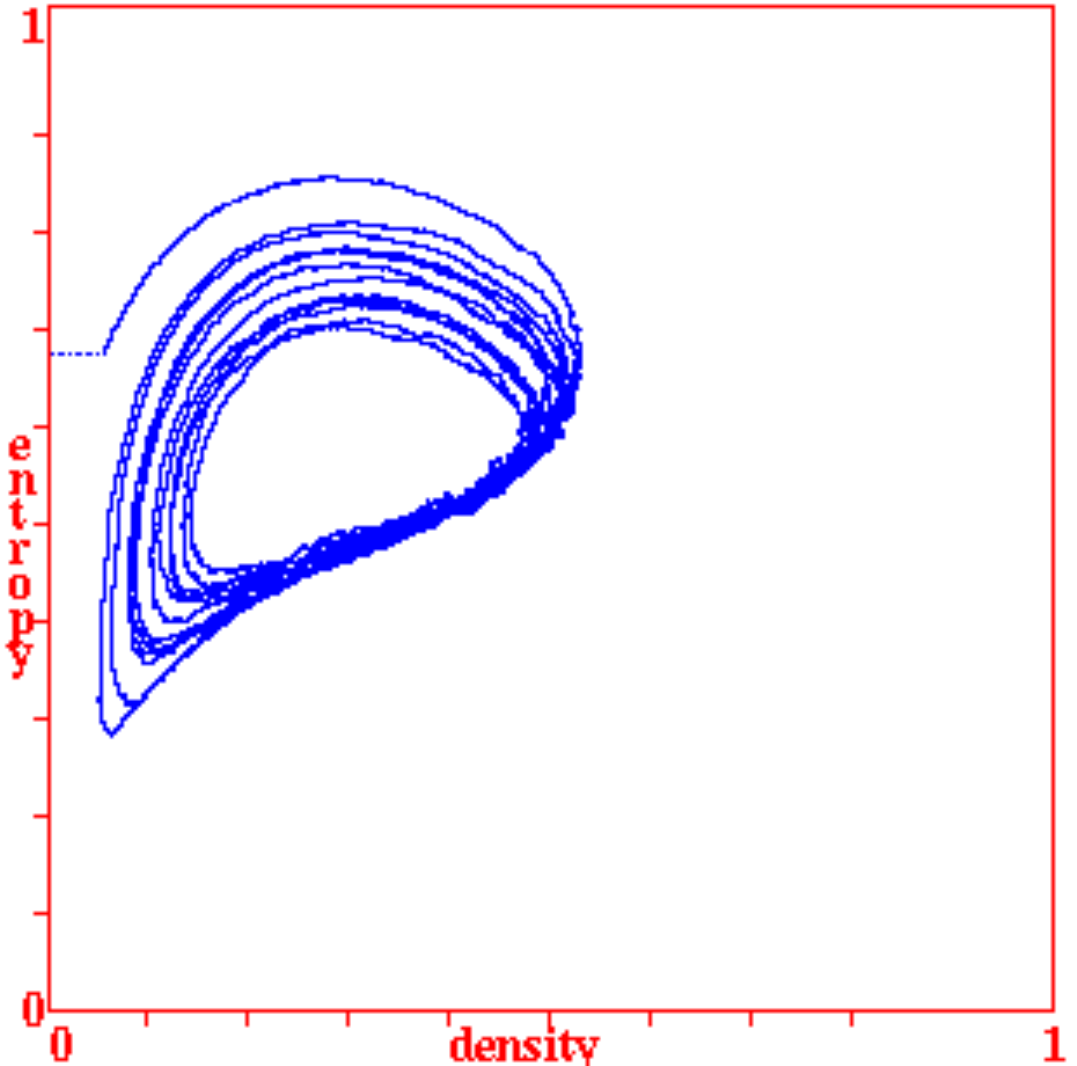}\\[1ex]
(a) $v3k7$ g3 $wl$$\approx$21 $wh$$\approx$0.6
\end{minipage}
\hfill
\begin{minipage}[c]{.45\linewidth}
\includegraphics[width=1\linewidth]{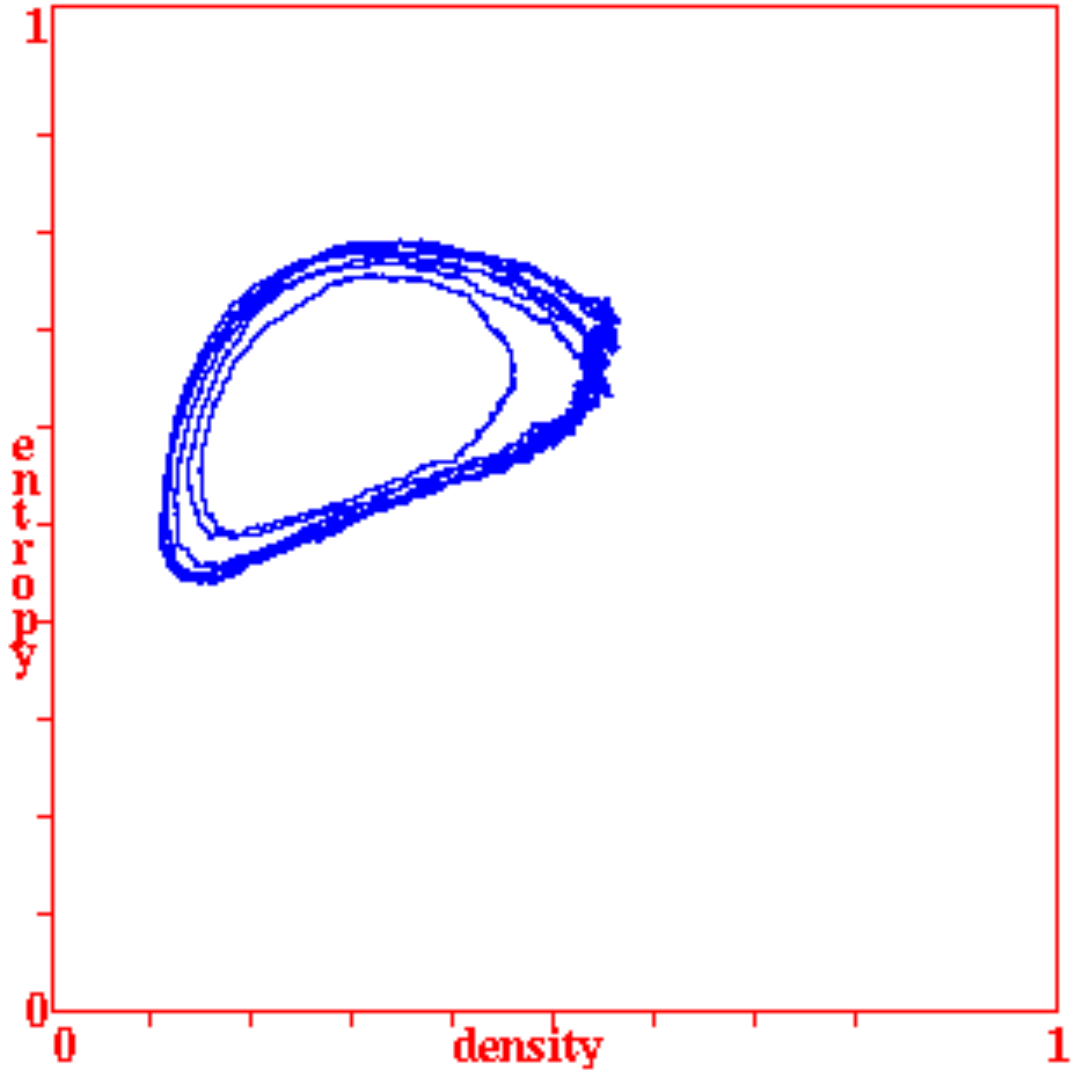}\\
(b)  $v3k7$ g35 $wl$,$wh$ see fig~\ref{Wave-length variable+jagged}
\end{minipage}
\end{minipage}
}
\end{center}
\vspace{-2ex}
\caption[one cell at a time]
{\textsf{
Entropy-density plots for sequential updating one cell at a time at random 
positions\cite{Wuensche-pulsingCA} result in pulsing. The network is 100$\times$100,
so $n$=10000 such updates are required to approximate one synchronous time-step. 
(a) $v3k7$ g3 (see fig~\ref{DDLab screen}), 
and (b)  $v3k7$ g35 (see fig~\ref{Wave-length variable+jagged}).
}}
\label{One cell at a time}
\end{figure}

Pulsing in the CAP model subject to asynchronous and noisy updating turns
out to be robust\cite{Wuensche-pulsingCA}, but perhaps the most intriguing
and unexpected result is that pulsing continues with a recognisable waveform when
random singe cells are updated one at a time. 
Experiments for $k$=6 and $k$=7 glider rules\footnote{For other rules and $k$ values,
the results are still under investigation.}  
Figures~\ref{One cell at a time}) confirm these results,  where data is plotted for each
cell update so experiments take 10000 times longer (for 100$\times$100) than
synchronous updating. To seed up computation the data can be sampled at longer intervals by
partial order or sequential updating, or a combination of both.
\clearpage

Although the CAP is a computer simulation, the fact that the pulsing
waveform is preserved for randomly sequential singe cells are updating
is significant in the sence that the CPU timer can be ruled out as an
external time-keeper.

\section{Experiments with DDLab}
\label{Experiments with DDLab}

\begin{figure}[htb]
\begin{center}
\textsf{\small
\begin{minipage}[c]{1\linewidth}
\begin{minipage}[c]{.30\linewidth}
\includegraphics[width=1\linewidth]{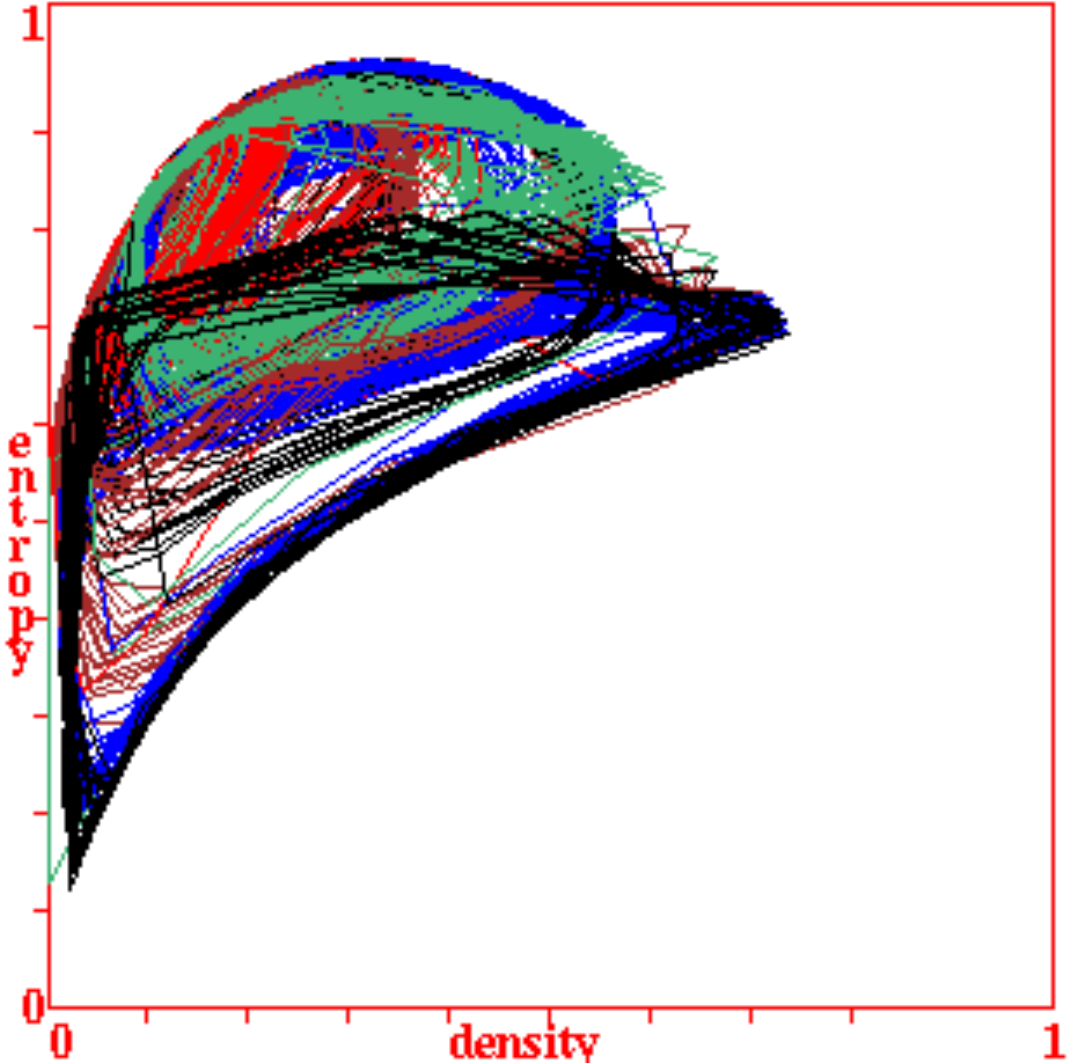}\\
$k$=3
\end{minipage}
\hfill
\begin{minipage}[c]{.30\linewidth}
\includegraphics[width=1\linewidth]{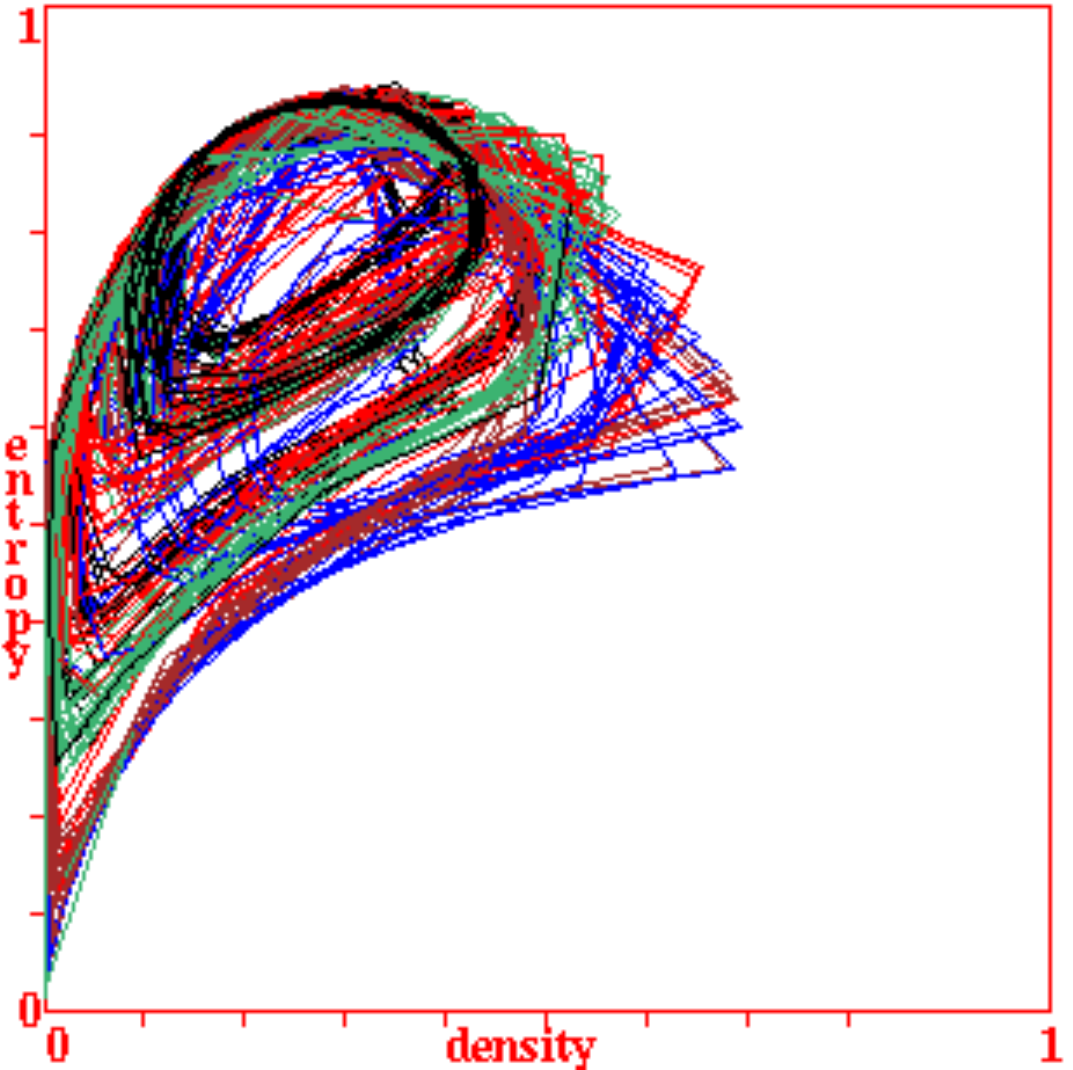}\\
$k$=4
\end{minipage}
\hfill
\begin{minipage}[c]{.30\linewidth}
\includegraphics[width=1\linewidth]{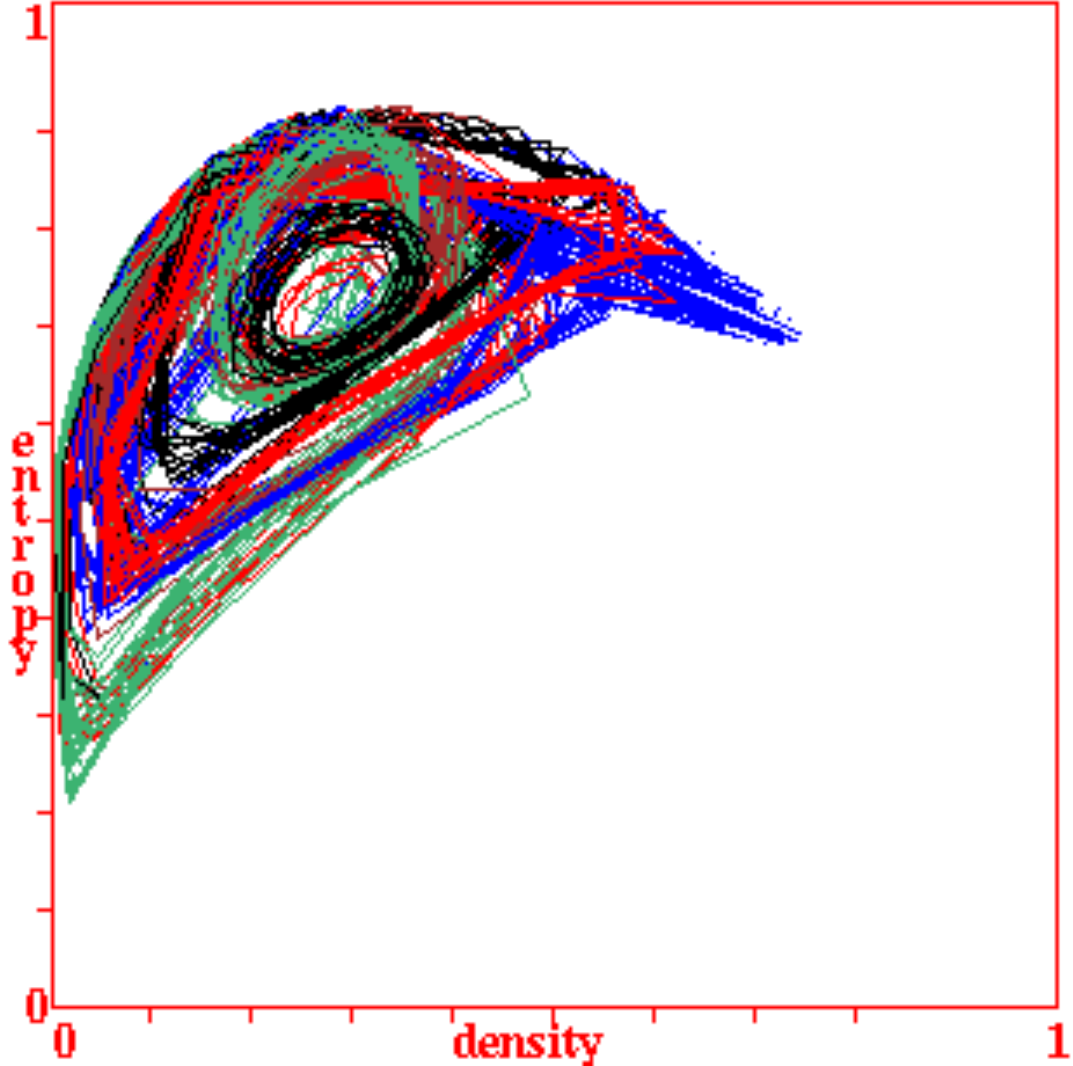}\\
$k$=5
\end{minipage}\\
\end{minipage}
}
\end{center}
\vspace{-4ex}
\caption[overlaid entropy-density plots]
{\textsf{
For $k$= 3, 4, and 5, 20 overlaid entropy-density plots
for just a few pulsing cycles each, and linking successive dots giving a time-history.
}}
\label{overlaid entropy-density plots}
\end{figure}

Using the DDLab software\cite{Wuensche-DDLab},
the results presented in \cite{Wuensche-pulsingCA}
and in this paper can be reproduced, and many other rules and aspects of pulsing dynamics
investigated. Pre-assembled collections of glider rules are available, and can be activated
on-the-fly (key {\bf g}) while space-time patterns are active, 
the wiring can be toggled between CA and random (key {\bf 7}),
and the dynamics observed, measured and recorded with other keys and interactive functions.
The older collections for $k$= 6 and 7 relating to  \cite{Wuensche-pulsingCA}
include complex rules as well as pulsing rules.
For the $k$= 3, 4, and 5 collections, only pulsing rules have been included --- 
figure~\ref{overlaid entropy-density plots}
shows an overlay of all 20 entropy-density plots for each of these rule collections. 


We summarise below the steps in DDLab to run the experiments,
referring to chapters and sections (denoted by \#x.x) from the book 
``Exploring Discrete Dynamics -- Second Edition''\cite{Wuensche2016}, --- its pdf
is kept \mbox{updated} online\cite{Wuensche-DDLab}. 
Having installed DDLab\footnote{Download the latest compiled version of
  DDLab (Nov 2018 or later)
for Linux or Mac from \url{www.ddlab.org} to
a directory called \url{ddlab}, and the extra files in \mbox{\url{dd_extra.tar.gz}}
to a subdirectory called
\url{ddlab/ddfiles} (directory names are arbitrary). \#3 gives further guidance.
For Microsoft Windows use a Linux or Mac emulator.
The DDLab code, written in C, can also be recompiled following instructions
in \url{readme} files
and \url{Makefiles} provided.},  
from a terminal in the directory \url{ddlab/ddfiles},
enter {\bf ../ddlabz07 -w \&} to start in a white screen. 

Read \#4.1 (Quick Start Examples)
for details about user input and control.
Briefly, flashing cursor prompts are presented in turn; respond with input followed
by {\bf Return} to step forward,  
{\bf q} to backtrack or interrupt. {\bf Return} without input selects a default.

\subsection{Basic steps to set up a CAP model}
\label{Basic steps to set up a CAP model}

For CAP pulsing experiments, we will
set up a 100$\times$100 2D network where $v$=3, 
and $k$=7 (for example) and totalistic rules (\#4.9.3).

\begin{figure}[htb]
\begin{center}
\includegraphics[width=1\linewidth]{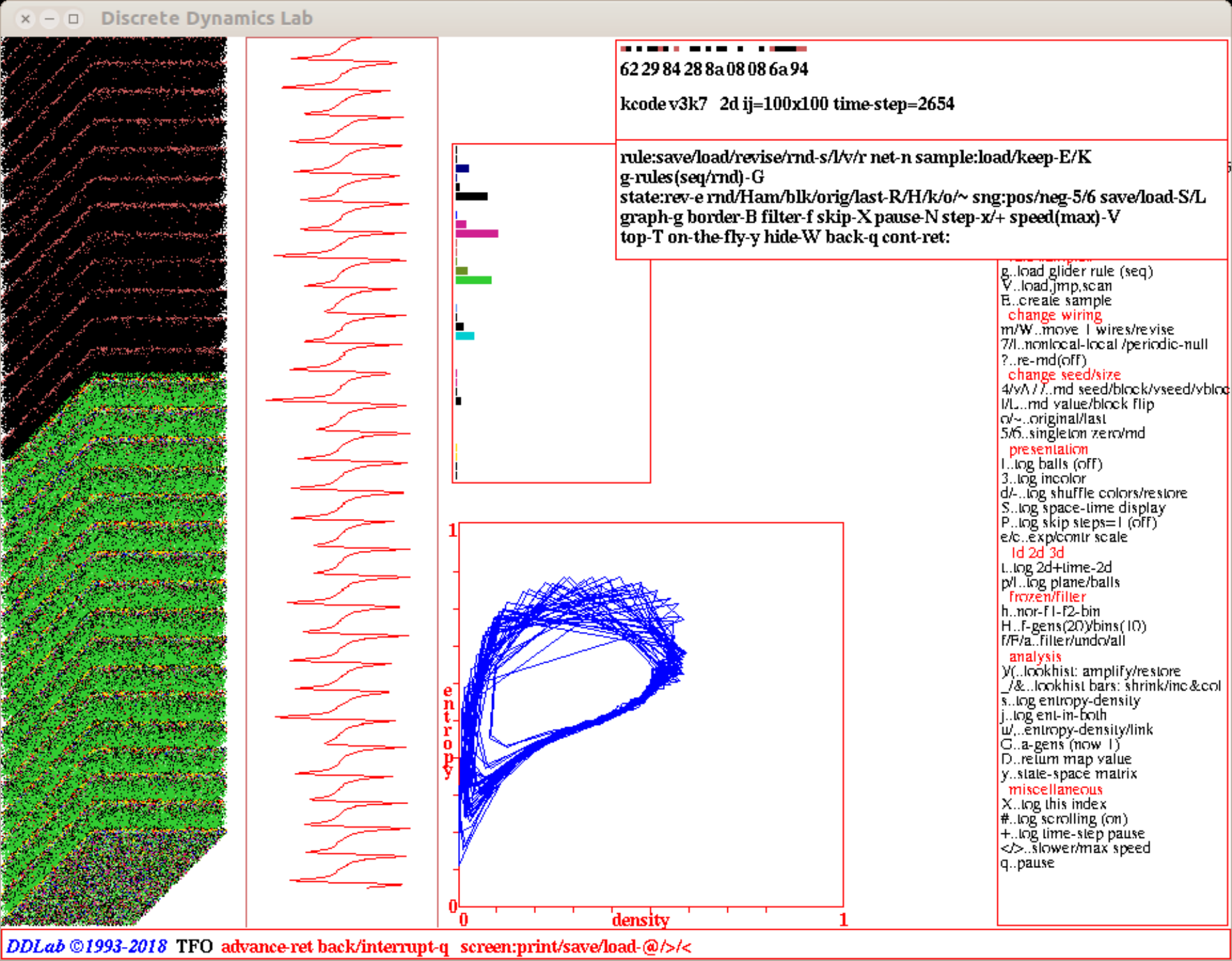}
\end{center}
\vspace{-2ex}
\caption[DDLab screen]
{\textsf{
The DDLab screen showing 2D space-time patterns, 100$\times$100, 
scrolling vertically upwards on the left.
The present time-step is at the bottom. The dark time-steps at the top are colored by value,
and below by neighborhood colors. To the right of space-time patterns
is the entropy plot and the input-histogram, below center the entropy-density
plot with linked dots.
Interrupt prompts are shown top-right, and on-the-fly reminder on the right. 
The rule is $v3k7$ g3 (hex) 622984288a08086a94.
}}
\label{DDLab screen}
\end{figure}

\begin{s-enumerate} 

\item At the first prompt, enter {\bf t} for TFO-mode (Totalistic Forward Only, \#6.2.1). 

\item At the next {\bf Value range} prompt enter {\bf 3} (\#7.1). 

\item Enter {\bf return} until a top-right {\bf WIRING} prompt window appears  (\#11.1).
Enter {\bf 2x} for hexagonal 2d ({\bf 2s} for square/orthogonal).
 
\item  At the next top-right prompt to set the 2d size (\#11.6.1) enter {\bf 100}
 for both {\bf i} and {\bf j}.

\item At the next top-right prompt, {\bf Neighborhood size k:} (\#11.7) enter~{\bf 7}.

\item At the next top-right {\bf 2d network ... wiring} prompt (\#17.1), enter {\bf 2} 
to display a 2D wiring graphic bottom left, showing CA neighborhoods,
together with a top right reminder (\#17.4).

\item Now set random wiring: enter {\bf b} for a 2d block (\#17.7.5), 
then {\bf a} to outline the ``block'' as the whole network (\#17.4 reappears),
then {\bf r} for random, which shows links for the
highlighted cell as in figure~\ref{Random wiring}(a). 
Click other locations to show the wiring for other cells.

\item Enter {\bf return} until the top-right prompt
  \color{BrickRed}{\bf revise from:}\color{black} (\#31.1), enter
  {\bf e}, then at the
  \color{BrickRed}{\bf entropy/density: }\color{black} prompt (\#31.5), enter
  {\bf e} again.

\item At the next top-right prompt, enter {\bf d} to skip further
  special options and start the 2d space-time patterns iterating in
  the top-left of the screen for a random rule, with the input-entropy
  plot (averaged for a moving window of 10 time-steps) and the
  input-histogram alongside. The rule was set at random so
  oscillations are unlikely.  To revise any of the above, backtrack
  with {\bf q}.
\end{s-enumerate}
\vspace{-3ex}

\subsection{Space-time patterns on-the-fly options}
\label{Space-time patterns on-the-fly options}

While 2D space-time patterns are running,
on-the-fly key-press options may be activated/toggled (\#32.3) ---
a reminder appears on the right of the screen.
To pause (or backtrack) at any time key-press {\bf q}, which gives a top-right prompt 
window with further options (\#32.16). 
For the CAP model experiments, the following options are the most relevant.

\begin{s-enumerate}

\item For a random rule from a glider 
  collections\footnote{In the $k$=6 and $k$=7 collections,
    not all rules are glider rules, so not all will pulse.},
key-press {\bf g} (\#32.6.1).

\item For a specific rule (thereafter consecutively) enter {\bf q} to 
backtrack to the pause prompt (\#32.16).
\begin{s-itemize}
\item enter {\bf G} for a top-right prompt showing the number of rules in the collection (\#31.2.9).
\item enter {\bf 1}, or any valid number, to select the rule (not yet activated).
\item once space-time patterns resume, key-press {\bf g} to activate the rule.
\item For the next rule enter {\bf g}, eventually cycling back to rule 1.
\end{s-itemize}

\item Waveform output is most pronounced when set to each single time-step, 
not to the default 10 time-step moving average. To change this, 
key-press {\bf G} and at a top-right prompt (\#32.12.7) enter enter {\bf 1}.

\item Important on-the-fly key-presses:
\begin{s-itemize}
\item toggle with {\bf 7} between random-wiring and regular CA.
\item try the 3-way toggle {\bf j} for input-histogram values plotted together with the input-entropy plot.
\item key-press {\bf 4} for a new random initial state. This is also required if the dynamics
stops --- reaches and attractor\cite{Wuensche92}.
\item key-press {\bf c} and {\bf e} to contract/expand space-time patterns. Contract to allow
more room for the entropy-density scatter plot below.
\end{s-itemize}

\item Toggle showing the entropy-density scatter plot with {\bf u} (\#32.12.6). While this is running,
\begin{s-itemize}
\item key-press {\bf ,} (comma) to toggle linking successive dots as in figure~\ref{input-entropy}(c).
\item key-press {\bf ?} (question mark)  to toggle re-randomising at each time-step, which slows down iteration.
\item key-press {\bf ``} (inverted coma) to toggle showing running data of the wave-length (wl) 
  and wave-height (wh) in the terminal as in section~\ref{Wave-length and wave-height data}.
  To save the data refer to \#32.12.6.2.
\end{s-itemize}

\item  To change to value-density instead of entropy, toggle with {\bf s}. 
The input-histogram (and entropy-density scatter plot) will stop updating. 
\begin{s-itemize}
\item  key-press {\bf ;} (semi-colon) to toggle showing the density return-map (\#32.12.7).
\item key-press {\bf ,} (comma) to toggle linking successive dots as in figure~\ref{value-density}(c).
\item To save the density return-map data refer to \#32.12.7.2.  
\end{s-itemize}

\item Other useful on-the-fly key-presses: 
\begin{s-itemize}
\item key-press {\bf 3} to toggle space-time patterns colored by value
or by neighborhood/input-histogram colors.
\item key-press {\bf t} to toggle between a 2D movie and 2D vertical space-time patterns as in
figure~\ref{DDLab screen}, then {\bf \#} to toggle upward scrolling.
\item when in normal 2D, key-press {\bf \#} to toggle space-time patterns scrolling diagonally.
\item key-press \boldmath{$<$} to slow down iteration, \boldmath{$>$} to revert to normal speed.
\end{s-itemize} 

\item Key-press {\bf q} to pause at any time for a top-right window providing options described in
\#32.16, including,
\begin{s-itemize}
\item  enter net-{\bf n} to re-set the random wiring (\#17.7) as in
section~\ref{Basic steps to set up a CAP model}(7), or with biases as in figure~\ref{Random wiring}.  
\item key-press {\bf q} as necessary to backtrack up the DDLab prompt sequence.
\item key-press {\bf d} to save wave-length data,
  or density return-map data (\#32.12.6.2 and \#32.12.7.2). 
\end{s-itemize} 

\end{s-enumerate}

\subsection{One cell at a time sequential updating}
\label{One cell at a time sequential updating}

To run an entropy-density plot for sequential updating one cell at a time at random 
positions\cite{Wuensche-pulsingCA}, as for the $k$=7 rules in figure~\ref{One cell at a time},
amend steps in section~\ref{Basic steps to set up a CAP model} as follows,

\begin{s-itemize}

\item After step (7) enter {\bf return} until the top-right prompt 
\color{BrickRed}{\bf revise from:}\color{black}\\
(\#31.1), enter {\bf u} for a top-right updating window (\#31.4).

\item Enter {\bf p} for a top-right partial order updating window (\#31.4.3).
For both {\bf min:} and {\bf max:}, enter {\bf 1}.
\item at the \color{BrickRed}{\bf entropy/density: }\color{black} 
prompt (\#31.5), enter {\bf e} as in step (8).

\end{s-itemize}

Follow further steps as listed, then toggle showing the entropy-density scatter plot with {\bf u}
as in step (5) in sections~\ref{Space-time patterns on-the-fly options}.
You will see single-cell updates in the 2D pattern, and a very slow trace of the plot.
For other ``asynchronous and noisy updating'' options refer to (\#31.4).   
There can be any combination of these settings, which can be toggled on-the-fly (\#32.4).

\section{Issues to explain the CAP model}
\label{Issues to explain the CAP model}

Explaining the CAP model is work in progress.  The emergence of
gliders in CA cannot be predicted directly from a rule-table, so in this
sense the mechanisms are unresolved --- they can only be observed by experiment, but 
must entail feedbacks driving a glider's head and eroding its tail.
A general theory to resolve this question would shed light on the underlying principles of
self-organisation. Pulsing when the wiring is randomised must utilise the same feedbacks, 
but distributed throughout the network instead of
localised in a regular neighborhood to create and move a glider.

Future work should also address the following issues,

\begin{s-itemize}
\item Because the pulsing waveform (shape/phase, wavelength, waveheight) is observed to be diverse,
how does the type of glider dynamics relate to the waveform.
\item Study how the pulsing waveform breaks down as the random wiring reach is reduced ---
is there a phase transition?
\item There is a very high probability that gliders imply pulsing, 
but the few observed exceptions should be examined,
gliders/no pulsing, and pulsing/no gliders.
\item Study how the sequential updating of one cell at a time at random 
positions\cite{Wuensche-pulsingCA}
can results in pulsing.
\end{s-itemize}

\section{The need for a bio-oscillation model}
\label{The need for a bio-oscillation model}

In a world where much biology is produced by reproducing patters,
produces reproducing patterns, or recognises these patterns, there has
historically been focus on the chemistry and physics of thermodynamic
equilibrium more so than on the bio-physics of collective oscillatory
phenomena. Oscillations can be found in all forms of life\cite{cao},
but we have focused on mammalian biology, and aspects of human
physiology where oscillations play a crucial
role\cite{Wuensche-pulsingCA}.

Athough differential equation models of ocsillations in single cells
have been proposed, such as the Hodgkins Huxley equations, negative
feedback with a time delay, or coupled negative and positive
feedback\cite{TunableBO}, currently there is no satisfactory theory to
explain essential oscillations in whole organs, for example the heart
beat, uterine contractions in childbirth, and various rhythmic
behaviours controlled by the central pattern generators of the central
nervous system, such as breathing and
locomotion\cite{Wuensche-pulsingCA}.

We are left searching for pacemaker neurones, pondering how signalling
in biofilms can occur faster than diffusion, how synchronisation can
occur over considerable distance and how biology is so robust with
such inbuilt redundancy.

We propose that clusters of excitable tissue are able to oscillate
according to a their appropriate waveform because non-localised
connectivity\cite{TTUBE,TERM} is subjected to a specific rule of
communication, the bio-rule, analogous to a glider rule. The bio-rule
is based on three (or more) cellular states, firing, refractory, and
ready to fire, generated by chemical signalling\cite{CaSPARKS}, action
potentials, calcium and sodium ion channels and concentrations.  This
model is favoured by evolution because a variety of synchronized
waveforms can arise from different bio-rules. Furthermore, a given
waveform is independent of the exact connection network, is noise
tolerant, can be turned off and on by altering the reach of non-local
connectivity, and is robust to noise, cell loss, and functional
reserve.

The CAP model carries the ability and benefits of modelling multiple
cells simultaneously forming a platform to probe the relationship
between network connectivity at one level and collective behaviour at
another.  This summery of the more detailed reasoning presented in
\cite{Wuensche-pulsingCA} suggests that the CAP model is applicable as
a conceptual model for bio-oscillations and can provide a basis for
further development of the ideas.

\section{Summery}
\label{Summery}

We have presented further evidence and results for this surprising
phenomenon of spontaneous, sustained and robust rhythmic oscillations,
pulsing dynamics, when random wiring is applied to a 2D ``glider''
rule running in a 3-value totalistic CA.

We have reiterated the potential of the CAP model to provide a much
needed decentralised model for bio-oscillations in nature, specifically
in the case of mammalian excitable tissue.

We have defined the system's architecture, and identified the
behaviour for both glider dynamics and pulsing, which are intimately
related, and noted the issues that require explanation. A guide to the
relevant functions in the software DDLab to repeat and extend pulsing
experiments is provided.  However, the underlying mechanisms remain
unresolved and are open to further study and research.

\section{Acknowledgements}
\label{Acknowledgements}

Experiments and figures were made with DDLab (\url{http://www.ddlab.org/}) ---
where the rules and methods are available, so repeatable\cite{Wuensche2016}.
Thanks to \mbox{Inman Harvey} for conversations regarding asynchronous updating,
to Terry Bossomaier for exchanges regarding phase transitions,
and to \mbox{Paul Burt} and \mbox{Muayad Alasady} for comments regarding bio-oscillations.

\end{document}